%% file: drafts-v2.tex
\documentclass[
aps,
prx,
reprint,
superscriptaddress,
longbibliography,
floatfix,
nofootinbib
]{revtex4-2}
\input{andi_thesis_preamble.tex}
\usepackage{amsmath,amssymb,amsthm}
\usepackage{braket}
\usepackage{xcolor}
\usepackage{comment}
\usepackage{algpseudocode}
\usepackage{longtable}
\usepackage{tikz}
\usetikzlibrary{decorations.markings,decorations.pathreplacing,arrows.meta}
\usepackage[colorlinks=true,allcolors=blue!60!black]{hyperref}
\makeatletter
\def\place@bibnumber#1{\textcolor{blue!60!black}{[#1]}}
\makeatother
\usepackage{bbold, bm, mathtools}
\usepackage{mathrsfs}%
\usepackage{enumitem}

\newtheorem{theorem}{Theorem}
\newtheorem{lemma}{Lemma}
\newtheorem{claim}{Claim}
\newtheorem{proposition}{Proposition}
\newtheorem{definition}{Definition}
\newtheorem{corollary}{Corollary}

\makeatletter
\newcommand{\slabel}[2]{\def\@currentlabel{#1}\label{#2}}
\makeatother

\newcounter{algorithm}
\makeatletter
\newenvironment{algorithm}[1][]{%
  \refstepcounter{algorithm}%
  \@beginparpenalty=10000\relax
  \@endparpenalty=10000\relax
  \par\medskip
  \hrule height 0.6pt\relax
  \nobreak\vspace{0.45ex}%
  \renewcommand{\caption}[1]{%
    \noindent\textbf{Algorithm \thealgorithm.} ##1\par\nobreak
    \vspace{0.35ex}\hrule height 0.4pt\relax\nobreak\vspace{0.45ex}%
  }%
}{%
  \par\nobreak\vspace{0.35ex}\hrule height 0.6pt\relax
  \par\medskip
}%

\makeatother

\algrenewcommand\algorithmicrequire{\textbf{Input:}}
\algrenewcommand\algorithmicensure{\textbf{Output:}}

\newcounter{appendixproof}

\newcommand{\intG}{\mathbb G}

\DeclareMathOperator{\rad}{rad}

\newcommand{\C}{\mathsf{C}}
\renewcommand{\S}{\mathsf{S}}
\newcommand{\G}{\mathsf{G}}
\newcommand{\F}{\mathsf{F}}
\newcommand{\R}{\mathsf{R}}
\newcommand{\T}{\mathsf{T}}%
\newcommand{\D}{\mathsf{D}}%
\newcommand{\BN}{\ensuremath{\mathscr{B}}}%
\newcommand{\SB}{\BN^{\S}}%
\newcommand{\CB}{\BN^{\C}}%
\newcommand{\GB}{\BN^{\G}}%
\newcommand{\Cong}{\mathsf{Cong}}
\newcommand{\RB}{\BN^{\R}}%
\newcommand{\RT}{t^{\R}}%

\newcommand{\LB}{\mathsf{LB}}
\newcommand{\LD}{\mathsf{LD}}%

\newcommand{\abs}[1]{\lvert #1\rvert}

\definecolor{Crimson}{RGB}{220,20,60}
\definecolor{Teal}{RGB}{0,128,128}
\definecolor{DarkMagenta}{RGB}{139,0,139}
\definecolor{SyndromeDarkBlue}{RGB}{0,92,165}
\definecolor{ReconnectCZColor}{RGB}{96,72,150}
\definecolor{FableRose}{RGB}{225,60,120}

\usepackage{xcolor}

\newif\ifABnotes \ABnotestrue %

\newcommand{\Fail}{\mathsf{Fail}}

\def\pfusion{P_{\rm fusion}}
\newcommand\myparagraph[1]{\noindent\textbf{#1.}}%
\begin{document}
\title{Provable circuit-level thresholds for non-abelian topological phases}
\author{Yuanjie (Collin) Ren}
\affiliation{Department of Physics, Massachusetts Institute of Technology, Cambridge, Massachusetts 02139, USA}

\author{Margarita Davydova}
\affiliation{Walter Burke Institute for Theoretical Physics and Institute for Quantum Information and Matter, California Institute of Technology, Pasadena, CA 91125, USA}

\author{Shankar Balasubramanian}
\affiliation{Walter Burke Institute for Theoretical Physics and Institute for Quantum Information and Matter, California Institute of Technology, Pasadena, CA 91125, USA}

\author{Andreas Bauer}
\affiliation{Department of Mechanical Engineering, Massachusetts Institute of Technology, Cambridge, MA 02139, USA}
\affiliation{IBM Quantum, IBM T. J. Watson Research Center, Yorktown Heights, NY 10598, USA}
\email{andreas.bauer@ibm.com}

\begin{abstract}
We provide an explicit scheme to realize a quantum memory in any 2D non-chiral topological order. For this, we construct syndrome extraction circuits based on the Turaev--Viro path integral as well as a classical algorithm for decoding on the fly.  We prove that our scheme has a threshold under a local stochastic circuit-level noise model and discuss how to generalize it to implement topological logical gates. 
\end{abstract}

\maketitle
\tableofcontents

\section{Introduction}
\myparagraph{Background}
Topological quantum error correction has been a promising route towards fault-tolerant universal quantum computation due to its 2D connectivity requirements and high thresholds.
Historically, there have been two routes towards topological quantum computation.
The first direction is to use a simple \emph{abelian} topological phase as a quantum memory, such as the toric code~\cite{Kitaev_2003,dennis2001}.
This phase can be implemented as a Pauli stabilizer code, and therefore has a low-depth syndrome extraction circuit using 2-qubit $CNOT$ gates and single-qubit Pauli measurements, as well as  efficient decoders with provable thresholds, such as minimum-weight matching.
Logical Clifford operators can be performed using topological methods, such as transversal gates, code deformation, or lattice surgery~\cite{horsman, bombin2009quantummeasurementsgatescode, fowler2012surfacecodespracticallargescale}, but non-Clifford gates generally require non-topological methods, for example, magic state distillation~\cite{bravyi2004universalquantumcomputationideal}.

The second direction is to use a more complicated \emph{non-abelian} topological phase.
The logical information can then be encoded into a ground state with prescribed anyon locations, and for special kinds of topological phases, universal quantum computation can be performed purely by braiding these anyons~\cite{Kitaev_2003, freedman2000modularfunctoruniversalquantum, nayak2008nonabeliananyonstopologicalquantum}.
In addition to braiding, computation can be performed by any 3D spacetime diagram describing some deformation of a 2D geometry that includes defects, such as boundaries, domain walls, corners between boundaries, anyons, or twist defects where domain walls end.

More recently, it was realized that a special class of ``solvable'' non-abelian topological phases, which can be obtained by applying a procedure called ``gauging'' to certain abelian topological phases, can inherit many features of error correction with abelian phases.
In particular, these non-abelian phases admit relatively efficient syndrome extraction circuits and decoders, and thresholds have been specifically proven for $D_4$ and $S_3$ quantum doubles~\cite{davydova2025universalfaulttolerantquantum, lyons2026quantumcomputinganyonsfault}.  In addition, when interfaced with toric codes, these non-abelian phases can be used to achieve an alternative implementation of non-Clifford gates, which provides a promising alternative to distillation methods. 

However, the fault-tolerance aspect of memory and computation is much less understood for general non-abelian topological phases that are not necessarily ``solvable''.
In particular, no rigorous end-to-end circuit-level threshold proof exists for arbitrary non-abelian topological phases.  In addition, no convenient machinery exists for constructing explicit circuits for syndrome extraction and decoding in general non-abelian topological phases.

\myparagraph{Literature review}
Ref.~\cite{Koenig_2010} suggested encoding logical information in the ground space of a Turaev--Viro/string-net model as we do here, but did not explicitly discuss syndrome extraction and decoding, and did not prove fault tolerance.
Refs.~\cite{Wootton2013,Hutter2015,Wootton_2016} provided some initial ideas on decoding in the $S_3$ and Ising topological phases.
Ref.~\cite{Dauphinais_2017} provided a threshold proof sketch for a specific renormalization-group (RG) decoder (see~\cite{harrington2004analysis, ducloscianci2013faulttolerantrenormalizationgroupdecoder, Bravyi_2013, balasubramanian2026local}) applied to the restricted class of non-cyclic anyon models assuming a phenomenological noise model with no underlying microscopic circuit.  Their proof could not be shown to generalize beyond non-cyclic anyon models.

In Ref.~\cite{schotte2021quantumerrorcorrectionthresholds}, the authors proposed a syndrome extraction circuit for the double-Fibonacci phase constructed from its string-net presentation.
Their Fibonacci circuit is closely related to our general circuit as discussed below.
They provide tentative numerical evidence for a threshold, but they assume no measurement errors and that only a single error occurs between two syndrome extraction rounds.
In Ref.~\cite{schotte2022faulttoleranterrorcorrectionuniversal}, the authors provided further numerical evidence for a threshold in the double-Fibonacci phase under a phenomenological noise model using an RG decoder.
This time, measurement errors were included, but the noise model remained phenomenological and no mathematical proof of fault-tolerance was provided.

\myparagraph{Contributions}
The core contributions of this work are as follows:
\begin{itemize}
\item We construct a quantum circuit combining syndrome extraction and correction for any non-chiral intrinsic 2D bosonic topological phase, including non-Abelian ones.
A path-integral perspective gives us more freedom to construct shallow circuits than earlier constructions in special cases, and allows us to directly prove circuit-level fault tolerance, see Section~\ref{sec:circuit}.
\item We propose a decoder that corrects errors on the fly by clustering anyon worldlines in spacetime and annihilating clusters by moving the anyons within them to a common base point one step at a time, see Section~\ref{sec:decoding}.
\item We prove a threshold with respect to circuit-level local stochastic noise, see Section~\ref{sec:proof}.
\item We discuss how the circuit, decoder, and proof generalize to topological protocols for logical operations, see Section~\ref{sec:logical_operation}.
\end{itemize}
Note that the decoder and proof in Sections~\ref{sec:decoding} and \ref{sec:proof} can be understood independently of the circuit in Section~\ref{sec:circuit}, with the path-integral perspective providing a rigorous link between them.

We now discuss these points in more detail.  We show that the ground-state quantum memory based on an arbitrary non-abelian model associated with a Turaev-Viro state sum can be realized fault-tolerantly by local quantum operations.  In particular, for a circuit-level $\epsilon$-bounded noise model (also known as a local stochastic model) with the noise rate $\epsilon$ below an $O(1)$ threshold, our fault tolerance scheme preserves logical information for a quasi-stretched-exponentially long time in the system size. The circuit consists of local quantum channels, and the classical decoder takes the current spacetime syndrome history (or, at least, the history in an $O(d)$ time window where $d$ is the code distance) and determines a correction in polynomial time.  We make the standard assumption that classical processing is fast and cheap. We also discuss how to generalize our results to fault-tolerant non-abelian topological quantum computation using arbitrary defect-decorated 3D spacetime diagrams.

Our circuits are based on the Turaev--Viro path integral~\cite{TURAEV1992865, Barrett1993}, whose spatial equivalent is the string-net model~\cite{Levin_2005,kirillov2011stringnetmodelturaevviroinvariants}, with prescribed anyon worldlines~\cite{Bonesteel_2012, Hu_2018,thesis}.
The circuits vastly generalize Ref.~\cite{schotte2021quantumerrorcorrectionthresholds} to more general non-chiral topological phases, and provide a general and unifying perspective.
In particular, instead of trying to extract a syndrome for error correction by deforming the 2D string-net diagrams, we directly implement the 2+1D Turaev--Viro path integral as a spacetime circuit, following and generalizing Refs.~\cite{Bauer_2024,
Bauer_2025,
bauer2025xyfloquetcodesimple,
davydova2025universalfaulttolerantquantum,
twisted_color_circuits,
delafuente2026highthresholddecodingnonpaulicodes}.
Measurements are then merely a consequence of turning non-unitary operators in the circuit into Kraus operators of proper trace-preserving quantum channels.

The spacetime path-integral perspective has two advantages:
First, it allows for a direct and completely self-contained proof at the circuit level without using any physics knowledge of the properties of anyons in the model.
In particular, the absence of logical faults for contractible damage and the neutrality of isolated damage are usually not mathematically proven in the literature but instead physically motivated. The circuit-to-path-integral correspondence allows us to explicitly prove these statements from the mathematical axioms of the path integral.
Second, it gives the user more flexibility in the microscopic circuit design.
In particular, we give a circuit of simple local quantum channels that extracts the syndrome in parallel for every unit cell of a square lattice, which is conceptually similar to the standard north-south-east-west syndrome extraction circuit for the toric code.

One important distinction from abelian phases is that decoding and corrections in a non-abelian phase need to happen ``continuously'' or ``on the fly'', a fact that has already been appreciated in early work~\cite{Wootton_2016, Dauphinais_2017, schotte2022faulttoleranterrorcorrectionuniversal}, and is known in the context of solvable anyon models as the ``just-in-time decoding''~\cite{bombin2019, Brown_2020, davydova2025universalfaulttolerantquantum, lyons2026quantumcomputinganyonsfault, twisted_color_circuits, delafuente2026highthresholddecodingnonpaulicodes}.  Here, we construct such a decoder for general non-abelian models. In our approach, the circuit combines syndrome extraction and correction layers and runs a classical decoder on the current spacetime syndrome history to determine where to move anyons at each time step. 
In general, non-abelian anyons can only be moved at a \emph{finite speed}, which puts a limit on the speed of cleaning up the syndrome.  Although this makes the analysis more difficult, such a property is important for fault tolerance: if at one time step the decoder makes an incorrect decision (due to measurement errors) about the movement directions, the speed constraint limits how much of an effect these errors can have, allowing them to be later undone.

Concretely, our decoding algorithm is based on the idea of clustering and is related to RG decoders~\cite{Duclos_Cianci_2010,
ducloscianci2013faulttolerantrenormalizationgroupdecoder,
Bravyi_2013, balasubramanian2026local}.
Namely, at each timestep $t$ we consider the full spacetime history of all recorded syndrome measurement outcomes up to that time. Here, we extend the definition of the syndrome to all nontrivial measurement outcomes, and not just violated detectors.
The decoder computes the cluster decomposition of the measured syndrome and determines a base point for each cluster toward which the anyons will be moved. After each round of syndrome extraction, the circuit implements active local corrections which move each syndrome point one step towards the base point of its cluster.
This procedure is robust to measurement errors as well: false-positive anyon detections results in false anyon movements, but due to the $O(1)$ speed of movement, their effect only increases error clusters linearly.
Similarly, the procedure is also robust to errors that forget to perform a correction step.

Note that there is an important difference with how one would usually implement a clustering step of the decoder in an abelian phase, such as the toric code: 
Our definition of syndrome includes anyon worldlines and false positives due to measurement errors. In the abelian case, one usually defines syndrome to be the detector violations, which are the endpoints of anyon worldlines and measurement error strings. Thus, instead of only clustering such endpoints, we cluster all observed syndrome.

Our fault-tolerance proof consists of three parts.
In the first part, we show that there is no logical fault in a given circuit realization if the \emph{damage} cannot contain a non-contractible loop. We define the damage to be the sum of the actual errors and detected syndrome, including the effect of active correction on its evolution.
In our work, this is a direct mathematical consequence of the path-integral perspective for our circuits and can be proved rigorously.

The second part of the proof shows that if the error configuration admits a ``burst decomposition'' where the largest burst is sufficiently smaller than some fraction of the system size, then the damage will not contain noncontractible loops and thus there are no logical errors. The burst decomposition of the error configuration is defined in the same way as the cluster decomposition of the syndrome, except that now the required separations between bursts can scale faster than linearly with the burst size.
It is easy to show that if the noise consists of an isolated burst that is not too large, there will not be a logical error:
As anyon worldlines must be continuous except in the vicinity of errors, the anyon worldlines created by the errors in the burst become longer and longer as time progresses.
After their time-like length becomes larger than their space-like distance, the decoder is guaranteed to correctly assign them all to a single cluster together, and after linear time they will be brought to a common point and annihilated, since individual clusters must be charge-neutral. 
Because of this, the total damage caused by a single isolated error burst must be contained in a somewhat larger enclosing box.
The difficulty in the proof comes when bursts of different sizes interact:
Smaller bursts can disturb the shrinking process of a larger burst, such that the speed of shrinking can be significantly reduced.
To ensure that it remains bounded from below, we use burst decomposition with a separation function super-linear in the burst size.

The third part of the proof shows that the probability that the error configuration does not admit a correctable burst decomposition vanishes quasi-stretched-exponentially in the system size, even with a super-linear separation function. This result has been known in the literature due to G\'acs~\cite{Gacs1983,gacs}, and such a burst decomposition with super-linear separation was used to prove the fault-tolerance of Toom's rule~\cite{gacs1988simple, gacsrecife}.

\section{Error-correction circuit}
\label{sec:circuit}
In this section, we describe the quantum circuit we use to fault-tolerantly store logical information in the ground space of a non-chiral topological phase on a torus or some other 2D manifold.
The circuit consists of projective measurements, unitaries, and classically controlled unitaries.
That is, it is a circuit of quantum channels with additional classical inputs and outputs.
The outputs correspond to syndrome measurement outcomes, whereas the inputs determine the corrections that need to be applied during the execution of the circuit to stabilize the ground space.
These inputs are chosen depending on the previous measurement outcomes by a non-local but efficient classical decoding algorithm, which we discuss in Section~\ref{sec:decoding}.

\subsection{The Turaev--Viro path integral}
The circuit is based on the Turaev--Viro path integral~\cite{TURAEV1992865, Barrett1993}, representing the topological phase.
The Hamiltonian version of the Turaev--Viro path integral is known as the string-net model~\cite{Levin_2005}.
Here, we follow a tensor-centric perspective, for more details see Ref.~\cite{thesis}.
The Turaev--Viro path integral is constructed from a \emph{unitary fusion category}.
Concretely, a unitary fusion category is specified by the following:
\begin{itemize}
\item There is a set $I$ of \emph{string labels}.
\item For every string label $i\in I$ there is a quantum dimension $d_i\in \mathbb R$.
\item For every triple of string labels $i,j,k$, there is a fusion multiplicity $N_{ij}^k\in \mathbb{N}_0$.
\item There is an $F$-tensor, which is a tensor whose indices are assigned to the edges and faces of a ``standard tetrahedron'' as
\begin{equation}
\label{eq:ftensor}
\begin{tikzpicture}
\atoms{vertex}{0/p=-150:2, 1/p=90:2, 2/p=-30:2, 3/}
\draw (0)edge[or,mark={slab=$a$}](1) (1)edge[or,mark={slab=$b$}](2) (2)edge[or,mark={slab=$c$}](3) (0)edge[or,mark={slab=$e$,r}](2) (1)edge[or,mark={slab=$f$}](3) (0)edge[or,mark={slab=$d$}](3);
\node at (150:0.65){$\alpha$};
\node at (30:0.65){$\beta$};
\node[black!60] at (30:0.3){$\gamma$};
\node at (-90:0.65){$\delta$};
\end{tikzpicture}
\quad\rightarrow\quad
{F^{abc}_d}_{e\gamma\delta}^{f\alpha\beta}\;.
\end{equation}
The indices $a,b,c,d,e,f$ assigned to the edges are string labels and take values in $I$.
The indices $\alpha$, $\beta$, $\gamma$, and $\delta$ are \emph{fusion labels}, and the values they can take depend on the string labels on the three edges of the corresponding triangle.
Namely, the fusion index $\alpha$ at a triangle
\begin{equation}
\begin{tikzpicture}
\atoms{vertex}{0/p=-150:1, 1/p=90:1, 2/p=-30:1}
\draw (0)edge[or,mark={slab=$a$}](1) (1)edge[or,mark={slab=$b$}](2) (0)edge[or,mark={slab=$c$,r}](2);
\node at (0,0){$\alpha$};
\end{tikzpicture}
\end{equation}
takes values $0\leq \alpha < N_{ab}^c$.
Note that the standard tetrahedron is equipped with edge orientations which are acyclic around every triangle,  known as \emph{branching structure}.
These orientations make it possible to uniquely identify the subsimplices of the tetrahedron.
In particular, the four vertices can be labeled by $0,1,2,3$ such that each edge points towards the higher-numbered vertex.
Edges and triangles can then be labeled by the 2 or 3 numbers of their vertices, respectively.
\item The $F$-tensor has to fulfil some polynomial/tensor-network equations, in particular the \emph{pentagon equation} and \emph{unitarity equation}, which we discuss below.
\item There are some further structures and axioms related to a notion of dual for string labels, which we will not use explicitly in the circuit construction and therefore omit for simplicity.
\end{itemize}

Using the data of a fusion category, we can define the \emph{Turaev--Viro path integral} as follows.
Consider a 3-dimensional ``spacetime'' triangulation with boundary where each tetrahedron is identified with the ``standard tetrahedron'' shown in Eq.~\eqref{eq:ftensor}.
Equivalently, every edge carries an orientation such that the orientations are acyclic around every triangle.
The Turaev--Viro path integral assigns to each such triangulation a tensor $Z$ by the formula
\begin{equation}
\begin{multlined}
Z_{\vec i_\partial, \vec \alpha_\partial}
\hspace{-0.05cm}=\hspace{-0.05cm}\frac{1}{D^{|S_0|}} \sum_{\vec i \in I^{S_1}} \sum_{(0\leq \vec\alpha_f<N_{\vec i(f)})_{f\in S_2}}\\
\prod_{t\in S_3} \big(F_{\vec i(t), \vec\alpha(t)}\big)^{*\sigma(t)}%
\prod_{e\in S_1} d_{i_e}\;.
\end{multlined}
\end{equation}
Here, $S_x$ denotes the set of interior $x$-simplices of the triangulation, i.e., those that are not contained in the 2D boundary triangulation.
$\vec i_\partial$ is a configuration of one string label at every edge of the 2D boundary triangulation.
$\vec\alpha_\partial$ denotes a configuration of one fusion label at every triangle of the 2D boundary triangulation.
We sum over all configurations of string labels $\vec i$ associated to the internal edges of the 3D triangulation.
Then we sum over all configurations of fusion labels $\vec\alpha$ on the internal triangles.
The range of the second summation depends on the string-label configuration $\vec i$ on the edges.
The summand is a product of one $F$-tensor for every tetrahedron, in addition to some normalization factors related to $d$ and $D\coloneqq \sum_a d_a^2$.
So $F_{\vec i(t), \vec\alpha(t)}$ denotes the $F$-tensor whose string and fusion labels correspond to $\vec i$ and $\vec\alpha$ on the corresponding edges and faces of the tetrahedron $t$.
Depending on the edge orientation we can assign to each tetrahedron $t$ an orientation $\sigma(t)\in \{\pm 1\}$, and $(F_{\ldots})^{*\sigma(t)}$ denotes $F_{\ldots}$ if $\sigma(t)=1$ and the complex conjugate $F_{\ldots}^*$ if $\sigma(t)=-1$.
Note again that we only sum over configurations of internal edges and faces, while the boundary configuration is fixed and corresponds to the indices of the resulting tensor $Z_{\vec i_\partial, \vec \alpha_\partial}$.

The dimension of the fusion-label indices of the path-integral tensors depends on the configuration of the string labels.
We can embed all the fusion-label vector spaces into one single vector space of dimension $N_{\max}\coloneqq \max_{i,j,k}(N_{ij}^k)$ by mapping the configurations $[0,\ldots,N_{ij}^k-1]$ onto the first $N_{ij}^k$ entries of $[0,\ldots,N_{\max}-1]$.
For any triangle of the 2D cellulation, the path-integral tensor is then only supported on the configurations of the string labels $i,j,k$ at the edges and fusion label $\alpha$ such that $0\leq \alpha<N_{ij}^k\leq N_{\max}$.
We call this subspace the \emph{fusion subspace} and denote the projector onto it by $\pfusion$.
Even though this is not very natural mathematically, the embedding into $N_{\max}$ turns the path-integral tensor into a vector of a proper local-tensor-product vector space.
We can interpret this tensor as the coefficient tensor for a state with one $|I|$-dimensional qudit on each edge, and one $N_{\max}$-dimensional qudit at each face of the 2D boundary triangulation.
This state is a ground state of the Levin-Wen string-net model on the 3-valent graph dual to the 2D boundary triangulation.

As mentioned above, the $F$-tensor must fulfill certain equations.
All of these equations are instances of the following three axioms:
\begin{itemize}
\item \emph{Topological invariance axiom}:
Consider any two 3D triangulations with (1) the same boundary triangulation, and (2) the same bulk topology.
Then the Turaev--Viro path-integral tensor for these two triangulations is the same.
\item \emph{Irreducibility axiom}:
Consider a triangulation with boundary of topology $S_2\times [0,1]$.
The path-integral tensor is a tensor product of two tensors, with respect to a division of the degrees of freedom into those at $S_2\times 0$ and at $S_2\times 1$.
\item \emph{Normal form:}
Consider a 2D triangulation with boundary such that all vertices are part of the boundary.
Take two orientation-reversed copies of this 2D triangulation (one at the ``back'' and one at the ``front'') and identify their boundaries, yielding a 2D triangulation without boundary.
Now use this 2D triangulation as boundary of a 3D manifold equal to the 2D manifold times $[0,1]$, where $[0,1]$ is collapsed to a point on the boundary of the 2D manifold.
Then the path-integral tensor as an operator from the back to the front part of the boundary is equal to the identity, times a normalization factor of $(d_i)^{-1/2}$ for every edge with string label $i$ internal to the front or back triangulation.
\end{itemize}
The topological invariance axiom defines an infinite set of equations for the $F$-tensor.
However, all of these equations can be derived from essentially a single equation (apart from some ``smaller'' auxiliary ones).
This is an equation between two triangulations of a 3-ball, one consisting of 2 tetrahedra sharing a single face, and another one consisting of 3 tetrahedra sharing an edge.
The resulting equation for the $F$-tensor is known as the \emph{pentagon equation} in category theory.
The equation can be applied to a triangulation, removing the 2 or 3 tetrahedra of one side of the equation and replacing them with the 3 or 2 tetrahedra of the other side.
The effect of this substitution is known as a \emph{2-3 Pachner move}.

Similarly, the irreducibility axiom implies an infinite set of equations, but given topological invariance, it suffices to impose this axiom for one (possibly very minimal) triangulation of $S_2\times [0,1]$.
Physically, the irreducibility axiom rules out GHZ-like entanglement in the model.
It is also the core axiom that leads to fault tolerance when constructing a circuit from the path integral.

The infinitely many equations of the standard-form axiom are also generated from a single instance, namely
\begin{equation}
\label{eq:triangle_extender_def_geo}
\begin{gathered}
\begin{tikzpicture}
\atoms{vertex}{{0/}, {1/p={1.5,0}}, {2/p={60:1.5}}, {3/p={-60:1.5}}}
\draw (0)edge[or=0.4,bend left=20,glab=e](1) (0)edge[or](2) (1)edge[or](2) (0)edge[or](3) (1)edge[or](3);
\node at (0.10,0.81){$a$};
\node at (1.40,0.81){$b$};
\node at (0.10,-0.81){$c$};
\node at (1.40,-0.81){$d$};
\draw[opacity=0.35] (0)edge[or,bend right=40,backedge,glab=f](1);
\node at (0.6,0.6){$\alpha$};
\node[opacity=0.35] at (0.9,0.6){$\beta$};
\node at (0.6,-0.6){$\gamma$};
\node[opacity=0.35] at (0.9,-0.6){$\delta$};
\end{tikzpicture}\\
= \frac{1}{\sqrt{d_ed_f}}\delta_{ef}\delta_{\alpha\beta}\delta_{\gamma\delta}\quad\forall a,b,c,d
\;.
\end{gathered}
\end{equation}
If we interpret the above tensor as an operator from the back to the front acting trivially on the labels $a,b,c,d$, ignore normalization factors, and embed fusion indices into the $N_{\max}$-dimensional vector space, this is equal to the fusion space projector $\pfusion$ corresponding to the pair of triangles involved.
The standard-form axiom has no real physical meaning but is mostly a mathematical convenience:
If we did not impose it, this would correspond to also allowing irreducible \emph{multi-fusion categories}, which can always be reduced to ordinary fusion categories.
In fact, this equation also implies the irreducibility axiom:
We can glue a closed 3D triangulation of $S_2\times S_1$ from two copies of it (with different arrow directions) which yields the number $1$, which also corresponds to the trace of a projector associated with $S_2\times [0,1]$, and therefore this projector has rank $1$.

\myparagraph{Anyon worldlines}
Next, we describe the Turaev--Viro path integral with \emph{anyon worldlines}.
An anyon worldline is defined by a \emph{representation of the tube algebra} of the fusion category defined by $F$.
Concretely, an anyon worldline (of a single fixed anyon type) is fully specified by the following:
\begin{itemize}
\item For each string label $a$, there is an ``anyon multiplicity'' 
\begin{equation}\label{eq:anyon_multiplicity}
A_a\in \mathbb{N}.
\end{equation}
\item There is a constant $d\in\mathbb R$ called the \emph{quantum dimension} of the anyon.
\item There is a $\rho$-tensor whose indices are associated with the edges and faces of the \emph{standard anyon cylinder},
\begin{equation}
\label{eq:anyon_segment_bdiag}
\begin{tikzpicture}
\atoms{vertex}{0/, {1/p={0,1.4}}}
\atoms{cod2vertex}{{0x/p={$(0)+(0:0.5)$},vback,dlab={\mu}{0:0.25}}, {1x/p={$(1)+(0:0.5)$},dlab={\nu}{0:0.25}}}
\draw (0)edge[mark={arr,p=0.8},out=-45,in=-90,mark={slab=$a$,e,r}]++(0:1) (0)edge[backedge,out=45,in=90]++(0:1);
\draw (1)edge[mark={arr,p=0.8},out=-45,in=-90,mark={slab=$b$,e,r}]++(0:1) (1)edge[out=45,in=90]++(0:1);
\draw (0)edge[or,glab=c](1) (0)edge[out=20,in=-90,mark={arr,p=0.8},mark={slab=$d$,e,r}](1,0.7) (1,0.7)edge[backedge,out=90,in=-20](1);
\draw[extend map] (0)edge[back](0x) (1)--(1x);
\draw[cod2ind,midedge] (0x)--(1x);
\node at (0.3,0.8){$\beta$};
\node[black!60] at (0.8,0.7){$\alpha$};
\end{tikzpicture}
\quad\rightarrow\quad
{\rho_{ab}^{cd\alpha\beta}}_\mu^\nu
\;.
\end{equation}
The 1-gon faces at the bottom and top are called \emph{anyon 1-gons}, and the corresponding indices $\mu$ and $\nu$ are \emph{multiplicity labels}.
$c$, $d$, $\alpha$ and $\beta$ are the string labels and fusion labels assigned to the two edges and two triangles forming the side of the cylinder.
Note that above we have also drawn the core of the cylinder as a thick dashed line, and the core of the anyon 1-gons as a thick dot.
For a 1-gon with string label $a$,
\begin{equation}
\label{eq:anyon_multiplicity}
\begin{tikzpicture}
\atoms{vertex}{0/}
\atoms{cod2vertex}{{1/p={0.4,0},dlab={\mu}{0:0.25}}}
\draw[or,looseness=2,glabr=a] (0)to[out=-60,in=-90]++(0:1)to[out=90,in=60](0);
\draw[extend map] (0)--(1);
\end{tikzpicture}
\;,
\end{equation}
the multiplicity label $\mu$ takes values $0\leq \mu<A_a$.
\item The $\rho$-tensor satisfies some polynomial tensor-network equations involving copies of $\rho$ and $F$-tensors as discussed below.
\end{itemize}

Next, we define the Turaev--Viro path integral with anyon worldlines of one fixed anyon type.
Consider a cellulation with boundary consisting of tetrahedra and anyon cylinders.
The anyon 1-gons of anyon cylinders must be glued with other anyon 1-gons, or be part of the 2D boundary.
Thus, the cores of the anyon cylinders (drawn as thick dashed lines in Eq.~\eqref{eq:anyon_segment_bdiag}) form a 1D cellulation.
The boundary is a 2D triangulation consisting of triangles and anyon 1-gons.

The Turaev--Viro path integral with anyon worldlines assigns a tensor to every 3D cellulation with boundary as discussed above.
Like for the path integral without anyon worldlines, we sum over all configurations of string labels on the interior edges, and fusion labels on the interior triangles.
In addition, we sum over one multiplicity label for each anyon 1-gon.
The summand is given by a product of one $F$-tensor for every tetrahedron and one $\rho$-tensor for every anyon cylinder.
The resulting tensor has one string label at every edge of the 2D cellulation, one fusion label for every triangle, and additionally one multiplicity label for every anyon 1-gon.

The $\rho$ and $F$-tensors have to satisfy a set of equations, all of which are instances of the following three axioms:
\begin{itemize}
\item Topological invariance axiom:
The path integral must yield the same tensor for any two cellulations with anyon worldlines with the same 2D boundary cellulation and the same 3D bulk topology, including the topology of the ``knot'' formed by the 1D anyon worldlines embedded into the 3D bulk.
\item Irreducibility axiom:
Consider a cellulation whose bulk topology is $S_2\times [0,1]$, and which includes two anyon worldlines connecting the sphere at $S_2\times 0$ with $S_2\times 1$.
The path-integral tensor assigned to this cellulation is a tensor product with respect to the division into $S_2\times 0$ and $S_2\times 1$.
\item Normal form:
Consider a 3-manifold whose boundary consists of a ``back'' and a ``front'' copy of the same 2D cellulation of triangles and anyon 1-gons such that (1) there are no internal vertices in the 2D cellulation and (2) the 3D topology is equal to the 2D cellulation times an interval (meaning anyon points in the 2D cellulation become anyon worldlines in the 3-manifold).
Then the associated tensor interpreted as a linear operator from the indices at the back boundary to the indices at the front boundary is equal to the identity.
\end{itemize}
As in the case without anyon worldlines, the infinitely many topological-invariance equations can be derived from a single equation:
There is one anyon cylinder on the left side and two stacked anyon cylinders on the right side, corresponding to a 1-2 Pachner move of a 1D cellulation.
The one cylinder on the left is padded with 3 tetrahedra such that the 2D boundary cellulations are the same on both sides.
The three padding tetrahedra form a triangulation of a solid torus obtained from identifying the top and bottom of a triangle prism as shown in Eq.~\eqref{eq:anyon_compactification}, whose boundary is composed of three times the triangulated annulus forming the side of the anyon cylinder.
The tensor assigned to this solid torus is equal to the structure coefficients of the tube algebra, and the equation makes the anyon cylinder $\rho$ a representation of the tube algebra, see section 6.2 of Ref.~\cite{thesis}.

Given topological invariance, it suffices to impose the irreducibility axiom for a single cellulation of the corresponding topology.
Mathematically, this condition corresponds to the fact that $\rho$ is an \emph{irreducible} representation of the tube algebra.

\myparagraph{Multiple anyon types}
For constructing our error-correcting circuit, we will not only use one type of anyon worldline but a \emph{complete set} of anyon types.
Each anyon type has its own $\rho$-tensor, anyon multiplicities, and quantum dimension, and for an anyon type called $\phi$ we denote them by $\rho^\phi$, $A^\phi$, and $d_\phi$.
We can also define the Turaev-Viro path integral with multiple types of anyon worldline.
This assigns a tensor to every 3D cellulation with anyon cylinders where every segment or loop of anyon worldlines is labeled by one anyon type.
The path integral is constructed simply by using the tensor $\rho^\phi$ for a segment labeled by anyon type $\phi$.

The complete set of anyon worldlines has to satisfy two further axioms:
\begin{itemize}
\item Anyon exclusion: This is an extension of the irreducibility axiom for each separate anyon worldline. Consider a cellulation whose bulk topology is $S_2\times [0,1]$, and which includes two anyon worldlines connecting the sphere at $S_2\times 0$ with $S_2\times 1$, one of type $\phi$ and one of type $\chi$.
The path-integral tensor assigned to this cellulation is a tensor product with respect to the division into $S_2\times 0$ and $S_2\times 1$ if $\phi=\chi$.
\item Anyon completeness:
Consider any 3-cellulation with boundary $M$ containing an anyon-worldline loop.
Let $Z^\phi(M)$ be the path-integral tensor for $M$ where the anyon loop has anyon type $\phi$ whereas the types of all potential other anyon worldlines are fixed.
Let $M'$ be the 3D cellulation obtained from performing a \emph{2-surgery} along the anyon loop, i.e., replacing an $S_1\times B_2$ neighborhood of the anyon loop with $B_2\times S_1$.
Then there exists a \emph{quantum dimension} $d_\phi$ for each anyon type $\phi$ such that
\begin{equation}
\sum_\phi \frac{d_\phi}{D} Z^\phi(M) = Z(M')\;.
\end{equation}
\end{itemize}
The anyon exclusion axiom makes sure that every $\rho$-tensor corresponds to a properly distinct anyon type.
Physically, it corresponds to the invariance of the topological charge, that is, the anyon type.
The anyon completeness axiom on the other hand ensures that every anyon type is represented by a $\rho$-tensor.
Mathematically, it implies that the direct sum of the tube-algebra irreps corresponding to all different anyon types is faithful.
This is imposed by the fact that tracing over products of a faithful representation of a semi-simple algebra (the tube algebra) yields the structure coefficients of that algebra itself, with correctly chosen normalizations.
Physically, the anyon completeness axiom is related to the principle of remote detectability.

Note that one anyon type can always be set to the trivial anyon.
The anyon 1-gon of the trivial anyon is just a bulk triangle with two edges glued such that it folds into a 1-gon,
\begin{equation}
\begin{tikzpicture}
\atoms{vertex}{0/}
\atoms{cod2vertex}{{1/p={0.4,0},dlab={b\alpha}{0:0.35}}}
\draw[or,looseness=2,glabr=a] (0)to[out=-60,in=-90]++(0:1.2)to[out=90,in=60](0);
\draw[extend map] (0)--(1);
\end{tikzpicture}
=
\begin{tikzpicture}
\atoms{vertex}{0/, {1/p={0.6,0},dlab={\alpha}{0:0.25}}}
\draw[or,looseness=2,glabr=a] (0)to[out=-60,in=-90]++(0:1.2)to[out=90,in=60](0);
\draw (0)edge[or,glab=b](1);
\end{tikzpicture}
\;,
\end{equation}
such that the multiplicity label of the trivial anyon 1-gon is formed by the internal edge label $b$ and the fusion label $0\leq \alpha<N_{ab}^b$ (though for a fusion category in its normal form we always have $N_{ab}^b = \delta_{b=1}\delta_{a=1}$ and the multiplicity label of the trivial anyon 1-gon has dimension $\delta_{a=1}$).
The anyon cylinder of the trivial anyon can be obtained by gluing two sides of the triangle prism shown in Eq.~\eqref{eq:anyon_compactification}, yielding a 1-gon prism triangulated with three tetrahedra.
So the $\rho$-tensor for the trivial anyon is a contraction of three $F$-tensors.
Then, the path integral with a trivial-anyon worldline is equal to a path integral consisting of only 3D bulk tetrahedra, implementing the idea that having a trivial anyon worldline is the same as having no anyon worldline.

\myparagraph{Anyon fusion}
Finally, anyons can fuse, corresponding to points in spacetime where three anyon worldlines of types $\phi$, $\textcolor{col2}{\chi}$, and $\textcolor{col3}{\psi}$ meet.
A fusion event between three anyon worldlines is specified by a \emph{fusion tensor} $S$ whose indices are assigned to the boundary edges, anyon 1-gons and triangles of the following \emph{fusion connector} volume whose boundary is a sphere cellulated by three anyon 1-gons and one triangle,
\begin{equation}
\label{eq:fusion_point_defect_volume_bdiag}
\begin{tikzpicture}
\atoms{vertex}{0/}
\atoms{cod2vertex}{{0a/p={$(0)+(-105:0.6)$},astyle=col3,dlab={\lambda}{-105:0.3}}, {0b/p={$(0)+(15:0.6)$},dlab={\mu}{15:0.25}}, {0c/p={$(0)+(135:0.6)$},astyle=col2,dlab={\nu}{135:0.25}}}
\atoms{cod3vertex}{{4/p={-70:1.3}, dlab={\alpha}{60:0.5}}}
\draw[or,looseness=1.5,glabr=a] (0)to[out=-15,in=-75]++(15:1.1)to[out=105,in=45](0);
\draw[or,looseness=1.5,glabr=b] (0)to[out=105,in=45]++(135:1.1)to[out=-135,in=165](0);
\draw[ior,looseness=1.5,glabr=c](0)to[out=-135,in=165]++(-105:1.1)to[out=-15,in=-75](0);
\draw[extend map] (0)--(0a) (0)--(0b) (0)--(0c);
\draw[extend map] (4)--(0) (4)--(0a) (4)--(0b) (4)--(0c);
\end{tikzpicture}
\quad\rightarrow\quad
(S_{\phi\textcolor{col2}{\chi}}^{\textcolor{col3}{\psi}})^{abc\alpha}_{\mu\textcolor{col2}{\nu}\textcolor{col3}{\lambda}}
\;.
\end{equation}
We have also drawn the core vertex of the volume as a large black dot.
The core vertices of the anyon 1-gons corresponding to $\phi$, $\textcolor{col2}{\chi}$, and $\textcolor{col3}{\psi}$ are drawn as smaller dots colored in black, blue, and red, respectively, such that the corresponding multiplicity labels are  $\mu$, $\textcolor{col2}{\nu}$, and $\textcolor{col3}{\lambda}$.
The label $\alpha$ corresponds to the degenerate triangle bounded by the three looping edges with string labels $a$, $b$, $c$, whose three corner vertices are all identified.
The fusion tensor has to fulfill certain equations as we discuss below.
The set of fusion tensors fulfilling these equations forms a vector space, so we can choose a basis and label the basis elements by attaching an extra index $\epsilon$ to $S$, $(S_{\phi\textcolor{col2}{\chi}}^{\textcolor{col3}{\psi}})^{abc\alpha}_{\mu\textcolor{col2}{\nu}\textcolor{col3}{\lambda}\epsilon}$.
The dimension of the index $\epsilon$ is known as the \emph{fusion space dimension} of $\phi$, $\textcolor{col2}{\chi}$, and $\textcolor{col3}{\psi}$ and denoted as $\widetilde{N}_{\phi\textcolor{col2}{\chi}}^{\textcolor{col3}{\psi}}$ such that $0\leq \epsilon<\widetilde{N}_{\phi\textcolor{col2}{\chi}}^{\textcolor{col3}{\psi}}$.

We can now define the Turaev--Viro path integral with anyon worldlines and fusion events.
This path integral assigns a tensor to every cellulation with boundary, consisting of tetrahedra, anyon cylinders, and fusion connectors as shown above.
The anyon 1-gons of fusion connectors may be identified with anyon 1-gons of anyon cylinders, anyon 1-gons of other fusion connectors, or anyon 1-gons in the 2D boundary cellulation.
The anyon cylinders and fusion connectors form a 3-valent string network embedded into the 3D bulk manifold.
The path-integral tensor is obtained as usual by summing over all configurations of internal string, fusion, and multiplicity labels, with the summand being a product of $F$, $\rho$, and $S$-tensors.
Note that we do not sum over the $\epsilon$-indices of the $S$-tensors but consider them part of the resulting path-integral tensor.

The $S$ tensor has to satisfy equations corresponding to the following axioms:
\begin{itemize}
\item Topological invariance:
As usual, two cellulations have equal path-integral tensors if their 2D boundary cellulations and their 3D bulk topologies are the same, including the ``homotopy class'' of the embedded 3-valent anyon network.
\item Fusion space completeness:
Consider a 3D cellulation $M$ with two fusion connectors whose anyon types are the same.
Let $Z^{\epsilon,\epsilon'}(M)$ denote the path integral associated with $M$ with the $\epsilon$-indices of the two fusion connectors fixed to $\epsilon$ and $\epsilon'$. 
Consider the 3D cellulation $M'$ where we remove both fusion connectors and identify their 2D boundaries (consisting of three anyon 1-gons and a triangle each).
Then we have
\begin{equation}
\sum_\epsilon Z^{\epsilon,\epsilon}(M) = Z(M')\;.
\end{equation}
\end{itemize}
The topological invariance axiom is generated by a single equation:
On the left side there is a single fusion connector, and on the right side there is a fusion connector wrapped by a volume of topology $S_2\times S_1$ consisting of three anyon cylinders and 3 tetrahedra.
Since the $S$-tensor occurs once on each side, this equation is linear and its solutions form a vector space.
The fusion space completeness axiom ensures that the full vector space of solutions is covered by $(S_{\phi\textcolor{col2}{\chi}}^{\textcolor{col3}{\psi}})^{abc\alpha}_{\mu\textcolor{col2}{\nu}\textcolor{col3}{\lambda}\epsilon}$.

\myparagraph{Example: Fibonacci}
Let's discuss a simple example, namely the Fibonacci fusion category.
There are two string labels, called $1$ and $\tau$.
The fusion multiplicity $N_{\bullet\bullet}^{\bullet}$ is $1$ if two, three, or none of the string labels are $\tau$, and $0$ if exactly one of the string labels is $\tau$.
The quantum dimensions are $d_1=1$ and $d_\tau=\phi=\frac{1+\sqrt{5}}{2}$, such that $D=1+\phi^2$.
Note that if the string labels are such that $N=0$ around any triangle, then the $F$-tensor has no entry for this string-label configuration.
If $N=1$ at every triangle, then the entry is a number.
Since $N\leq 1$ always, the fusion labels are trivial.
We can draw a configuration of string labels on the tetrahedron by marking all edges with a $\tau$ label by a small circle.
The $F$-tensor is invariant under symmetries of the tetrahedron, and is fully specified by:
\begin{equation}
\begin{gathered}
\begin{tikzpicture}
\markedtetrahedron{mke0}{mke0}{mke0}{mke0}{mke0}{mke0}%
\end{tikzpicture}
=1
\;,\quad
\begin{tikzpicture}
\markedtetrahedron{mke1}{mke1}{mke1}{mke0}{mke0}{mke0}%
\end{tikzpicture}
=\phi^{-1/2}
\;,\quad
\begin{tikzpicture}
\markedtetrahedron{mke1}{mke1}{mke0}{mke0}{mke1}{mke1}%
\end{tikzpicture}
=\phi^{-1}
\;,\\
\begin{tikzpicture}
\markedtetrahedron{mke1}{mke1}{mke1}{mke0}{mke1}{mke1}%
\end{tikzpicture}
=\phi^{-1}
\;,\quad
\begin{tikzpicture}
\markedtetrahedron{mke1}{mke1}{mke1}{mke1}{mke1}{mke1}%
\end{tikzpicture}
=-\phi^{-2}
\;.
\end{gathered}
\end{equation}
There are four anyon types, labeled by $1$, $\tau_0$, $\tau_1$, and $\tau_{01}$.
For the anyon $1$, the anyon multiplicity is $A^1_1=1$ and $A^1_\tau=0$, where the superscript denotes the anyon type $1$ and the subscript denotes a string label. The tensor $\rho^1$ (superscript denoting the anyon type $1$) is given by
\begin{equation}
\markedcylinder{mke0}{mke0}{mke0}{mke0}= 1\;,\quad
\markedcylinder{mke0}{mke0}{mke1}{mke1}= 1\;.
\end{equation}
For the anyon $\tau_0$, the anyon multiplicity is $A^{\tau_0}_1=0$ and $A^{\tau_0}_\tau=1$, and the tensor $\rho^{\tau_0}$ is given by
\begin{equation}
\begin{gathered}
\markedcylinder{mke1}{mke1}{mke0}{mke1}= \phi^{-1}\;,\quad
\markedcylinder{mke1}{mke1}{mke1}{mke0}= \phi^{-1}e^{4\pi i/5}\;,\\
\markedcylinder{mke1}{mke1}{mke1}{mke1}= \phi^{-1}e^{-3\pi i/5}\;.
\end{gathered}
\end{equation}
For the anyon $\tau_1$, the anyon multiplicity is $A^{\tau_1}_1=0$ and $A^{\tau_1}_\tau=1$, and the tensor $\rho^{\tau_1}$ is given by
\begin{equation}
\begin{gathered}
\markedcylinder{mke1}{mke1}{mke0}{mke1}= \phi^{-1}\;,\quad
\markedcylinder{mke1}{mke1}{mke1}{mke0}= \phi^{-1}e^{-4\pi i/5}\;,\\
\markedcylinder{mke1}{mke1}{mke1}{mke1}= \phi^{-1}e^{3\pi i/5}\;.
\end{gathered}
\end{equation}
For the anyon $\tau_{01}$, the anyon multiplicity is $A^{\tau_{01}}_1=1$ and $A^{\tau_{01}}_\tau=1$, and the tensor $\rho^{\tau_{01}}$ is given by
\begin{equation}
\begin{gathered}
\markedcylinder{mke0}{mke0}{mke0}{mke0}= 1\;,\quad
\markedcylinder{mke0}{mke0}{mke1}{mke1}= -\phi^{-2}\;,\quad
\markedcylinder{mke1}{mke1}{mke0}{mke1}= \phi^{-1}\;,\\
\markedcylinder{mke1}{mke1}{mke1}{mke0}= \phi^{-1}\;,\quad
\markedcylinder{mke1}{mke1}{mke1}{mke1}= \phi^{-3}\;,\\
\markedcylinder{mke0}{mke1}{mke1}{mke1}= \phi^{-2}\sqrt{D}\;,\quad
\markedcylinder{mke1}{mke0}{mke1}{mke1}= \phi^{-2}\sqrt{D}\;.
\end{gathered}
\end{equation}

\myparagraph{Ground state space}
Finally, we discuss the ground state space, which is the subspace of the vector space associated with a 2D triangulation into which we encode the logical information.
On a 2D manifold triangulation $T$, the ground state space is defined as follows:
Consider an arbitrary cellulation of $T\times [0,1]$ such that $T\times 0$ and $T\times 1$ are both copies of $T$.
Then the associated path-integral tensor can be viewed as an operator $P$ from $T\times 0$ to $T\times 1$.
Now take two copies of $T\times [0,1]$ and glue $T\times 0$ of the first copy with $T\times 1$ of the second copy, corresponding to the operator $P^2$ by the topological invariance axiom.
This yields another 3D triangulation of type $T\times [0,1]$ associated with $P$, so $P=P^2$ is a projector.
The ground state space is the support of this projector.
It follows from the topological invariance axiom that the dimension of the ground state space only depends on the topology of $T$, but not on its triangulation.

\subsection{Constructing the circuit from the path integral}
In this section, we describe how to construct a circuit of quantum instruments from the path integral representation of a topological phase.
First of all, we define the Hilbert space on which the circuit acts:
To every 2D cellulation consisting of triangles and anyon 1-gons, we can assign qudits as follows:
\begin{itemize}
\item One $|I|$-dimensional qudit at every edge 
\item and one $N_{\max}\coloneqq \max_{i,j,k}(N_{ij}^k)$-dimensional qudit at every triangle.
\end{itemize}
Note that there are no multiplicity labels at the anyon 1-gons -- these labels will correspond to measurement outcomes rather than qudits.
The circuit maps between Hilbert spaces of different 2D cellulations, but the Hilbert spaces of all different 2D cellulations can be embedded into one set of local static degrees of freedom if desired.
We will start by describing the elements of the circuit abstractly for any 2D cellulation, and later pick a concrete cellulation.

The quantum circuit consists of three different layers which are repeated alternately.
\begin{itemize}
\item The \emph{fusion-space recovery} layer consists of local channels that map the quantum state on the 2D cellulation into the fusion subspace.
\item The \emph{syndrome extraction} layer corresponds to linear operators that are equal to the path-integral tensors $F$ or $\rho$, mapping from some of the tensor indices to the remaining indices.
\item The \emph{correction} layer again corresponds to path-integral tensors -- this time their task is to move the measured anyons.
\end{itemize}

\myparagraph{Fusion-space recovery}
The fusion-space recovery is a circuit of quantum channels that (1) acts as the identity within the fusion subspace and (2) maps any state outside the fusion subspace back into the fusion subspace.
These quantum channels can be understood as a sequence of measurement and local unitary feedback based on the measurement outcome.
Concretely, the fusion-space recovery consists of applying the following channel $C^t_e$ to every triangle $t$ of the 2D cellulation, where $e$ is a chosen \emph{correction edge} of $t$.
$C^t_e$ acts on the string labels $a$, $b$, and $c$ at the three edges of $t$, and the fusion label $\alpha$ of $t$:
We start by making a 2-outcome measurement, which returns $0$ for all configurations where $0\leq \alpha<N_{ab}^c$, and $1$ for all other configurations.
In other words, we measure whether the triangle is in the fusion subspace or not.
In particular, if $N_{ab}^c=0$, then we get outcome $1$ (not in the fusion subspace) no matter what $\alpha$ is.
If the outcome is $0$, we do nothing.
If the outcome is $1$, then we measure the configuration of $a$ and $b$.
Depending on the outcome, we then reinitialize $c$ to any value $c'$ such that $N_{ab}^{c'}\neq 0$, and reinitialize $\alpha$ to $0$.

It is easy to see that $C^t_e$ does nothing if $t$ is in the fusion subspace, and maps $t$ back to the fusion subspace if not.
However, applying $C^t_e$ may map the other triangle $t'$ adjacent to $e$ out of the fusion subspace, so we need to apply $C^{t'}_e$ afterwards.
This cycle can be broken if $e$ is not part of another triangle $t'$ but instead part of an anyon 1-gon.

\myparagraph{Syndrome extraction}
Next, we discuss the syndrome extraction layer, which consists of $F$ and $\rho$-tensors applied by contracting some of their indices with the current state and leaving the rest of the indices uncontracted.
This operation corresponds to gluing some of the triangles and anyon 1-gons in the boundary of the tetrahedron or anyon cylinder to the 2D cellulation, such that the rest of the triangles and 1-gons remain in the resulting 2D cellulation:
The gluing changes the 2D cellulation in the same way as the application of the linear operator changes the Hilbert space.
The syndrome extraction layer corresponds to gluing a set of tetrahedra and anyon cylinders to the 2D cellulation such that these tetrahedra and cylinders form a proper layer of 3D cellulation, with non-zero thickness everywhere.

We now concretely look at the linear operators corresponding to $F$ and $\rho$-tensors.
\begin{proposition}
View the $F$-tensor as a linear operator $\bm F^{(22)}$ from the string and fusion labels associated with the $013$ and $123$ triangles of the tetrahedron to the labels associated with the $012$ and $023$ triangles:
\begin{equation}
\bm F^{(22)}_{(abcde\gamma\delta),(a'b'c'd'f\alpha\beta)}
= \delta_{aa'}\delta_{bb'}\delta_{cc'}\delta_{dd'} {F^{abc}_d}_{e\gamma\delta}^{f\alpha\beta} \sqrt{d_e d_f}
\end{equation}
$\bm F^{(22)}$ is a partial isometry, whose support is the fusion subspace,
\begin{equation}
\label{eq:f22_operator_unitary}
(\bm F^{(22)})^\dagger \bm F^{(22)}= \pfusion\;.
\end{equation}
Here, $\pfusion$ denotes the fusion space projector on the pair of triangles involved.
Thus $\{\mathbf F^{(22)}, 1-\pfusion\}$ forms a pair of Kraus operators defining a 2-outcome measurement.
\footnote{Alternatively, we could turn $\mathbf F^{(22)}$ into a proper unitary by extending it to outside the fusion subspace.
Both the measurement and the unitary version yield a fault-tolerant circuit.}
\end{proposition}
\begin{proof}
Evaluating the summation in Eq.~\eqref{eq:f22_operator_unitary} yields the path integral on two tetrahedra that are glued at two of their triangles.
This gluing yields the volume in Eq.~\eqref{eq:triangle_extender_def_geo}, which by the normal-form axiom is equal to the fusion-space projector $\pfusion$.
\end{proof}

Physically, the quantum instrument defined by the Kraus operators $\{\mathbf F^{(22)}, 1-\pfusion\}$ corresponds to an ordinary measurement of the observable $\pfusion$ followed by a feed-forward unitary $\mathbf F^{(22)}$ (extended arbitrarily outside of $\pfusion$) conditioned on the $+1$ measurement outcome.
Applying the $\bm F^{(22)}$ measurement to the degrees of freedom associated with two neighboring triangles on the 2D cellulation corresponds to gluing the $013$ and $123$ triangles of the tetrahedron to these triangles.
If we imagine the 2D (spatial) triangulation as the boundary of some 3D (spacetime) triangulation, adding the tetrahedron to the 3D spacetime replaces the $013$ and $123$ with the $012$ and $023$ triangles of the 2D boundary, which has the following effect on the 2D triangulation known as a 2-2 Pachner move:
\begin{equation}
\label{eq:22pachner_move}
\begin{tikzpicture}
\atoms{vertex}{{0/}}
\atoms{vertex}{{1/p={0.5,-0.5}}}
\atoms{vertex}{{2/p={1,0}}}
\atoms{vertex}{{3/p={0.5,0.5}}}
\draw (0)edge[or](3) (3)edge[or](2) (2)edge[or](1) (0)edge[or](1) (3)edge[or](1);
\end{tikzpicture}
\quad\rightarrow\quad
\begin{tikzpicture}
\atoms{vertex}{{0/}}
\atoms{vertex}{{1/p={0.5,-0.5}}}
\atoms{vertex}{{2/p={1,0}}}
\atoms{vertex}{{3/p={0.5,0.5}}}
\draw (0)edge[or](3) (3)edge[or](2) (2)edge[or](1) (0)edge[or](1) (0)edge[or](2);
\end{tikzpicture}
\;.
\end{equation}

Analogously, we can define isometries from interpreting $F$ as a linear map from any two of its triangles to the remaining two.
We will refer to all of these operators as ``an $\bm F^{(22)}$ operator'' even though strictly speaking they are different operators.
Next, we can also view the $F$-tensor as a linear operator from one of its triangles to the remaining three.
\begin{proposition}
View the $F$-tensor as a linear operator $\bm F^{(13)}$ from all indices associated with the $012$ face to those associated with the $013$, $023$, and $123$ faces,
\begin{equation}
\bm F^{(13)}_{(a'b'e'cdf\alpha\beta\delta),(abe\gamma)}
= \delta_{aa'}\delta_{bb'}\delta_{ee'} {F^{abc}_d}_{e\gamma\delta}^{f\alpha\beta}
\sqrt{\frac{d_c d_d d_f}{D}}
\;.
\end{equation}
$\bm F^{(13)}$ is a partial isometry whose support is the fusion subspace of the single triangle,
\begin{equation}
\label{eq:f13_isometry}
(\bm F^{(13)})^\dagger \bm F^{(13)}= \pfusion\;.
\end{equation}
So $\{\bm F^{(13)}, 1-\pfusion\}$ forms a pair of Kraus operators defining a valid quantum instrument.
\end{proposition}
\begin{proof}
The left-hand side of Eq.~\eqref{eq:f13_isometry} corresponds to gluing two tetrahedra at three face pairs, which results in a triangulation of a 3-ball whose boundary consists of two triangles with pairwise identified edges.
According to the normal form axiom, the associated path-integral tensor is equal to $\pfusion$.
\end{proof}

Applying $\bm F^{(13)}$ corresponds to gluing the $012$ triangle of the tetrahedron to the 2D cellulation, such that the $023$, $013$, and $123$ faces remain, which corresponds to the following transformation of the 2D cellulation known as a 1-3 Pachner move:
\begin{equation}
\label{eq:2d_31pmove_geo}
\begin{tikzpicture}
\atoms{vertex}{1/p={-30:0.6}, 2/p={90:0.6}, 3/p={-150:0.6}}
\draw (3)edge[or](2) (2)edge[or](1) (3)edge[or](1);
\end{tikzpicture}
\quad\rightarrow\quad
\begin{tikzpicture}
\atoms{vertex}{0/, 1/p={-30:0.6}, 2/p={90:0.6}, 3/p={-150:0.6}}
\draw (0)edge[or](1) (0)edge[or](2) (0)edge[or](3) (3)edge[or](2) (2)edge[or](1) (3)edge[or](1);
\end{tikzpicture}
\;.
\end{equation}
Analogously, we can also get isometries from interpreting $F$ as a linear map from any other triangle to the remaining three triangles, and we will refer to all of these operators as ``an $\bm F^{(13)}$ operator''.

We should note the $F$-tensor does not give rise to a partial isometry when viewed as a linear operator from three faces to the single remaining face.
The simple reason for this is that the overall Hilbert space dimension decreases when we decrease the number of triangles in the 2D cellulation (in this case by 2).
Gluing a proper layer of 3D cellulation mapping one 2D slice to the next must necessarily involve gluings that remove vertices of the 2D slice and lower the number of triangles.
In order to achieve this, we need proper measurements which we can construct from anyon cylinders as follows.

\begin{proposition}
View the $\rho$-tensor as a linear map from the indices associated with the two triangles and the bottom anyon 1-gon to those associated with the top anyon 1-gon, parameterized by (1) the multiplicity label $\mu$ of the bottom 1-gon, (2) the corresponding string label $a$ of the edge wrapping around that 1-gon, (3) the multiplicity label $\nu$ at the top anyon 1-gon, and (4) the anyon type $\phi$,
\begin{equation}
\label{eq:anyon_measurement}
\bm \rho_{(b'),(cd\alpha\beta a b)}^{(\phi,a',\mu,\nu)}
= \delta_{bb'}\delta_{aa'} {(\rho^\phi)_{ab}^{cd\alpha\beta}}_\mu^\nu
\sqrt{d_cd_d}
\sqrt{\frac{d_\phi}{D}}\;.
\end{equation}
Then we have
\begin{equation}
\label{eq:anyon_worldline_instrument}
    \sum_{\phi,a',\mu,\nu} (\bm\rho^{(\phi,a',\mu,\nu)})^{\dagger} \bm\rho^{(\phi,a',\mu,\nu)} = \pfusion\;.
\end{equation}
Thus the collection $\{\bm\rho^{(\phi,a',\mu,\nu)}\}_{\phi,a',\mu,\nu}\cup \{1-\pfusion\}$ forms a valid set of Kraus operators defining a quantum instrument.
\end{proposition}
\begin{proof}
Plugging Eq.~\eqref{eq:anyon_measurement} into Eq.~\eqref{eq:anyon_worldline_instrument} yields
\begin{equation}
\begin{gathered}
\sum_{\phi,a',\mu,\nu} \big((\bm\rho^{(\phi,a',\mu,\nu)})^{\dagger} \bm\rho^{(\phi,a',\mu,\nu)}\big)_{(cd\alpha\beta ab)(\tilde c\tilde d\tilde\alpha\tilde\beta \tilde a\tilde b)}\\
= \sum_{\phi,a',\mu,\nu,b'}
\frac{d_\phi}{D}
\delta_{bb'}\delta_{aa'} {(\rho^\phi)_{ab}^{cd\alpha\beta}}_\mu^\nu
\sqrt{d_cd_d}\\\cdot
\delta_{\tilde b b'}\delta_{\tilde aa'} {((\rho^\phi)^*)_{\tilde ab}^{\tilde c\tilde d\tilde\alpha\tilde\beta}}_\mu^\nu
\sqrt{d_{\tilde c}d_{\tilde d}}\\
= \delta_{b\tilde b}\delta_{a\tilde a}
\sqrt{d_cd_d d_{\tilde c}d_{\tilde d}}
\sum_{\phi,\mu,\nu}  \frac{d_\phi}{D}
{(\rho^\phi)_{ab}^{cd\alpha\beta}}_\mu^\nu
{((\rho^\phi)^*)_{ab}^{\tilde c\tilde d\tilde\alpha\tilde\beta}}_\mu^\nu\\
=\pfusion
\;.
\end{gathered}
\end{equation}
In the last equation, we have used the anyon completeness axiom:
The sum over $\nu$ and $\mu$ corresponds to evaluating the path integral on a cellulation consisting of two anyon cylinders that are glued at their anyon 1-gons into a ring.
This is a simple cellulation of a solid torus $S_1\times B_2$ with a $\phi$ anyon worldline along $S_1$.
By anyon completeness, summing over $\phi$ on $S_1\times B_2$ is equal to the path integral on the dual solid torus $B_2\times S_1$,
\begin{equation}
\label{eq:anyon_measurement_solid_torus}
\begin{tikzpicture}
\atoms{vertex}{0/, 1/p={-0.6,0}}
\draw[ior,looseness=2] (0)to[out=-60,in=-90]++(0:0.7)to[out=90,in=60](0);
\draw[ior,looseness=2] (1)to[out=-60,in=-90]++(0:1.9)to[out=90,in=60](1);
\draw[or,looseness=2] (0)to[out=75,in=90]++(0:0.9)to[out=-90,in=-35](1);
\draw[or,looseness=2,dashed] (0)to[out=85,in=90]++(0:1.1)to[out=-90,in=-45](1);
\draw (0)edge[or](1);
\end{tikzpicture}
\;.
\end{equation}
The boundary of this solid torus $B_2\times S_1$ is a torus glued from two annuli, one of which is in the back and one of which is in the front in the drawing above.
Each annulus is triangulated by two triangles, and the triangulations are orientation-reversed copies.
The triangulated annulus has no internal vertices.
So by the standard-form axiom, as an operator mapping from the back to the front this is equal to $\pfusion$.
\end{proof}

Applying the $\bm \rho$ measurement corresponds to gluing the bottom anyon 1-gon and both triangles to the 2D cellulation, which deforms the 2D cellulation as follows,
\begin{equation}
\label{eq:2dgs_00pachner_move}
\begin{tikzpicture}
\atoms{vertex}{0/, 1/p={-0.6,0}}
\atoms{cod2vertex}{0x/p={0.4,0}}
\draw[ior,looseness=2] (0)to[out=-60,in=-90]++(0:0.7)to[out=90,in=60](0);
\draw[ior,looseness=2] (1)to[out=-60,in=-90]++(0:1.9)to[out=90,in=60](1);
\draw[or,looseness=2] (0)to[out=80,in=90]++(0:1)to[out=-90,in=-40](1);
\draw (0)edge[or](1);
\draw[extend map] (0)--(0x);
\end{tikzpicture}
\quad\rightarrow\quad
\begin{tikzpicture}
\atoms{vertex}{0/}
\atoms{cod2vertex}{1/p={0.4,0}}
\draw[ior,looseness=2] (0)to[out=-60,in=-90]++(0:0.7)to[out=90,in=60](0);
\draw[extend map] (0)--(1);
\end{tikzpicture}
\;.
\end{equation}
So this removes two triangles that are ``wrapped around'' an anyon 1-gon.

Given the operators $\mathbf F^{22}$, $\mathbf F^{13}$ and $\rho$, we can construct the syndrome extraction layer as follows:
We consider a 3D spacetime triangulation with anyon cylinders foliated by disjoint 2D time slices.
We then consider a way of getting from one time slice to the next one by gluing the tetrahedra and anyon cylinders in between the two slices such that each gluing is of the type in Eq.~\eqref{eq:22pachner_move}, \eqref{eq:2d_31pmove_geo} or \eqref{eq:2dgs_00pachner_move}.
The syndrome extraction layer then consists of the associated $\mathbf F^{22}$, $\mathbf F^{13}$ and $\mathbf \rho$ operators.
We present a concrete choice in Eq.~\eqref{eq:syndrome_extraction_layer} below.
Note that the $\mathbf \rho$-tensors are essential for this:
If the two 2D time slices were separated only by tetrahedra, then one of the gluings would need to be of 3-1 type (like Eq.~\eqref{eq:2d_31pmove_geo} but in the opposite direction), and the associated operator is not a valid quantum instrument.

\myparagraph{Corrections}
The third layer of the circuit is the corrections layer.
The job of this layer is to move each anyon of the current state by some $O(1)$ distance and fuse anyons that end up at the same location.
This layer depends on classical input from the decoder, which determines the directions in which the anyons are moved.
This layer of the circuit is most easily viewed as an adaptive circuit that changes its geometry depending on the instructions from the decoder, but it can also be formulated as a fixed-geometry circuit with additional classical inputs.

Like the syndrome extraction layer, the correction layer consists of path-integral tensors that are applied as linear operators.
This time, we use the anyon cylinder $\rho$ and the fusion connector $S$.
\begin{proposition}
For any fixed anyon type $\phi$ and fixed multiplicity label $\mu$ for the bottom 1-gon, view the $\rho$-tensor as an operator from the bottom 1-gon and the attached triangle to the top 1-gon and its attached triangle, parametrized by the multiplicity label $\nu$ of the top anyon 1-gon,
\begin{equation}
(\bm\rho^{(22),\phi,\mu})^{(\nu)}_{(c'd'b\beta),(cda\alpha)}
\coloneqq
\delta_{cc'} \delta_{dd'}
{(\rho^\phi)_{ab}^{cd\alpha\beta}}_\mu^\nu
\sqrt{d_a d_b}
\;.
\end{equation}
Then we have
\begin{equation}
\label{eq:anyon_movement_krauss}
\sum_{\nu} ((\bm\rho^{(22),\phi,\mu})^{(\nu)})^\dagger (\bm\rho^{(22),\phi,\mu})^{(\nu)} = \pfusion\;.
\end{equation}
Thus, for any fixed $\phi$ and $\mu$, the collection $\{(\bm\rho^{(22),\phi,\mu})^{(\nu)}\}_{\nu} \cup \{1-\pfusion\}$ defines a valid set of Kraus operators giving rise to a quantum instrument.

\end{proposition}
\begin{proof}
Performing the summation over $\nu$ and $b$ in  Eq.~\eqref{eq:anyon_movement_krauss} is the same as gluing two anyon cylinders at a pair of 1-gons and a pair of triangles, which yields the following volume,
\begin{equation}
\sqrt{d_ad_{\tilde a}}\quad
\begin{tikzpicture}
\atoms{vertex}{0/, 3/p={2,0}}
\atoms{cod2vertex}{{1/p={-20:0.65},dlab={\mu}{-20:0.25}}, {2/p={20:0.65},vback,dlab={\mu}{20:0.25}}}
\draw[or,mark={slab=$\tilde a$,r,p=0.55}] (0)to[out=-60,in=-110]++(-20:1.1)to[out=70,in=20](0);
\draw[or,dashed,mark={slab=$a$,r,p=0.45}] (0)to[out=-20,in=-70]++(20:1.1)to[out=110 ,in=60](0);
\draw (0)edge[looseness=1.5,bend left=75,or,glab=c](3) (0)edge[looseness=1.5,bend right=75,or,glabr=d](3);
\draw[extend map] (0)--(1) (0)edge[dashed](2);
\node at (1.5,0.3){$\alpha$};
\node at (1.5,-0.3){$\tilde\alpha$};
\end{tikzpicture}
= \delta_{\mu\mu} \delta_{a\tilde a} \delta_{\alpha\tilde \alpha}\;.
\end{equation}
This is a volume whose boundary consists of the same 2D cellulation at the back and the front without any internal vertices.
So by the normal-form axiom of the path integral with anyon worldlines, the associated path-integral tensor is $\pfusion$.
\end{proof}

Gluing the bottom 1-gon and attached triangle of the anyon cylinder to a 1-gon and attached triangle (corresponding to the operator $\bm\rho^{(22)}$) deforms the 2D spatial cellulation like this:
\begin{equation}
\label{eq:anyon_hopping}
\begin{tikzpicture}
\atoms{vertex}{0/, 1/p={1.5,0}}
\atoms{cod2vertex}{2/p={0.5,0}}
\draw (0)edge[bend left=45,or](1) (0)edge[bend right=45,or](1);
\draw[extend map] (0)--(2);
\draw[or] (0)to[out=-45,in=-90]++(0:0.8)to[out=90,in=45](0);
\end{tikzpicture}
\quad\rightarrow\quad
\begin{tikzpicture}
\atoms{vertex}{0/, 1/p={1.5,0}}
\atoms{cod2vertex}{2/p={1,0}}
\draw (0)edge[bend left=45,or](1) (0)edge[bend right=45,or](1);
\draw[extend map] (1)--(2);
\draw[or] (1)to[out=-135,in=-90]++(180:0.8)to[out=90,in=135](1);
\end{tikzpicture}
\;.
\end{equation}
So this moves the attached anyon 1-gon from one vertex to another.

We can also turn $S$ into an operator.
\begin{proposition}
For two fixed input anyon types $\phi$ and $\chi$ and multiplicity labels $\mu$ and $\nu$ of their anyon 1-gons, view the tensor $S$ as an operator from two of its boundary anyon 1-gons and the boundary triangle to the remaining anyon 1-gon, parametrized by (1) the outgoing anyon type $\psi$, (2) the corresponding multiplicity label $\lambda$, and (3) the anyon fusion label $\epsilon$,
\begin{equation}
(\bm S^{\phi,\mu,\chi,\nu})^{(\psi,\lambda,\epsilon)}_{(c'),(abc\alpha)}
\coloneqq
(S_{\phi\textcolor{col2}{\chi}}^{\textcolor{col3}{\psi}})^{abc\alpha}_{\mu\textcolor{col2}{\nu}\textcolor{col3}{\lambda}\epsilon}
\delta_{cc'}
\sqrt{d_ad_bd_\psi}\;.
\end{equation}
Then we have
\begin{equation}
\label{eq:anyon_fusion_kraus_property}
\sum_{\psi,\lambda,\epsilon} ((\bm S^{\phi,\mu,\chi,\nu})^{(\psi,\lambda,\epsilon)})^\dagger (\bm S^{\phi,\mu,\chi,\nu})^{(\psi,\lambda,\epsilon)} = \pfusion\;.
\end{equation}
So for any fixed $\phi$, $\mu$, $\chi$, $\nu$, the collection $\{(\bm S^{\phi,\mu,\chi,\nu})^{(\psi,\lambda,\epsilon)}\}_{\psi,\lambda,\epsilon}\cup \{1-\pfusion\}$ defines a valid set of Kraus operators.
\end{proposition}
\begin{proof}
The contraction in Eq.~\eqref{eq:anyon_fusion_kraus_property} corresponds to a 3-ball with two anyon fusion points, separated by a $\psi$ anyon worldline.
By the fusion space completeness axiom, summing over $\epsilon$ corresponds to attaching a 1-handle to the two fusion points, such that there is a $\psi$ loop wrapping around that 1-handle.
By the anyon completeness axiom, summing over $\psi$ corresponds to attaching a 2-handle to this $\psi$-loop, effectively removing the $\psi$-loop and the 1-handle.
The result is a 3-ball with a $\phi$ and a $\chi$ anyon worldline connecting the ``front'' and ``back'' disks of the 3-ball boundary.
Geometrically, the boundary of the 3-ball consists of two copies of the same 2D cellulation consisting of 2 anyon 1-gons and one triangle shown below in Eq.~\eqref{eq:anyon_fusion_move}.
This cellulation has no internal vertices, hence by the normal form axiom, the associated tensor is equal to $\pfusion$.
\end{proof}

Gluing two 1-gons and the triangle of the fusion volume deforms the spatial lattice like:
\begin{equation}
\label{eq:anyon_fusion_move}
\begin{tikzpicture}
\atoms{vertex}{0/}
\atoms{cod2vertex}{1/p={-30:0.5}, 2/p={30:0.5}}
\draw[or] (0)to[out=-60,in=-120]++(-30:0.8)to[out=60,in=0](0);
\draw[or] (0)to[out=0,in=-60]++(30:0.8)to[out=120,in=60](0);
\draw[or,looseness=2] (0)to[out=-70,in=-90]++(0:1.2)to[out=90,in=70](0);
\draw[extend map] (0)--(1) (0)--(2);
\end{tikzpicture}
\quad\rightarrow\quad
\begin{tikzpicture}
\atoms{vertex}{0/}
\atoms{cod2vertex}{2/p={0.8,0}}
\draw[extend map] (0)--(2);
\draw[or] (0)to[out=-60,in=-90]++(0:1.2)to[out=90,in=60](0);
\end{tikzpicture}
\;.
\end{equation}
So this move fuses two nearby anyon 1-gons at the same vertex into one.

Furthermore, we can ``rotate'' the location of an anyon 1-gon around a vertex by simply gluing a tetrahedron, corresponding to an $\bm F^{(22)}$ operator,
\begin{equation}
\label{eq:anyon_rotation}
\begin{tikzpicture}
\atoms{vertex}{0/, 1/p={1.5,0}, 3/p={60:1.5}}
\atoms{cod2vertex}{2/p={0.5,0}}
\draw (0)edge[bend left=45,or](1) (0)edge[bend right=45,or](1);
\draw (0)edge[or](3) (3)edge[or](1);
\draw[extend map] (0)--(2);
\draw[or] (0)to[out=-45,in=-90]++(0:0.8)to[out=90,in=45](0);
\end{tikzpicture}
\quad\rightarrow\quad
\begin{tikzpicture}
\atoms{vertex}{0/, 1/p={1.5,0}, 3/p={60:1.5}}
\atoms{cod2vertex}{2/p={60:0.5}}
\draw (0)edge[bend left=45,or](3) (0)edge[bend right=45,or](3);
\draw (0)edge[or](1) (3)edge[or](1);
\draw[extend map] (0)--(2);
\draw[or] (0)to[out=-45+60,in=-90+60]++(60:0.8)to[out=90+60,in=45+60](0);
\end{tikzpicture}
\end{equation}

Finally, we can create a trivial anyon 1-gon at a vertex, for example,
\begin{equation}
\begin{tikzpicture}
\atoms{vertex}{0/, 1/p={1.5,0}, 3/p={60:1.5}}
\draw (0)edge[or](1);
\draw (0)edge[or](3) (3)edge[or](1);
\end{tikzpicture}
\quad\rightarrow\quad
\begin{tikzpicture}
\atoms{vertex}{0/, 1/p={1.5,0}, 3/p={60:1.5}}
\atoms{vertex}{2/p={0.5,0}}
\draw (0)edge[bend left=45,or](1) (0)edge[bend right=45,or](1);
\draw (0)edge[or](3) (3)edge[or](1);
\draw[or] (0)--(2);
\draw[or] (0)to[out=-45,in=-90]++(0:0.8)to[out=90,in=45](0);
\end{tikzpicture}
\end{equation}
by gluing two tetrahedra.

Combining all three moves above, we can move an anyon 1-gon to a neighboring vertex and fuse it with the existing anyon 1-gon there.

\myparagraph{Square-lattice realization}
In the previous paragraphs, we have introduced a procedure to write down a fault-tolerant circuit for any 2D cellulation that is equipped with enough anyon 1-gons.
To make things more concrete, we now consider a square-lattice realization of the circuit.
We attach an anyon 1-gon to each square-lattice vertex, and triangulate the rest,
\begin{equation}
\label{eq:compactification}
\begin{tikzpicture}
\clip (1.2,1)rectangle++(5,4);
\foreach \x in {0,1,2,3,4}{
\foreach \y in {0,1,2,3,4,5}{
\atoms{vertex}{0\x\y/p={1.5*\x,1.5*\y}}
\atoms{cod2vertex}{1\x\y/p={1.5*\x+0.5,1.5*\y}}
\draw (0\x\y)edge[or,bend right=45]++(0:1.5) (0\x\y)edge[or]++(90:1.5);
\draw[or] (0\x\y)to[out=-45,in=-90]++(0:0.8)to[out=90,in=45](0\x\y);
\draw (0\x\y)edge[or,bend left=45]++(0:1.5) (0\x\y)edge[or]++(1.5,1.5);
\draw[extend map] (0\x\y)--(1\x\y);
}}
\end{tikzpicture}
\end{equation}
As one can see, we wrap each anyon 1-gon with a triangle yielding a 2-gon.
We then arrange these 2-gons horizontally and triangulate the remaining squares with two triangles each.

\myparagraph{Square lattice fusion-space recovery}
Let us now look at the fusion-space recovery layer of the circuit on the square lattice above.
This works by applying the $C^t_e$ operators to the three triangles of each unit cell in the following order:
\begin{equation}
\begin{tikzpicture}
\clip (1.2,1)rectangle++(2.2,2.2);
\foreach \x in {0,1,2}{
\foreach \y in {0,1,2}{
\atoms{vertex}{0\x\y/p={1.5*\x,1.5*\y}}
\atoms{cod2vertex}{1\x\y/p={1.5*\x+0.5,1.5*\y}}
\draw[or] (0\x\y)to[bend left=45]coordinate[pos=0.6](m\x\y)++(0:1.5);
\draw[or] (0\x\y)to++(90:1.5);
\draw[or=0.6] (0\x\y)to[out=-45,in=-90]++(0:0.8)to[out=90,in=45](0\x\y);
\draw (0\x\y)edge[or,bend right=45]++(0:1.5) (0\x\y)edge[or]++(1.5,1.5);
\draw[extend map] (0\x\y)--(1\x\y);
}}
\node[red,font=\small, inner sep=0,circle](x0) at (1.9,2.4){0};
\draw[red] (x0)edge[->]($(011)!0.4!(022)$);
\node[red,font=\small, inner sep=0,circle](x1) at (2.6,2.2){1};
\draw[red] (x1)edge[->](m11);
\node[red,font=\small, inner sep=0,circle](x2) at (2.6,1.5){2};
\draw[red] (x2)edge[->](2.3,1.5);
\end{tikzpicture}
\end{equation}
That is, we first apply $C^t_e$ for the triangle $t$ marked ``$0$'' with the edge $e$ that the red arrow is pointing to, then at $t$ marked $1$ with the arrow edge, and then at $t$ marked $2$ with the arrow edge.
We do this in parallel inside every unit cell, so the fusion-subspace recovery is a constant-depth circuit.

\myparagraph{Square lattice syndrome extraction}
Now take the product of the spatial lattice with a sequence of edges in the time direction.
This yields a 3D spacetime cellulation consisting of triangle prisms and anyon cylinders.
We further decompose each triangle prism into three tetrahedra,
\begin{equation}
\label{eq:anyon_compactification}
\begin{tikzpicture}
\atoms{vertex}{{0/}, {2/p={1.6,0}}, {1/p={1,0.5}}}
\atoms{vertex}{{0x/p={0,1.4}}, {2x/p={1.6,1.4}}, {1x/p={1,1.9}}}
\draw (0)edge[mark=arr,backedge](1) (1)edge[mark={arr},backedge](2) (0)edge[mark={arr}](2) (0x)edge[mark=arr](1x) (1x)edge[mark={arr}](2x) (0x)edge[mark={arr}](2x) (0)edge[mark=arr](0x) (1)edge[mark=arr,backedge](1x) (2)edge[mark=arr](2x) (0)edge[mark=arr,backedge](1x) (1)edge[mark=arr,backedge](2x) (0)edge[mark=arr](2x);
\end{tikzpicture}\;.
\end{equation}
The syndrome extraction layer of the circuit is defined by going from one 2D time slice to the next by gluing all tetrahedra and anyon cylinders in the depth-1 layer of 3D cellulation in between.

\begin{proposition}
There is an ordering of gluing the spacetime tetrahedra and anyon cylinders of the depth-1 3D cellulation layer to the 2D lattice, such that each gluing is of the form in Eqs.~\eqref{eq:22pachner_move}, \eqref{eq:2d_31pmove_geo}, or \eqref{eq:2dgs_00pachner_move}.
\end{proposition}
\begin{proof}
The ordering is illustrated in the following picture:
\begin{equation}
\label{eq:syndrome_extraction_layer}
\begin{tikzpicture}
\clip (1.2,1)rectangle++(2.2,2.2);
\foreach \x in {0,1,2}{
\foreach \y in {0,1,2}{
\atoms{vertex}{0\x\y/p={1.5*\x,1.5*\y}}
\atoms{cod2vertex}{1\x\y/p={1.5*\x+0.5,1.5*\y}}
\draw (0\x\y)edge[or,bend right=45]++(0:1.5) (0\x\y)edge[or]++(90:1.5);
\draw[or] (0\x\y)to[out=-45,in=-90]++(0:0.8)to[out=90,in=45](0\x\y);
\draw (0\x\y)edge[or,bend left=45]++(0:1.5) (0\x\y)edge[or]++(1.5,1.5);
\draw[extend map] (0\x\y)--(1\x\y);
}}
\node[red,font=\small]at (1.9,2.4){035};
\node[red,font=\small]at (2.6,2.2){146};
\node[red,font=\small]at (2.6,1.5){278};
\node[red,font=\small]at (1.75,1.5){9};
\end{tikzpicture}
\end{equation}
Each triangle in the 2D cellulation corresponds to a triangle prism in the 3D layer, which is triangulated by three tetrahedra that are stacked in the time direction.
So marking a triangle in the 2D triangulation means gluing the next of the three tetrahedra in the stack corresponding to this triangle.
Marking the anyon 1-gon means gluing the corresponding anyon cylinder in the 3D layer.
There are 9 tetrahedra and 1 anyon cylinder per unit cell, giving rise to a 10-step gluing process.
In the above drawing, we have put each step number (0 to 9) at the triangle or anyon 1-gon where the gluing happens.
Step 0 corresponds to a gluing as in Eq.~\eqref{eq:2d_31pmove_geo}.
Steps 1-8 are gluings as in Eq.~\eqref{eq:22pachner_move}.
Step 9 is a gluing as in Eq.~\eqref{eq:2dgs_00pachner_move}.
Overall, step 0 creates two additional triangles, and step 9 removes two triangles, such that we are back at the original 2D cellulation.
\end{proof}

\myparagraph{Square lattice corrections}
Finally, we discuss the correction layer on the square lattice.
The classical input for the correction layer determined by the decoder consists of one \emph{movement direction} $\mu_{x,y}\in \{0,+1,-1\}\times \{0,+1,-1\}$ for every vertex coordinate $(x,y)$ of the 2D square lattice (which is $(0,0)$ if the measurement outcome of the previous syndrome extraction layer yielded the trivial anyon type).
The circuit then consists in using Eqs.~\eqref{eq:anyon_hopping} and \eqref{eq:anyon_rotation} to move the anyon at $(x,y)$ to $(x+\mu_{x,y}^{(0)}, y+\mu_{x,y}^{(1)})$, where $\mu_{x,y}^{(0)}$ and $\mu_{x,y}^{(1)}$ are the first and second components of $\mu_{x,y}$.
After moving all the anyons, we fuse all the anyons whose 1-gons are based at the same vertex using Eqs.~\eqref{eq:anyon_rotation} and \eqref{eq:anyon_fusion_move}.

\subsection{Error model}
We view the overall circuit including fusion-subspace recovery, syndrome extraction, and correction as a quantum circuit with classical output (measurement results) and classical inputs (correction directions).
An individual error event consists of replacing any (classically controlled) channel in the circuit with any other channel.
An error configuration is a configuration of error events happening at different spacetime locations.
The noise model assigns a probability to each error configuration.
Here, we assume that the noise model is \emph{$\epsilon$-bounded} (also called local stochastic), meaning that for any set of $w$ error locations, the probability that the error configuration contains errors at all of these locations is smaller than $\epsilon^w$.%

\section{Decoding algorithm}
\label{sec:decoding}
\subsection{Notation}%
\label{sec:spacetime-metric}
We work on a spatial square lattice \(\mathbb X\) and a cubic spacetime lattice \(\intG\).
For a spacetime point $p$, $\mathbb X(p)$ and $\mathbb T(p)$ give its spatial position and time, respectively.
We choose \(\mathbb X=(\mathbb Z/L\mathbb Z)^2\), where \(L\) is the integer
linear size of the spatial lattice.
For spacetime points \((x,t)\) and \((y,s)\), use the box metric
\[
\begin{aligned}
d_{\Box}((x,t),(y,s))=\max\left\{\max_{i=1,2}\min_{k\in\mathbb Z}|x_i-y_i+kL|,\ |t-s|\right\}.
\end{aligned}
\]
Thus spatial coordinates \(0\) and \(L-1\) have distance \(1\) for
\(L\ge2\), whereas the time distance is the ordinary absolute difference.
For finite nonempty sets \(A,A'\), define
\[
d_{\Box}(A,A')=\min_{a\in A,\ a'\in A'}d_{\Box}(a,a').
\]
Next, we define the box radius of a set $A$, which is the  maximum linear size of the set in spacetime:
\[
\rad_{\Box}(A)=
\frac{\max_{a,a'\in A}d_{\square}(a,a')+1}{2}.
\]
Let \(H(A)\) be the smallest integer-sized spacetime box enclosing \(A\).
For \(R\ge0\), define the $R$-neighborhood box:
\begin{equation}
\BN_R(A)=\bigl(H(A)+([-R,R]^3\cap\mathbb Z^3)\bigr)\cap\intG
\end{equation}
where sddition is modulo \(L\) in the two spatial coordinates and ordinary
in time.
If $A$ is a single point $x$, we may write $\BN_R(x)$ for $\BN_R(\{x\})$.

\subsection{General cluster decomposition}
In this subsection, we describe a cluster decomposition, whose input is a subset of vertices of the 3D cubic spacetime lattice.  This decomposition is used both in the decoding algorithm in the next subsection and the fault-tolerance proof in the next section.

We say that two sets of points are $g$-near for some function $g$ if the neighborhood of one intersects the other set, and vice versa.  The function $g$ indicates how large the neighborhood of a set is as a function of the radius of the set.%

\begin{definition}[Mutual \(g\)-nearness]
\label{def:only-g-nearness}
Let \(g:[\tfrac12,\infty)\to[0,\infty)\) satisfy
\[
g(r)>r,\qquad g(v)-g(u)\ge v-u
\quad(\tfrac12\le u\le v).
\]
For a finite nonempty set \(A\), write
\[
\BN_g(A)=\BN_{g(\rad_{\Box}(A))-\rad_{\Box}(A)}(A).
\]
Two subsets $A\subset \intG$ and $B\subset \intG$ are mutually $g$-near if
\begin{align*}
A\cap  \BN_g(B) &\ne\varnothing\text{ and }\\
B\cap \BN_g(A) &\ne\varnothing\;,
\end{align*}
written as 
\(A\sim_g B\).
\end{definition}

Using the notion of $g$-nearness, we can now define the  cluster decomposition of a subset $A\subset \intG$.

\begin{definition}\label{def:only-g-cluster-closure}
Let \(A\subset \intG\) be a finite subset of points of the cubic lattice.
The cluster decomposition of $A$ is the finest partition of $A$ into a collection of subsets $\mathcal K_g(A)$ (called clusters) such that (1) no two distinct subsets in $\mathcal K_g(A)$ are mutually $g$-near, and (2) any nonempty proper subset $c\subsetneq C\in \mathcal K_g(A)$ is $g$-near to its complement $C\setminus c$. We call $g$ the separation function of the cluster decomposition.
\end{definition}

The cluster decomposition is unique and is the output of the clustering algorithm below.
This follows quite straightforwardly from the fact that $\BN_g(A)\subset \BN_g(A\cup B)$ for any $A,B\subset \intG$; for more details see the discussion around Proposition~\ref{prop:only-clustering-termination}.
The pseudocode in Algorithm~\ref{alg:clustering} can be described quite simply in words: Start with each point forming its own cluster, and repeatedly fuse two clusters if they are mutually $g$-near, until there are no remaining pairs of clusters that are mutually $g$-near.

\begin{algorithm}[t]
\caption{\(g\)-cluster decomposition
(Definition~\ref{def:only-g-cluster-closure})}
\label{alg:clustering}
\begin{algorithmic}[1]
\Require A finite subset \(A\subset\intG\) of points of the cubic lattice
and a separation function \(g\) as in Definition~\ref{def:only-g-nearness}.
\Ensure The cluster decomposition \(\mathcal K_g(A)\). %
\State \(\mathcal K_g(A)\gets\{\{x\}:x\in A\}\).
\While{there exist distinct \(C_0,C_1 \in \mathcal K_g(A)\) that are mutually
\(g\)-near, $C_0\sim_g C_1$}
  \State \(\mathcal K_g(A)\gets
  \bigl(\mathcal K_g(A)\setminus\{C_0,C_1\}\bigr)
  \cup\{C_0\sqcup C_1\}\).
\EndWhile
\State \Return \(\mathcal K_g(A)\).
\end{algorithmic}
\end{algorithm}

\subsection{Decoding algorithm}
The input to the decoding algorithm at time step $t$ is the collection of all measurement outcomes obtained before, and including, time $t$.
The protocol is still fault-tolerant if we restrict the input to the last $\sim L$ time steps for a distance-$L$ code; assuming that the decoder uses the full history of measurement outcomes simplifies the discussion slightly.
Each such measurement outcome can be associated with one vertex in the cubic spacetime lattice using the square-lattice spacetime construction following Eq.~\eqref{eq:compactification}; we associate an anyon cylinder with the spacetime vertex of its ``older'' anyon 1-gon.
The output consists of a set of movement directions $\mu_{x,y}\in \{0,+1,-1\}\times\{0,+1,-1\}$, one for each spatial vertex with coordinate $(x,y)$. These directions are then fed as classical control data into the correction layer of the circuit.  The anyon type itself need not be specified by the decoder, since it is supplied locally by the preceding syndrome-measurement outcome.

\begin{figure}[!htbp]
\centering
\begin{tikzpicture}[x=0.288cm,y=0.288cm,>=stealth]
  \definecolor{MoveBlue}{RGB}{0,65,115}
  \definecolor{MoveRed}{RGB}{175,35,45}

  \draw[step=1,gray!35,line width=0.3pt]
    (0,0) grid (\dimexpr\columnwidth-0.3pt\relax,15);
  \draw[gray!35,line width=0.3pt]
    (\dimexpr\columnwidth-0.3pt\relax,0) -- (\dimexpr\columnwidth-0.3pt\relax,15);

  \foreach \sx/\sy in {
    2/9,3/5,4/12,5/2,6/7,8/13,9/9,11/8,
    11/4,14/10,15/3,16/5,8/2}
  {
    \draw[->,MoveBlue,line width=0.425pt,opacity=0.5]
      (\sx,\sy) -- ({\sx+sign(13-\sx)},{\sy+sign(5-\sy)});
    \fill[MoveBlue] (\sx,\sy) circle (1.65pt);
  }
  \foreach \sx/\sy in {23/8,24/5,26/9,28/8,27/5,25/4}
  {
    \draw[->,MoveRed,line width=0.425pt,opacity=0.5]
      (\sx,\sy) -- ({\sx+sign(25-\sx)},{\sy+sign(7-\sy)});
    \fill[MoveRed] (\sx,\sy) circle (1.65pt);
  }

  \fill[MoveBlue] (13,5) ++(-2.5pt,-2.5pt) rectangle ++(5pt,5pt);
  \node[MoveBlue,above right=2pt,font=\small] at (13,5) {\(b\)};
  \fill[MoveRed] (25,7) ++(-2.5pt,-2.5pt) rectangle ++(5pt,5pt);
  \node[MoveRed,above right=2pt,font=\small] at (25,7) {\(b'\)};

  \node[MoveBlue,above=2pt,font=\scriptsize] at (3.5,8.5)
    {\(\mu_{x,y}\)};
\end{tikzpicture}
\caption{
An example of movement directions $\mu_{x,y}$ for a given subset of anyons on the 2D spatial lattice.
The points are elements of $S_{=t}$ where non-trivial anyons have been measured in the last syndrome extraction layer.
The blue and red points are a part of two different spacetime clusters, whose (spatially projected) base points $b$ and $b'$ are represented by squares of the same color.
The arrows indicate the movement directions.
}
\label{fig:only-full-slice-inward-move}
\end{figure}
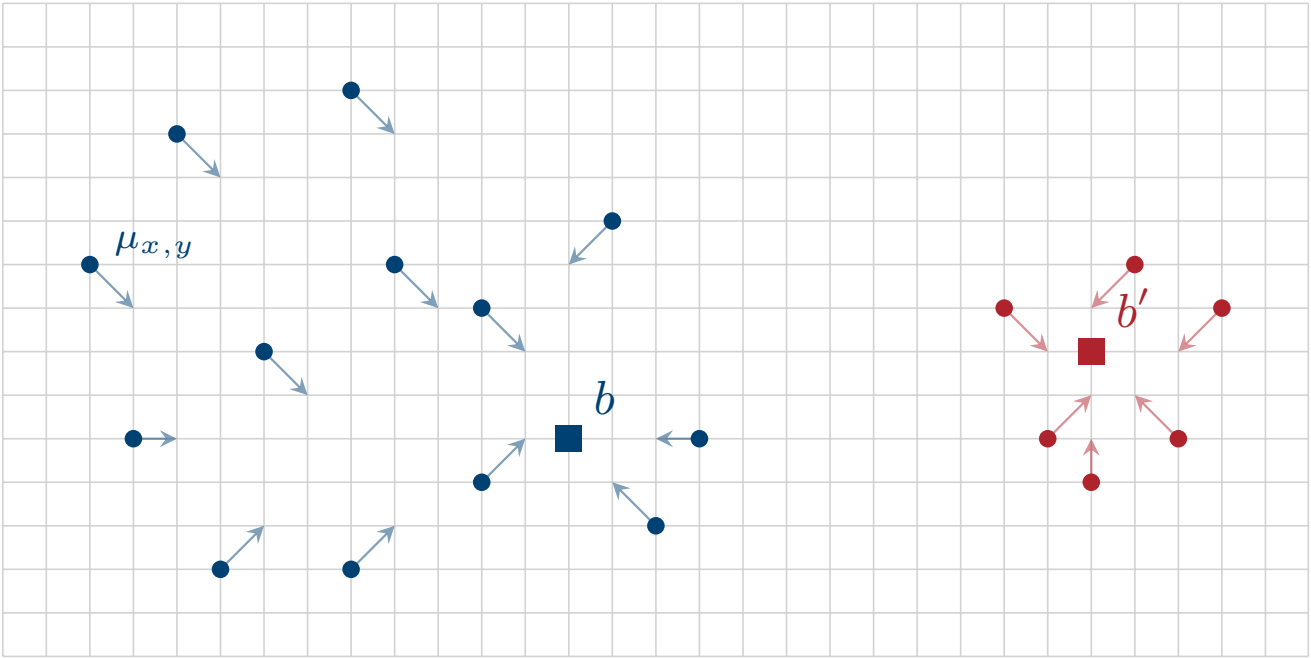

The algorithm operates by clustering the set of all spacetime points up to time $t$ containing non-trivial measurement outcomes (corresponding to a non-trivial anyon type).
The clustering is applied with the separation function \(\beta(r)\coloneqq C_\beta r\) for some constant $C_\beta$ defined in Lemma~\ref{lem:only-auxiliary-constraints}.
Write $\Delta_L(a)$ for the representative of $a\pmod L$ of smallest absolute value, choosing the positive representative in a tie. Movement directions use these signed cyclic displacements.
It determines a \emph{base point} for each cluster, which is the vertex in the cluster with the lexicographically smallest coordinate where we compare the time component first.
Then, it chooses the movement directions $\mu_{x,y}$ such that each spatial point with a non-trivial anyon measurement in the previous syndrome extraction layer is moved towards the spatial projection of the base point of its cluster.
The movement towards the base point is illustrated in Figure~\ref{fig:only-full-slice-inward-move}.
The algorithm is summarized in Algorithm~\ref{alg:only-error-correction}.
\begin{algorithm}[t]
\caption{Decoding at time \(t\)}
\label{alg:only-error-correction}
\begin{algorithmic}[1]
\Require Subset $S_{\leq t}$ of spacetime vertices with non-trivial measurement outcomes.
\Ensure Movement directions \(\mu_{x,y}\) at time \(t\).
\State Apply Algorithm~\ref{alg:clustering} to $S_{\leq t}$ with the separation function $\beta(r)\coloneqq C_\beta r$ to obtain the cluster decomposition $\mathcal K$. %
\State For each cluster $C\in \mathcal K$, let the base point $b(C)\in C$ be the vertex with the lexicographically smallest $(t,x,y)$ coordinate. %
\For{each spatial coordinate $(x,y)$}
\State If $(x,y)\in S_{=t}$ (the time-$t$ slice of $S_{\leq t}$), then let $C\in \mathcal K$ be the cluster containing $(x,y)$.
Set $\mu_{x,y} \gets (\operatorname{sgn}(\Delta_L(b(C)_x-x)), \operatorname{sgn}(\Delta_L(b(C)_y-y)))$, where $\operatorname{sgn}(a)$ is $+1$ if $a>0$, $-1$ if $a<0$ and $0$ if $a=0$.
\State If $(x,y)\notin S_{=t}$, set $\mu_{x,y} \gets (0,0)$.
\EndFor
\end{algorithmic}
\end{algorithm}

\section{Fault-tolerance proof}
\label{sec:proof}

\subsection{Overview and statement}
In this section, we prove that the probability for a logical memory error realized by the circuit from Section~\ref{sec:circuit} using the decoder introduced in Section~\ref{sec:decoding} decays quasi-stretched-exponentially in the system size. %
The proof consists of three parts:
First, we show in Section~\ref{sec:contractible-no-fault} that for a spacetime trajectory with a noise realization containing only a contractible \emph{damage configuration} no logical fault occurs.
This is a direct consequence of the path-integral picture for the circuit.
Here, the ``damage'' consists of both error events and non-trivial measurement outcomes.

Second, we show in Section~\ref{sec:burst-decomposable} that spacetime trajectories with noise realizations whose error configuration admits a \emph{burst decomposition} have a contractible damage configuration.
This is the most difficult part of the proof which requires induction over different sizes of ``bursts'', which are clusters of errors.
The anyon worldlines created by error bursts may interact with other bursts before they can be detected and fused to vacuum.
We thus need to consider larger ``conglomerates'' of bursts and their anyon worldlines, and bound the size of these conglomerates.
Third, we show in Section~\ref{sec:probabilistic} that error configurations that do not admit a burst decomposition are quasi-stretched-exponentially unlikely in the system size.
This is a fact known in the literature, but we provide a simple proof in our context.

Let us now briefly state what exactly we mean by the circuit being a fault-tolerant memory.
The logical information is encoded in the ground-state space of the corresponding Turaev-Viro/string-net model.
\begin{definition}
Consider the quantum channel $C_{L\times L\times T}$ corresponding to the noisy circuit from Section~\ref{sec:circuit} on an $L\times L$ lattice run for $T$ periods of time, where corrections/movement directions are chosen by the classical decoding algorithm based on the previous measurement results.
Let $E$ be the encoding isometry exactly embedding the logical Hilbert space into the physical Hilbert space.
Let $D$ be the decoding channel consisting of running the \textbf{noiseless} circuit for another $L$ time steps, where the measurement results used by the decoder include those collected during $C_{L\times L\times T}$ (i.e., formally this is a channel with additional classical input).
Then the logical error rate $p_L$ is the smallest probability for which there exists a channel $C_{\text{error}}$ such that
\begin{equation}
D\circ C_{L\times L\times T} \circ E = (1-p_L)\mathbb 1 + p_L C_{\text{error}}\;.
\end{equation}
\end{definition}
In the main Theorem~\ref{thm:only-main-threshold}, we will show that the logical error rate decays quasi-stretched-exponentially with $L$.

So we do not show fault-tolerance for a complete logical circuit including fault-tolerant state preparation and readout.
We merely show that there exists some way to encode and decode logical information in and out of the circuit such that it preserves the logical information for a long time.
For the proof, we concretely choose the decoding map to be implemented by hypothetical $O(L)$ extra noise-free syndrome-extraction and correction steps.
In Section~\ref{sec:logical_operation} we briefly discuss how the proof can be adapted not only to fault-tolerant state preparation and readout, but to any logical gate corresponding to a spacetime 3-manifold with boundaries, corners, domain walls, etc., each labeled by a topological superselection sector.

\subsection{Contractible damage means no logical fault}
\label{sec:contractible-no-fault}
\begin{definition}[Damage]
Expand the noisy circuit as
\begin{equation}
C_{L\times L\times T} = \sum_{\vec e, \vec m} P(\vec e, \vec m) \cdot C_{L\times L\times T}^{\vec e,\vec m}\;,
\end{equation}
where $\vec e$ is a configuration of errors and $\vec m$ is a configuration of measurements, and we call the pair $(\vec e,\vec m)$ a \textbf{circuit realization}.
$P(\vec e,\vec m)$ is the joint probability distribution of errors and measurement results (including the fact that the circuit is adaptive) and $C_{L\times L\times T}^{\vec e,\vec m}$ denotes the linear operator corresponding to the circuit with a fixed $(\vec e,\vec m)$ configuration.

We call the subset of vertices of the cubic spacetime lattice that either contain an error in $\vec e$ or a non-trivial measurement outcome in $\vec m$ \textbf{the damage} of $(\vec e,\vec m)$.
We also call all nontrivial measurement outcomes in $\vec m$ the \textbf{syndrome}, which includes both anyon worldlines and (false positive) measurement errors. The damage includes the syndrome as well as the (false negative) measurement errors.
\end{definition}

\begin{proposition}[Contractible damage leads to no logical error]
\label{prop:only-contractible-no-fault}
Consider a circuit realization $(\vec e,\vec m)$ whose damage can be bounded by a region with the topology of a 3-ball after adding a depth-1 padding on the cubic lattice.
We call such a realization \textbf{contractible}, since, roughly speaking, the damage does not contain any non-contractible loop.
Then, $C_{L\times L\times T}^{\vec e,\vec m}\circ E=\mathbb 1\circ E$, that is, no logical fault occurs in that circuit realization.
\end{proposition}
\begin{proof}
Note that the damage indirectly includes the effect of the feedback implemented by the correction layer of the circuit due to the decoding algorithm. 

The circuit was constructed such that, away from any errors and post-selected on a particular configuration of measurement outcomes, it is equal to the Turaev--Viro path integral with inserted anyon worldline cylinders based on the measurements.
If the damage is contained in a 3-ball ($B_3$) after depth-1 padding, then it is surrounded by a region of error-free and anyon-worldline-free path integral of topology $S_2\times [0,1]$.
By the irreducibility axiom, the path integral on $S_2\times [0,1]$ is an exterior product $\ket v\bra w$ with respect to the division of the boundary into $S_2\times 0$ and $S_2\times 1$.
So if $T$ is the circuit on the unpadded $B_3$ without any errors or anyon worldlines, and $T'$ is the circuit on the unpadded $B_3$ including the damage, then we trivially have $\braket{T|v}\bra w \propto \braket{T'|v}\bra w$.
Thus, the overall circuit on the full spacetime $S_1\times S_1\times [0,1]$ with given damage inside $B_3$ is proportional to the path integral on $S_1\times S_1\times [0,1]$ without any errors or non-trivial anyon worldlines.
By definition, the path integral on $S_1\times S_1\times [0,1]$ is the ground space projector on $S_1\times S_1$, which acts as the identity on the ground space.
So the post-selected circuit is equal to the identity, since its normalization is fixed by the trace-preserving property.
\end{proof}

The path-integral axioms also allow us to explicitly prove the following statement about the behavior of anyon worldlines when moved and fused by the correction layer.
\begin{proposition}[Damage clusters are neutral]
\label{prop:only-damage-neutrality}
Suppose part of the damage is enclosed in a spacetime 3-ball whose depth-1 padding contains no errors, including measurement errors. If the only nontrivial anyon worldline in the padding is a single worldline of type $\phi \neq 1$ crossing from the inside to the outside, then the circuit run has probability zero.
\end{proposition}
\begin{proof}
The damage is surrounded by an error-free path integral of topology $S_2\times [0,1]$ with a single $\phi$ anyon worldline connecting $S_2\times 0$ with $S_2\times 1$.
Thus, by the anyon exclusion axiom for $\chi=1$, the path integral is zero.
This implies that the probability for actually obtaining the measurement result $\phi$ along the $\phi$-worldline during the syndrome extraction layer is zero.
\end{proof}

Informally, the proposition above states that the total anyon charge of all anyons created by an isolated noise event is zero. Thus, fusing the anyon worldlines together (which is what the decoding algorithm does) will yield the trivial anyon.

\subsection{Burst-decomposable error configuration means contractible damage}
\label{sec:burst-decomposable}
We define the burst decomposition as follows:
\begin{definition}[Burst decomposition]
\label{def:only-burst-decomposition}
Let \(C_\gamma>0\) and \(\zeta>1\) be constants defined in Lemma~\ref{lem:only-auxiliary-constraints}.
The \textbf{burst decomposition} $\mathcal B$ is the cluster decomposition (Algorithm~\ref{alg:clustering}) applied to the error configuration with the separation function
\begin{align*}
\gamma(r) = C_\gamma r \log^\zeta(e+r),
\end{align*}
where $e$ is the Euler constant.
We call the maximal radius of any burst in $\mathcal B$ the \textbf{maximal burst size}.
\end{definition}
Below is one of the two main theorems of this section:
\begin{theorem}[Burst-decomposable noise is correctable]
\label{thm:only-main-deterministic}
Consider the circuit and recovery procedure of Proposition~\ref{prop:only-contractible-no-fault}, run with an error realization whose maximal burst size is at most 
\begin{equation} \label{eq:maximal}
    \rad_\Box(B) \leq \frac{L-2}{800 \lambda^\C Q},
\end{equation}
where $\lambda^{\C}>1$ and $Q>1$ are the constants defined in Lemma~\ref{lem:only-auxiliary-constraints}. Then, with probability $1$, the damage (including the errors and the effect of the decoder's corrections) is contractible, and no logical fault occurs.
\end{theorem}
The rest of this subsection establishes the results needed to prove this theorem. Throughout it, we will only assume error configurations that obey this maximum burst size (where errors can either be physical or measurement).  Later, we show that under the local stochastic error model, the probability of violating this maximum burst property is small, thus establishing fault-tolerance.

\subsubsection{Bounding damage by conglomerates}
The first step of the proof is to bound the connected components of the damage by potentially larger components, which we call ``conglomerates of dominant bursts''.
Informally, a conglomerate consists of all the anyon worldlines caused by the dominant burst, as well as all smaller bursts that are connected to it via anyon worldlines or are not sufficiently isolated from it, together with the anyon worldlines coming out of them.
It is most convenient to formalize this idea by dividing bursts into different ``levels'' corresponding to their size.
\begin{definition}[Levels of bursts]
\label{def:only-admissible-scale-schedule}%
Let $Q > 1$ be a constant specified in Lemma~\ref{lem:only-auxiliary-constraints}, and define
\begin{equation}
\label{S:ratio}%
r_k =Q^{k}\quad(k\ge-1).
\end{equation}
For the burst decomposition $\mathcal B$, define
\[
\mathcal B_k =
\{B\in\mathcal B:r_{k-1}<\rad_{\Box}(B)\le r_k\}.
\]
We refer to the bursts in $\mathcal B_k$ as \textbf{$k$-bursts} or bursts at \textbf{level} $k$.
In particular, \(\mathcal B_0\) contains all the radius-\(\tfrac12\) and radius-\(1\) bursts.
\end{definition}
For each $k$-burst $B$, fix an enclosing lattice cube $\overline{B}$ of geometric radius
$r_k$, whose temporal faces are denoted as $\mathbb T_\pm(\overline{B})$.

\begin{definition}[Syndrome and cluster box]
\label{def:syndrome_cluster_box}
Let $\lambda^{\C}>\lambda^{\S}>0$ be constants defined in Lemma~\ref{lem:only-auxiliary-constraints}, with $\lambda^{\C}$ satisfying condition (k)
of the Lemma.
For a $k$-burst $B$, call
\begin{align*}
\SB(B)&=\BN_{(\lambda^{\S}-1)r_k}(\overline{B})\;,\\
\CB(B)&=\BN_{(\lambda^{\C}-1)r_k}(\overline{B}),
\end{align*}
the \textbf{syndrome box} and \textbf{cluster box} of a $k$-burst $B\in\mathcal B_k$, respectively.
\end{definition}

\begin{figure}[!htbp]
\centering
\includegraphics[width=\columnwidth]{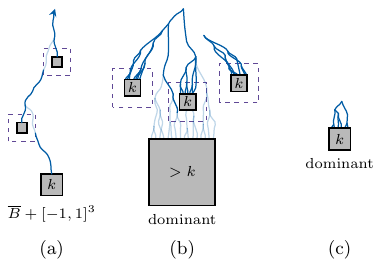}
\caption{(a) A syndrome line starting from a $k$-burst $B$. (b,c)~Illustration of conglomerates. Bursts are drawn as gray boxes, anyon worldlines are drawn as blue lines, and cluster boxes for non-dominant bursts are shown as dashed purple outlines. In (b,c), there are four bursts of level $k$ (smaller boxes) and one burst of level $>k$, and bursts that are dominant are indicated. None of the three level-$k$ bursts in (b) are dominant as their cluster boxes overlap with anyon worldlines originating at the level-$>k$ burst.}
\label{fig:syndrome-conglomerates}
\end{figure}

\begin{definition}[Syndrome line]\label{def:only-syndrome-line}
We say that a syndrome point $s$ \textbf{follows} a syndrome point $s'$ if the anyon worldline segment for $s$ is after $s'$ in time and is connected to $s'$ either directly via a shared anyon 1-gon or via a fusion connector.

We also allow connections through cluster boxes of bursts; whenever a syndrome point $s$ lies in the cluster box $\CB(Y)$ of a burst $Y$, we say that $Y$ follows $s$ and that $s$ follows $Y$.

A \textbf{syndrome line} $\ell$ starting from a $k$-burst $B\in \mathcal B_k$ is a sequence whose first element is $B$, second element is a syndrome point that is directly neighboring (distance 1 from) an error in $B$, and every consecutive element is either (1) a syndrome point or (2) a $j$-burst $Y\in \mathcal B_j$ with $j<k$, following the previous element.
See Fig.~\ref{fig:syndrome-conglomerates}(a). If $B$ produces no syndrome points, we still associate a syndrome line with it. 

For the syndrome line $\ell$ starting from a $k$-burst, we will call the lower-level bursts included in it its ``interrupting bursts''.
\end{definition}

\begin{definition}[Conglomerate] \label{def:conglomerate}
Consider a $k$-burst \(B\in\mathcal B_k\).
The \textbf{conglomerate} \(\Cong(B)\) of $B$ is the union of all syndrome lines starting from \(B\),
\[
\Cong(B)=
\bigcup\{\ell:\ell\text{ is a syndrome line from }B\}.
\]
\end{definition}

\begin{definition}[Dominant bursts]
We say that a $k$-burst \(B\in\mathcal B_k\) is {\bfseries dominant} if it is not contained in the conglomerate of another burst.
\end{definition}
In other words, a $k$-burst $B$ is not dominant if there is some ($>k$)-burst $B'$ such that a syndrome worldline starting from $B'$ enters the cluster box $\CB(B)$.

\subsubsection{Bounding conglomerates of \texorpdfstring{dominant}{dominant} bursts}%
Next, we will show that the conglomerates of dominant bursts are contained inside their syndrome box, whose size is only linear in the size of the dominant burst.
Namely, by the end of this section, we prove the following lemma.
\begin{lemma}
\label{thm:cong_in_syndrome_box}
With probability one, the conglomerate of every dominant burst is contained in its syndrome box (see Definition~\ref{def:syndrome_cluster_box}),
\begin{equation}
\Cong(B)\subset \SB(B)
\;.
\end{equation}
\end{lemma}

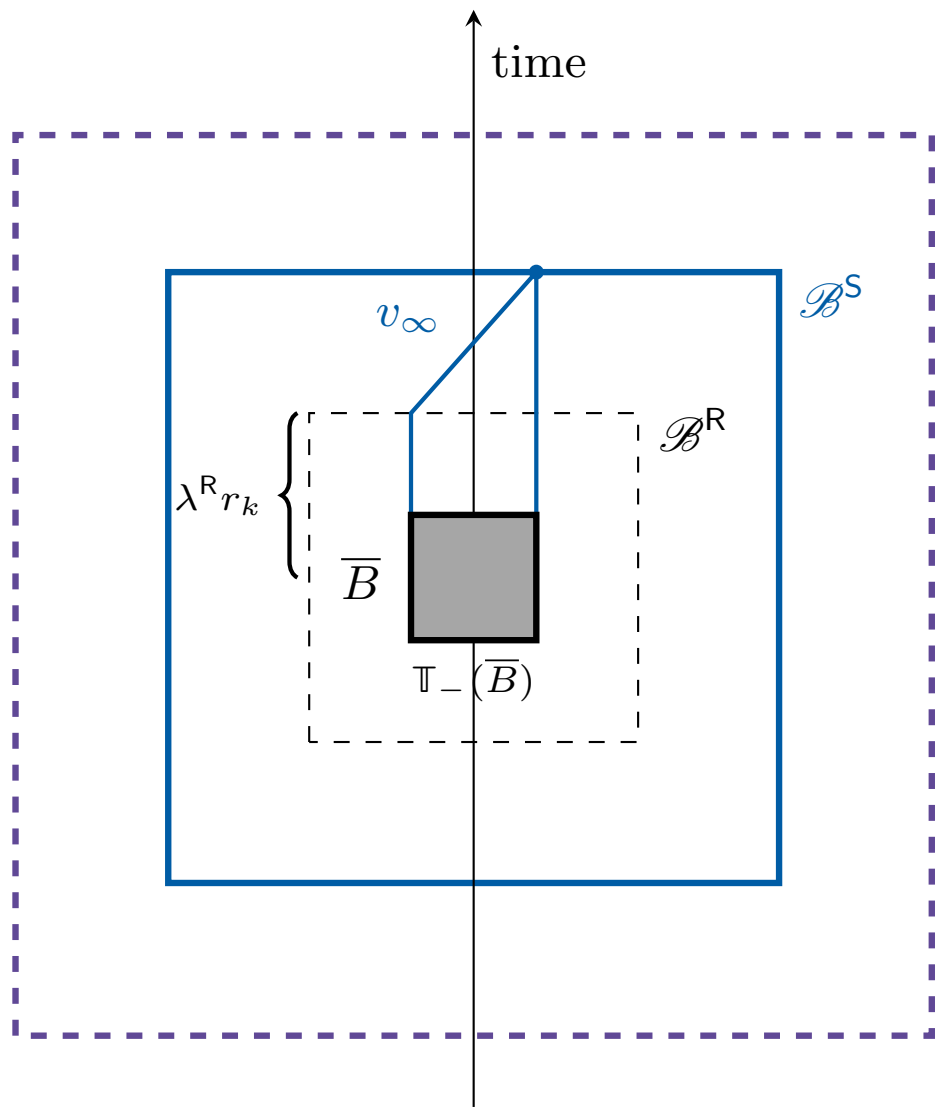
\begin{figure}[!htbp]
\centering
\begin{tikzpicture}[x=0.518cm,y=0.518cm,>=stealth]%
  \definecolor{ReconnectSZColor}{RGB}{0,92,165}
  \draw[ReconnectCZColor,very thick,dashed] (-5.85,-4.70) rectangle (5.85,6.80);
  \node[ReconnectCZColor,right] at (5.92,6.08) {\(\CB\)};%
  \draw[ReconnectSZColor,very thick] (-3.90,-2.75) rectangle (3.90,5.05);
  \node[ReconnectSZColor,right] at (3.93,4.75) {\(\SB\)};
  \draw[->] (0,-5.61) -- (0,8.40);%
  \node[right] at (0,7.74) {time};
  \draw[dashed] (-2.10,-0.95) rectangle (2.10,3.25);
  \node[right] at (2.14,3.04) {\(\RB\)};
  \coordinate (worlddot) at (0.80,5.05);%
  \draw[thick,ReconnectSZColor] (0.80,1.95) -- (worlddot);%
  \draw[thick,ReconnectSZColor] (-0.80,1.95) -- (-0.80,3.25) -- (worlddot)
    node[pos=0.42,above left=1pt] {\(v_{\infty}\)};%
  \fill[ReconnectSZColor] (worlddot) circle (1.4pt);%
  \draw[very thick,fill=gray!70] (-0.80,0.35) rectangle (0.80,1.95);
  \node[align=center] at (-1.44,1.15) {\(\overline{B}\)};%
  \node[below,font=\scriptsize] at (0,0.35) {\(\mathbb T_-(\overline{B})\)};
  \draw[dashed] (-0.80,1.95) -- (0.80,1.95);
  \draw[decorate,decoration={brace,amplitude=3pt},thick]
    (-2.25,1.15) -- (-2.25,3.25)
    node[midway,left=3pt,font=\scriptsize] {\(\lambda^{\R}r_k\)};%
\end{tikzpicture}
\caption{Reconnection, syndrome, and cluster boxes $\RB(B)\subseteq\SB(B)\subseteq\CB(B)$ associated with a  $k$-burst $B$. The later time boundary of $\RB(B)$ marks the reconnecting time $\RT(B)$. At each decoder time $t\ge\RT(B)$, the time-$t$ syndrome points of $\Cong(B)$ outside smaller guard boxes belong to a common $\beta$-cluster (Corollary~\ref{lemma:only-reconnecting-box-property}), so the decoder directs the corresponding anyons toward that cluster's base point. The syndrome box $\SB(B)$, by Lemma~\ref{thm:cong_in_syndrome_box}, contains the conglomerate $\Cong(B)$. The cluster box $\CB(B)$ contains the $\beta$-neighborhood of $\SB(B)$, and thus, any decoding cluster that can merge with a cluster contained in $\SB(B)$ must intersect $\CB(B)$.}
\label{fig:figure_boxes}
\end{figure}

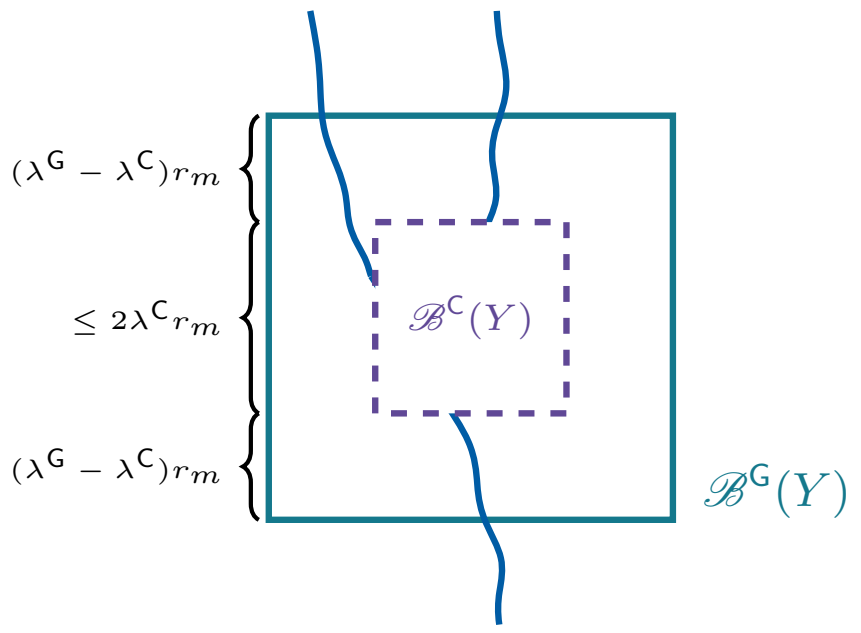
\begin{figure}[!htbp]
\centering
\begin{tikzpicture}[x=0.55cm,y=0.55cm,>=stealth]
  \definecolor{GZEdge}{RGB}{20,120,140}
  \definecolor{CZEdge}{RGB}{96,72,150}
  \definecolor{LineBlue}{RGB}{0,92,165}
  \definecolor{BridgeOrange}{RGB}{210,105,30}
  \definecolor{MutedGray}{RGB}{90,90,90}

  \draw[GZEdge,very thick]
    (-2.43,-2.43) rectangle (2.43,2.43);
  \node[GZEdge,anchor=west] at (2.58,-2.08) {\(\GB(Y)\)};

  \draw[LineBlue,very thick]
    (0.342,-3.690)
      .. controls (0.342,-3.687) and (0.341,-3.685) .. (0.341,-3.682)
      .. controls (0.322,-3.552) and (0.324,-3.422) .. (0.345,-3.291)
      .. controls (0.367,-3.160) and (0.391,-3.027) .. (0.361,-2.891)
      .. controls (0.331,-2.755) and (0.247,-2.615) .. (0.183,-2.469)
      .. controls (0.118,-2.322) and (0.089,-2.170) .. (0.078,-2.009)
      .. controls (0.066,-1.848) and (0.051,-1.678) .. (-0.027,-1.497)
      .. controls (-0.105,-1.316) and (-0.240,-1.123) .. (-0.371,-0.918)
      .. controls (-0.502,-0.712) and (-0.621,-0.492) .. (-0.747,-0.257)
      .. controls (-0.873,-0.021) and (-1.020,0.230) .. (-1.200,0.500);

  \draw[LineBlue,very thick]
    (-1.200,0.500)
      .. controls (-1.218,0.550) and (-1.230,0.607) .. (-1.256,0.671)
      .. controls (-1.281,0.736) and (-1.323,0.808) .. (-1.365,0.887)
      .. controls (-1.408,0.967) and (-1.445,1.054) .. (-1.464,1.150)
      .. controls (-1.484,1.245) and (-1.487,1.349) .. (-1.504,1.462)
      .. controls (-1.522,1.574) and (-1.560,1.695) .. (-1.612,1.825)
      .. controls (-1.665,1.955) and (-1.724,2.094) .. (-1.758,2.242)
      .. controls (-1.773,2.303) and (-1.783,2.366) .. (-1.790,2.430)
      .. controls (-1.800,2.522) and (-1.805,2.618) .. (-1.809,2.716)
      .. controls (-1.815,2.884) and (-1.825,3.061) .. (-1.853,3.249)
      .. controls (-1.874,3.390) and (-1.904,3.537) .. (-1.931,3.690);

  \draw[LineBlue,very thick]
    (0.000,0.000)
      .. controls (0.016,0.105) and (0.055,0.212) .. (0.083,0.323)
      .. controls (0.111,0.434) and (0.114,0.547) .. (0.111,0.665)
      .. controls (0.107,0.782) and (0.117,0.903) .. (0.164,1.029)
      .. controls (0.210,1.154) and (0.280,1.284) .. (0.300,1.418)
      .. controls (0.320,1.553) and (0.286,1.692) .. (0.265,1.838)
      .. controls (0.243,1.983) and (0.258,2.133) .. (0.307,2.290)
      .. controls (0.357,2.446) and (0.420,2.609) .. (0.426,2.778)
      .. controls (0.432,2.948) and (0.387,3.124) .. (0.353,3.307)
      .. controls (0.331,3.432) and (0.319,3.559) .. (0.312,3.690);

  \filldraw[fill=white,draw=CZEdge,very thick,dashed]
    (-1.15,-1.15) rectangle (1.15,1.15);
  \node[CZEdge,font=\scriptsize] at (0,0) {\(\CB(Y)\)};

  \draw[decorate,decoration={brace,amplitude=3pt},thick]
    (-2.55,-1.15) -- (-2.55,1.15)
    node[midway,left=3pt,font=\tiny] {\(\le 2\lambda^{\C}r_m\)};%
  \draw[decorate,decoration={brace,amplitude=3pt},thick]
    (-2.55,1.15) -- (-2.55,2.43)
    node[midway,left=3pt,font=\tiny]
      {\((\lambda^{\G}-\lambda^{\C})r_m\)};
  \draw[decorate,decoration={brace,amplitude=3pt},thick]
    (-2.55,-2.43) -- (-2.55,-1.15)
    node[midway,left=3pt,font=\tiny]
      {\((\lambda^{\G}-\lambda^{\C})r_m\)};
\end{tikzpicture}
\caption{Each guard box is chosen large enough so that all syndrome points on the syndrome lines cut by it are clustered together.}
\label{fig:only-guard-box-property}
\end{figure}

To prove Lemma~\ref{thm:cong_in_syndrome_box}, we need to show that the anyons created by a dominant burst are fused to vacuum before their syndrome lines can leave its syndrome box. Let us briefly give an outline of the proof which forms the remainder of this section.
We start with Lemma~\ref{prop:only-guard-box-property} and Corollary~\ref{lemma:only-reconnecting-box-property} showing that the anyon worldlines outside of isolated ``guard boxes'' end up in one common decoder cluster, at least after a certain ``reconnecting time".
Then we show in Lemmas~\ref{lem:only-motion-clearing} and \ref{lemma:terminal_time} that if, except for the guard boxes, all syndrome lines originating from the burst are moved by the decoder towards a common base point, then they will meet at the base point after a time linear in the size of the originating dominant burst.
Without the guard boxes of smaller bursts in the conglomerate, the syndrome lines would all move towards the base point at speed 1, but the interruptions by smaller guard boxes slow down the movement speed. At this stage, we have still not established that there is a single largest-level dominant burst in a conglomerate. 

Next, Claim~\ref{claim:only-stable-base} establishes that, assuming the decoder does not cluster the current conglomerate with any syndrome points outside of the conglomerate, the base point of the syndrome lines outside the guard boxes remains constant after the reconnecting time.
We already know that these syndrome points all share one common cluster, but we need to rule out that as time progresses and the cluster grows, the base point (which is the oldest point in the cluster) does not move around. 
Lemma~\ref{lemma:terminal_time} shows that all anyon worldlines caused by the burst meet at the common base point where they fuse to the vacuum, making the same assumption as Claim~\ref{claim:only-stable-base}.

Finally, Lemma~\ref{lem:only-confinement-separation} proves that (1) the conglomerate of a dominant burst remains in its linear-sized ``syndrome box", and (2) its syndrome points do not cluster with anything outside.
The second statement establishes that the assumption in Claim~\ref{claim:only-stable-base} and Lemma~\ref{lemma:terminal_time} holds, which is then used to show (1).  This shows the main statement of this subsection, Lemma~\ref{thm:cong_in_syndrome_box}.

We now begin with the first step.
A part of a syndrome line consisting of a sequence of anyon cylinders (not interrupted by bursts via their cluster boxes) is always contained in a single cluster. If it is additionally divided into segments because it encounters cluster boxes of a smaller-level burst, the segmented syndrome line will still belong to one cluster as viewed by the decoder.  The following lemma (Lemma~\ref{prop:only-guard-box-property}) formalizes this notion.
We will show that a syndrome line clusters together except for points that are ``guarded" inside linear-sized boxes surrounding the error bursts.
To prove this lemma, and for use in other proofs, we first establish the following claim bounding the size and separation between connected components of these guard boxes.

Define the separation scales
\begin{equation}
\gamma_0=\gamma(\tfrac12)-\tfrac12,\qquad
\gamma_j=\gamma(r_{j-1})-r_{j-1}.
\end{equation}
Distinct bursts, each of level at least $j$, are separated further than $\gamma_j$.

\begin{claim}[Connected components of guard boxes]
\label{claim:only-separated-boxes}
Let $\lambda^{\G} > \lambda^{\C}$ be a constant defined in Lemma~\ref{lem:only-auxiliary-constraints}.
For a $k$-burst $B\in\mathcal B_k$, call
\[\GB(B)=\BN_{(\lambda^{\G}-1)r_k}(\overline{B})\]
the \textbf{guard box} of $B$. %
Denote the set of all guard boxes of $k$-bursts with $k\leq j$ by $\GB_{\le j}$, with $\GB_{\le-1}=\varnothing$.
Consider connected components of $\GB_{\le j}$, where boxes are connected if their distance-1 padded supports overlap.
Then:
\begin{itemize}
    \item Each connected component has diameter at most $ar_j$, where $a=2\lambda^{\G}+8$, and contains a unique burst of its largest level;
    \item Distinct connected components whose largest level is $j$ are separated by at least
\[
\delta_j=\gamma_j-2ar_j>0.
\]
\end{itemize} 
\end{claim}
\begin{proof}
 We show the statement about guard boxes by induction over the level $j$. At $j=0$, the guard boxes are separated and each connected component consists of a single burst. Assuming the statement holds for $j-1$, let us now show that it holds for $j$.
 Connected components of $\GB_{\leq j}$ consist of level-$j$ guard boxes together with connected components of $\GB_{<j}$.
 If a connected component of $\GB_{<j}$ joined two distinct level-$j$ guard boxes of bursts $B$ and $B'$, then by Lemma~\ref{lem:only-auxiliary-constraints},
\[
d_{\Box}(B,B')\le2(\lambda^{\G}+2)r_j+ar_{j-1}+2<\gamma_j,
\]
a contradiction to the $\gamma$-separation between $B$ and $B'$.
 So a connected component of $\GB_{\leq j}$ that's not in $\GB_{<j}$ consists of a single level-$j$ guard box and attached connected components of $\GB_{<j}$.
 Using the induction statement for $j-1$, these connected components increase the diameter of the level-$j$ guard box to
\begin{equation}
2\lambda^{\G}r_j+2ar_{j-1}+4
\le 2\lambda^{\G}r_j+6r_{j}+4
\le ar_j.
\notag
\end{equation}
using basic arithmetic and Lemma~\ref{lem:only-auxiliary-constraints}.
The $\delta_j$ separation bound follows straightforwardly from subtracting the diameter bound $a r_j$ from the $\gamma$-separation bound of the bursts.
\end{proof}

\begin{lemma}[Clustering outside guard boxes]
\label{prop:only-guard-box-property}
Let \(\ell\) be a syndrome line starting from some $m$-burst \(B'\).
Then, at every time \(t\), all syndrome points in $\ell$ outside all guard boxes and the bounding box of $B'$,
\[
(\ell \cap S_{\leq t}) \setminus (\overline B' \cup \GB_{< m}),
\]
are assigned by the decoder at time $t$ to a single common \(\beta\)-cluster of \(\mathcal K_\beta(S_{\le t})\). 
%
\end{lemma}
\begin{proof}
We first prove the lemma for any syndrome line segment whose interrupting bursts have levels below $m$ and whose syndrome points all lie in $S_{\le t}$. The segment need not start from an $m$-burst; we justify the restriction to  segments at the end.

\myparagraph{Induction start}
For $m=0$, there are no interrupting bursts, and the syndrome points of $\ell$ form a forward chain of $1$-adjacent points. Its visible part therefore belongs to one $\beta$-cluster, since $\beta(\tfrac12)-\tfrac12=(C_\beta-1)/2\ge1$.

\myparagraph{Induction step $m-1\rightarrow m$}
We now assume that the induction statement is true for $m-1$, and show that it also holds for $m$.
Consider a segment $\ell$ with only level-$<m$ interrupting bursts, and let $Y$ be an interrupting burst of the maximal level $m-1$.
Consider the subsegments before and after $Y$ which themselves only have level-$<m-1$ interrupting bursts.
Assume that there are syndrome points $a_-$ and $a_+$ in the subsegments before and after that are outside all guard boxes, including the guard box of $Y$.
Let $A_-$ and $A_+$ denote the clusters containing all syndrome points outside the guard boxes in the subsegments before and after, which exist by the induction hypothesis, such that $a_-\in A_-$ and $a_+\in A_+$.
We now show that $A_-$ and $A_+$ are mutually $\beta$-near and thus in the same cluster.
As $A_-$ and $A_+$ contain everything outside level-$<m-1$ guard boxes, they must touch $\CB(Y)$ enlarged by a connected component of level-$<m-1$ guard boxes.
By Claim~\ref{claim:only-separated-boxes}, these connected components have size at most $ar_{m-2}$.
Thus, the distance between $A_-$ and $A_+$ is at most
\begin{align}
d_{\Box}(A_-,A_+)
\le 2\lambda^{\C} r_{m-1}+2ar_{m-2}.
\notag
\end{align}
On the other hand, both $A_-$ and $A_+$ have a syndrome point touching the enlarged cluster box of $Y$ and a syndrome point ($a_-$ or $a_+$) outside the guard box of $Y$.
So the diameter of $A_-$ and $A_+$ must be at least
\begin{align}
2\rad_{\Box}(A_\pm)
\ge \lambda^{\G}r_{m-1}-\lambda^{\C} r_{m-1}-a r_{m-2}.
\notag
\end{align}
If we choose $\lambda^{\G}$ large enough as in Lemma~\ref{lem:only-auxiliary-constraints}, then simple arithmetic yields
\begin{equation}
d_{\Box}(A_-,A_+)
\le (C_\beta-1)\rad_{\Box}(A_\pm).
\notag
\end{equation}
Thus $A_-\sim_\beta A_+$, so the two decoder clusters coincide.
Connected components of level-$\leq m-1$ guard boxes must be properly separated as shown in Claim~\ref{claim:only-separated-boxes}, so the $A_{\pm}$ cluster of any subsegment connecting to level-$m-1$ bursts $Y$ and $Y'$ must be nonempty.
Thus, since each $A_-$ subsegment clusters with the following $A_+$ subsegment, all $A_{\pm}$ clusters of all subsegments belong to one common cluster.
\end{proof}

Note that the proof above would almost work if we used cluster boxes instead of the larger guard boxes, since far in the past, the parts of the syndrome line outside cluster boxes do cluster together. However, close to the current time $t$, if there is a cluster box that recently interrupted the syndrome line, the new line segment coming out of it might not be long enough to be clustered with the rest of the syndrome line; to exclude it, we use a larger guard box.

\begin{corollary}[All syndrome lines cluster after reconnecting time]
\label{lemma:only-reconnecting-box-property}
Let $\lambda^{\R} > 0$ be the constant specified in row (j) of Lemma~\ref{lem:only-auxiliary-constraints}.  For a $k$-burst $B\in\mathcal B_k$, call 
\begin{equation}
    \RT(B)=\mathbb T_+(\overline{B})+(\lambda^{\R}-1)r_k
\end{equation} 
its \textbf{reconnecting time}.  Consider all syndrome points of \(\Cong(B)\) outside the bounding box of $B$ and outside all the guard boxes of the smaller bursts:
\begin{equation}
    \Cong(B)\setminus (\overline B\cup \GB_{<k})
    \label{eq:cong-guard}
\end{equation}
Then, at each decoder time $t\ge\RT(B)$, all syndrome points at time $t$ in the set~\eqref{eq:cong-guard} belong to a single $\beta$-cluster of $\mathcal K_\beta(S_{\le t})$.
\end{corollary}
\begin{proof}
Fix $t\ge\RT(B)$. For each syndrome point $s$ at time $t$ in the set~\eqref{eq:cong-guard}, choose a syndrome line $\ell$ starting from $B$ and containing $s$. Let
\begin{equation}
A_\ell=(\ell\cap S_{\le t})\setminus\GB_{<k}.
\end{equation}
By Lemma~\ref{prop:only-guard-box-property}, $A_\ell$ is contained in one decoder cluster. The starting point $s'$ of this line must also be in $A_\ell$ because of the burst decomposition assumption: $s'$ neighbors an error in $B$, and lower-level guard boxes must be sufficiently far from the errors in $B$. 

By row (j) of Lemma~\ref{lem:only-auxiliary-constraints}, we have $\lambda^{\R}\ge C_\beta+4$ and thus
\begin{equation}
    \RT(B)-\mathbb T_+(B)\ge (\lambda^{\R}-1)r_k\ge(C_\beta-1)r_k+4.
\end{equation}
Thus the two points $s',s\in A_\ell$ are at $d_\Box$-distance at least $(C_\beta-1)r_k+3$. Each set $A_\ell$ therefore has radius at least $((C_\beta-1)r_k+4)/2$. Since $C_\beta\ge3$, its $\beta$-neighborhood expands its bounding box by at least
$(C_\beta-1)((C_\beta-1)r_k+4)/2 \ge 2 r_k + 4$.
Any two chosen lines originating at $B$ start at most $2r_k+1$ apart. Therefore, their respective clusters are mutually $\beta$-near and must be clustered together.

Thus all the chosen syndrome points $s$ at time $t$ belong to a single decoder cluster.
\end{proof}

In the next step, we assume that the base point remains fixed for a syndrome line outside the guard boxes, which we later show is true after the reconnecting time for the cluster.
With this assumption, the syndrome line will be moved towards the base point by the decoder at speed 1 outside all guard boxes.
That is, an uninterrupted piece of anyon worldlines moves at speed $v_{-1}=1$ towards the base point.
If there are interrupting level-$0$ guard boxes, then the size of the deflection caused by each is $\sim\lambda^{\G}r_0$ whereas the minimal time between deflections is determined by the $\gamma$-separation of bursts.
These deflections reduce the movement speed to $v_0<v_{-1}$, but since $\lambda^{\G}r_0\ll \gamma(r_0)$ the speed reduction is small.
If we additionally allow for level-$1$ interrupting bursts, the speed is further reduced to $v_1<v_0<v_{-1}$, and so on.
However, as the separation function $\gamma$ grows super-linearly, the ratio $\frac{\lambda^{\G}r_k}{\gamma(r_k)}$ determining the speed reduction goes to $0$ sufficiently fast, and the total effect from deflections at all levels converges. Therefore, the movement speed $v_k$ remains lower bounded by some $v_\infty$ for any $k$.

In the technical proof below, the movement speed of a syndrome line interrupted by $\leq k$-bursts is given by
\begin{equation}
v_{-1}=1,\qquad v_k=1-\sum_{j=0}^k\frac{10\lambda^{\G}r_j}{\gamma_j}.
\label{eq:only-motion-constants}
\end{equation}
Note that $v_k\ge v_\infty=1/2$, which is checked in App.~\ref{app:constants}.

After some fixed time on the syndrome line, a deflection might occur immediately before movement can begin.
Thus, the movement bound includes an offset, which is determined by constants
\begin{equation}
E_{-1}=0,\qquad E_k=10\lambda^{\G}r_k.
\label{eq:only-motion-constants-E}
\end{equation}

For a syndrome line $\ell$, let $\ell_t$ consist of its syndrome points at time $t$ along with the time-$t$ cross-sections of the cluster boxes of all interrupting bursts on $\ell$.

For a spatial square $H$ and a spacetime set $A$, define
\begin{equation}
d_H(A)=\sup_{p\in A}d_{\Box}(\mathbb X(p),H).
\end{equation}

\begin{lemma}[Movement of syndrome lines]
\label{lem:only-motion-clearing}
Consider a syndrome line segment from time $t_0$ to time $t_1:= t_0+\Delta t$, with interrupting bursts of levels at most $k$. 

\begin{itemize}
    \item If all ordinary moves outside $\GB_{\le k}$ are directed towards base points in a fixed square $H$, then for $\Delta t\geq 0$
\begin{equation}
\label{eq:only-line-contraction}
d_H(\ell_{t_1})\le\max(0,d_H(\ell_{t_0})-v_k\Delta t)+E_k,
\end{equation}
where $v_k$ and $E_k$ are defined in Eqs.~\eqref{eq:only-motion-constants}  and~\eqref{eq:only-motion-constants-E}, respectively. 
\item Without the assumption on base points, for any spatial point $z$ and $\Delta t\ge0$,
\begin{equation}
\label{eq:only-line-growth}
d_{\{z\}}(\ell_{t_1})\le d_{\{z\}}(\ell_{t_0})+\Delta t/v_k+E_k.
\end{equation}
\item For $\Delta t<0$, a segment ending earlier than it starts has both spatial displacement and backward time displacement at most $E_k$. 
\end{itemize}
\end{lemma}
\begin{proof}
A syndrome line can enter and leave the same connected component of guard boxes several times. Replace the part between its first entrance and last exit by a straight line which we call a \emph{shortcut}. Keep repeating this procedure until each remaining connected component is visited at most once, and call the result the \emph{simplified path}. 
By Claim~\ref{claim:only-separated-boxes}, a connected component whose largest level is $j\le k$ has diameter at most $ar_j$, and distinct connected components are separated by at least $\delta_j$. Thus a shortcut through such a connected component has spatial and absolute time displacement at most $ar_j$.

Let $n_j$ be the number of shortcuts through connected components whose largest level is $j$, and let $g_{t_0;t_1}$ be the number of remaining ordinary steps where the syndrome line continues uninterrupted. Introduce notation
\[
X_{t_0;t_1}=\sum_{j=0}^k ar_jn_j,\qquad X_0=\sum_{j=0}^k ar_j,\qquad
\theta_k=\sum_{j=0}^k\frac{ar_j}{\delta_j}.
\]
Here $X_{t_0;t_1}$ bounds the total shortcut length, $X_0$ allows for one first encounter at each level, and $\theta_k$ controls how the shortcut bound grows with the length of the simplified path. 

Let $P_{t_0;t_1}$ be the length of the simplified path; 
then $P_{t_0;t_1}\le g_{t_0;t_1}+X_{t_0;t_1}$.
Between $n_j$ distinct components there are $n_j-1$ gaps of length at least $\delta_j$, so $n_j\le1+P_{t_0;t_1}/\delta_j.$
Every ordinary move corresponds to a single timestep, while the sum of the absolute time displacements of the shortcuts is at most $X_{t_0;t_1}$. Thus, $|g_{t_0;t_1}-\Delta t|\le X_{t_0;t_1}.$
Combining these gives
\begin{equation}
\label{eq:only-shortcut-bound}
X_{t_0;t_1}\le X_0+\theta_kP_{t_0;t_1}
\end{equation}
with
\begin{equation}
\label{eq:only-shortcut-bound2}
P_{t_0;t_1}\le\Delta t+2X_{t_0;t_1}
\le\frac{\Delta t+2X_0}{1-2\theta_k}.
\end{equation}
The constants in App.~\ref{app:constants} give $\theta_k\le(1-v_k)/4$ and $2X_0/(1-2\theta_k)\le E_k$. Since spatial displacement is at most $P_{t_0;t_1}$, this proves Eq.~\eqref{eq:only-line-growth}. If $\Delta t<0$, Eq.~\eqref{eq:only-shortcut-bound2} gives
\begin{equation}
\label{eq:only-backward-displacement}
\begin{aligned}
P_{t_0;t_1}
&\le\frac{2X_0}{1-2\theta_k}\le E_k.
\end{aligned}
\end{equation} Combining the first inequalities in Eq.~\eqref{eq:only-shortcut-bound} and Eq.~\eqref{eq:only-shortcut-bound2} also gives $|\Delta t|\le2X_0\le E_k$.

Now suppose the line is directed towards $H$. Every ordinary step decreases its distance to $H$ by one until it reaches $H$, while all the shortcuts together can increase this distance by at most $X_{t_0;t_1}$. If the line has not yet reached $H$, then
\begin{align*}
d_H(\ell_{t_1})-d_H(\ell_{t_0})
&\le-g_{t_0;t_1}+X_{t_0;t_1}\le-\Delta t+2X_{t_0;t_1}\\
&\le-\frac{1-4\theta_k}{1-2\theta_k}\Delta t
+\frac{2X_0}{1-2\theta_k}\\
&\le-v_k\Delta t+E_k.
\end{align*}
If the line reaches $H$, apply the same argument after the last point exiting $H$, taken in path order. Consider a segment starting from $H$. If its time displacement is nonnegative, this gives a final distance at most $E_k$; otherwise, Eq.~\eqref{eq:only-backward-displacement} gives the same bound. Together with the case where the line never reaches $H$, this proves Eq.~\eqref{eq:only-line-contraction}.
\end{proof}

Having bounded the movement speed of an individual syndrome line, we now look at multiple syndrome lines interrupted only by level-$\leq k$ bursts.
We assume that all those lines share the same base point outside of the guard boxes, and this assumption will be justified later.
Then we can show that if the endpoints of all the lines at time $t$ are contained in a size-$O(r_k)$ region, then they will all meet at a single point within time $\tau_k=O(r_k)$.
In the proof below, set $\lambda^{\T}=100\lambda^{\G}/v_\infty$ and define
\begin{equation}
\tau_{-1}=1,\qquad \tau_k=\lambda^{\T}r_k\quad(k\ge0).
\label{eq:only-motion-constants1}
\end{equation}

\begin{lemma}[Terminal time for fusion]
\label{lemma:terminal_time}
Let $t_*>\mathbb T_+(\overline B)+1$. Consider all syndrome lines from $B$, whose time-$t_*$ slice,  including cross-sections of cluster boxes of interrupting bursts, is contained in a spatial square $H_k$ of box diameter $20\lambda^{\G}r_k$. Assume that:
\begin{itemize}
\item at every time $t\in[t_*,t_*+\tau_k]$, all endpoints of these lines lying outside $\GB_{\le k}$ have the same fixed base point $b\in H_k$;
\item in the interval of times $[t_*,t_*+\tau_k]$, these syndrome lines are interrupted only by bursts of levels at most $k$.
\end{itemize}
Then there exists a time $t\in[t_*,t_*+\tau_k]$ such that the time-$t$ slice contains no cross-section of an interrupting cluster box, and all surviving anyons have been brought to $b$ and fused at $b$.
\end{lemma}
\begin{proof}
The proof involves induction on $k$. At $k=-1$, all the anyons are already at $b$, and one fusion round suffices. 

Now assume the lemma holds at level $k-1$. 
We apply Lemma~\ref{lem:only-motion-clearing}, Eq.~\eqref{eq:only-line-contraction} at level $k$, with $H = \{b\}$, to segments from the full time-$t_*$ slice that are contained in $H_k$. First note that the distance from any point to $b$ is at most $20\lambda^{\G}r_k$. Then, we obtain
\begin{equation} \label{eq:30}
    d_{\{b\}} (\ell_t) \leq \max \{ 0, 20 \lambda^{\G} r_k - v_k \Delta t) + E_k\leq 30 \lambda^\G r_k
\end{equation}
Thus, every point in the union of the slices $\ell_t$ remains within distance $30\lambda^{\G}r_k$ of $b$. At most one level-$k$ guard box can overlap with this region during the next $\tau_k$ rounds. Indeed, if two such boxes overlapped with it, the corresponding bursts would have separation at most
\begin{align}
    &\max \left (2 (30+1) \lambda^{\G}r_k, \tau_k + 2 \lambda^G r_k + 4 r_k\right) \nonumber
    \\
    &\leq \max \left (62 \lambda^{\G}r_k, (\lambda^T + 2 \lambda^\G  + 4)r_k \right) < \gamma_k
\end{align}
contradicting the separation requirement between bursts.

Suppose first that no level-$k$ guard box is encountered during the following contraction and fusion. We again apply Eq.~\eqref{eq:only-line-contraction} in Lemma~\ref{lem:only-motion-clearing}, this time  at $k-1$ with $H=\{b\}$. By an analogous equation to \eqref{eq:30},  everything in that slice comes within $E_{k-1}$ of $b$ in $\lceil20\lambda^{\G}r_k/v_\infty\rceil$ rounds. Since $2E_{k-1}\le20\lambda^{\G}r_{k-1}$, this full slice fits in $H_{k-1}$. By the induction hypothesis at level $k-1$, we have to wait at most another $\tau_{k-1}$ time before the full slice contains no cluster-box sections, and all surviving anyons have been fused at $b$.
Thus, if no level-$k$ guard box is encountered before this contraction and fusion finish, the conclusion follows. 

Now suppose there is one $k$-burst $Y$ whose guard box overlaps with the distance-$30\lambda^{\G}r_k$ box around $b$ in the next $\tau_k$ rounds.
Wait until just after the guard box of $Y$ ends.
This adds at most $2\lambda^{\G}r_k+2$ rounds. The gap between the cluster box and its temporal guard faces satisfies $T_+(\GB(Y)) - T_+(\CB(Y)) = (\lambda^\G - \lambda^\C)r_k > E_{k-1}$, so the third statement of Lemma~\ref{lem:only-motion-clearing}, applied at level $k-1$, rules out returning to that cluster box. The full slice remains within $30\lambda^{\G}r_k$ of $b$, so we again apply Eq.~\eqref{eq:only-line-contraction} in Lemma~\ref{lem:only-motion-clearing} at level $k-1$. The total time after that is
\begin{equation}
\left\lceil\frac{20\lambda^{\G}r_k}{v_\infty}\right\rceil
+2\lambda^{\G}r_k+2+\left\lceil\frac{30\lambda^{\G}r_k}{v_\infty}\right\rceil
+2\tau_{k-1}<\tau_k.
\end{equation}
At this time, the full slice contains no cross-section of an interrupting cluster box, and all surviving anyons have been brought to $b$ and fused there. This proves the lemma.
\end{proof}

We next consider how merging clusters can change base points of clusters throughout the decoding process. In Claim~\ref{claim:only-fixed-regions}, we bound the regions containing a cluster throughout such a merging process, assuming that all syndrome is contained within a given box and in guard boxes of smaller bursts. These regions will be defined using connected components of enlarged guard boxes.

We begin with the Claim below, which is an analogue of Claim~\ref{claim:only-separated-boxes} for these enlarged boxes. As before, boxes belong to the same connected component if they can be joined by a chain of boxes whose consecutive distance-1 padded supports overlap.
\begin{claim}
\label{claim:only-F-components}
Set $\lambda^{\F}=8\lambda^{\G}$. For $Y\in\mathcal B_h$, define 
\begin{align*}
\BN^{\F}(Y)&=\BN_{(\lambda^{\F}-1)r_h}(\overline Y),\\
\BN^{\F}_{\le j}&=\bigcup_{h\le j}\bigcup_{Y\in\mathcal B_h}\BN^{\F}(Y).
\end{align*}
Every component of $\BN^{\F}_{\le j}$ has diameter at most $3\lambda^{\F}r_j$ and a unique burst of its largest level.
\end{claim}
\begin{proof}
The same induction proof applies to $\BN^{\F}_{\le j}$. The separation of the bursts again prevents smaller components from joining two level-$j$ boxes. Using the induction hypothesis that each smaller component has diameter at most $3\lambda^{\F}r_{j-1}$, the diameter is at most $2\lambda^{\F}r_j+6\lambda^{\F}r_{j-1}+4\le3\lambda^{\F}r_j$. All these inequalities follow from the constants in App.~\ref{app:constants}.
\end{proof}

The enlarged box $\BN^{\F}(Y)$ provides room both for clusters near the guard box of $Y$ and for their $\beta$-neighborhoods, which determine where those clusters can merge.

\begin{definition} \label{def:11}
Fix levels $q\ge-1$ and $j \geq 0$. For a burst $Y \in \mathcal B_j$ and a  spacetime box $W$, we define $\mathscr Z(Y)$ and $\mathscr  Z_q(W)$ as:
\begin{itemize}
    \item $\mathscr{Z}(Y)$ is the union of $\GB(Y)$ and all connected components of $\BN^{\F}_{<j}$ whose box distance from $\GB(Y)$ is at most one.
    \item $\mathscr{Z}_q(W)$ is the union of $W$ and all connected components of $\BN^{\F}_{\le q}$ whose distance from $W$ is at most one.
\end{itemize}
In particular, $\mathscr{Z}(Y)=\GB(Y)$ when $j=0$, and $\mathscr Z_q(W)=W$ when $q=-1$.
\end{definition}

The region $\mathscr Z(Y)$ allows syndrome associated with the level-$j$ burst $Y$ to absorb nearby components from smaller levels. Similarly, $\mathscr  Z_q(W)$ enlarges $W$ to accommodate attached components from all allowed levels up to $q$.  The claim below states that these regions suffice to contain every cluster throughout the merging process. A merge event may place a cluster in a larger burst's region or in $\mathscr  Z_q(W)$, but the regions themselves never need to be enlarged.

\begin{claim}[Clustering in fixed regions]
\label{claim:only-fixed-regions}
Let $q\ge-1$, and let $W$ be a spacetime box.
\begin{itemize}
    \item For $q\ge0$ and $Y\in\mathcal B_j$ with $0\le j\le q$:
\begin{align}
\mathscr Z(Y)&\subseteq
\BN_{3\lambda^{\F}r_{j-1}+2}\bigl(\GB(Y)\bigr), \label{eq:ZY}\\
\mathscr  Z_q(W)&\subseteq
\BN_{3\lambda^{\F}r_q+2}(W). \label{eq:ZW}
\end{align}
\item 
For every such burst $Y$, the enlarged box $\BN^{\F}(Y)$ intersects all possible merge candidates under Algorithm~\ref{alg:clustering}: for every nonempty set $A\subseteq\mathscr Z(Y)$,
\begin{equation}
\BN_\beta(A)\subseteq\BN^{\F}(Y). \label{eq:BF}
\end{equation}
Thus anything merging with any cluster $A$ contained in $\mathscr Z(Y)$ must intersect $\BN^{\F}(Y)$.
\end{itemize}
\noindent Additionally, let $S'$ be a finite set of syndrome points satisfying $$S'\subseteq W\cup\GB_{\le q}.$$ Then
\begin{itemize}
\item Apply Algorithm~\ref{alg:clustering} to $S'$ with separation function $\beta$. At every time, every obtained cluster is contained in $\mathscr  Z_q(W)$ or in $\mathscr Z(Y)$ for some burst $Y$ of level $\leq q$.
\end{itemize}
\end{claim}

\begin{proof}
The bounds \eqref{eq:ZY} and \eqref{eq:ZW} on the extent of $\mathscr{Z}(Y)$ and $\mathscr Z_q(W)$ follow from the
diameter bounds on connected components from Claim~\ref{claim:only-F-components}. 
Put $R=\lambda^{\G}r_j+3\lambda^{\F}r_{j-1}+2$. By Eq.~\eqref{eq:ZY}, $\mathscr Z(Y)$ is contained in a box of radius $R$. Consider $A\subseteq\mathscr Z(Y)$; since $\rad_{\Box}(A)\le R+\tfrac12$ and $\beta(r)-r=2r$, its $\beta$-neighborhood lies in the concentric cube of radius at most
\begin{equation}
R+2\rad_{\Box}(A)\le3R+1<\lambda^{\F}r_j.
\end{equation}
where the last inequality follows from App.~\ref{app:constants}. Thus $\BN_\beta(A)\subseteq\BN^{\F}(Y)$, and in particular $\mathscr Z(Y)\subseteq\BN^{\F}(Y)$. Any merger partner must intersect $\BN_\beta(A)$ and hence $\BN^{\F}(Y)$. This proves eqs.~\eqref{eq:ZY} and ~\eqref{eq:BF}.

We prove the last point by induction over successive merging steps. The base case holds automatically, as each cluster is a single point that is contained in $S'$, and thus it is contained in $\mathscr  Z_q(W)$ or in $\mathscr Z(Y)$ for some burst $Y$ of level at most $q$.
Suppose that the statement holds before two clusters $A$ and $D$ merge, and choose a containing region for each using the induction hypothesis. If both are assigned to $\mathscr  Z_q(W)$, then $A\cup D\subseteq\mathscr  Z_q(W)$; likewise, if both are assigned to the same $\mathscr Z(Y)$, then $A\cup D\subseteq\mathscr Z(Y)$. It remains to consider clusters assigned to different regions.

First suppose $A\subseteq \mathscr{Z}(X)$ and $D\subseteq \mathscr Z_q(W)$. Then, there is a point $p\in D\cap\BN_\beta(A)$. By \eqref{eq:BF}, $\BN_\beta(A)\subseteq\BN^{\F}(X)$, so $p\in\BN^{\F}(X)$. 
Let $K$ be the connected component of $\BN^{\F}_{\le q}$ containing $\BN^{\F}(X)$.
Since $p\in D\subseteq\mathscr  Z_q(W)$, either $p\in W$ or $p$ lies in a connected component already attached to $W$. In both cases, $K\subseteq\mathscr  Z_q(W)$. Consequently,
\begin{equation}
A\subseteq\BN_\beta(A)\subseteq\BN^{\F}(X)\subseteq K\subseteq\mathscr  Z_q(W).
\end{equation}
Since also $D\subseteq\mathscr  Z_q(W)$, the merged cluster satisfies $A\cup D\subseteq\mathscr  Z_q(W)$.

Now suppose $A\subseteq \mathscr{Z}(X)$ and $D\subseteq \mathscr{Z}(Y)$, with $X$ of lower level than $Y$. Similarly as before, there is a point $p\in D\cap\BN_\beta(A)\subseteq \BN^{\F}(X)$. Thus $\BN^{\F}(X)$ meets either $\GB(Y)$ or a component already attached to it. Writing $j$ for the level of $Y$, the component of $\BN^{\F}_{<j}$ containing $\BN^{\F}(X)$ is therefore included in $\mathscr Z(Y)$. Since $A\subseteq\mathscr Z(X)\subseteq\BN^{\F}(X)$, the merged cluster lies in $\mathscr Z(Y)$. If $X$ and $Y$ have the same level but are distinct, a merger event would imply that $\BN^{\F}(X) \cap \BN^{\F}(Y) \neq \varnothing$, contradicting Claim~\ref{claim:only-F-components}.
This completes the induction step.
\end{proof}

Corollary~\ref{lemma:only-reconnecting-box-property} shows that, at each time $t\ge\RT(B)$, the syndrome from $\Cong(B)$ outside $\GB_{<k}$ belongs to a common decoder cluster. This does not yet show that its base point stays fixed. The decoder considers all past syndrome, including that from anyon worldlines that have already fused to vacuum, as well as syndrome points inside $\GB_{<k}$ and belonging to other clusters. As time goes on and new syndrome points are added, a merge event could bring an earlier point into the common cluster and change its base point. We will rule this out in Claim~\ref{claim:only-stable-base}, assuming that any cluster containing syndrome from $\Cong(B)$ has no syndrome from outside $\Cong(B)$. 
The possibility of clustering with syndrome outside $\Cong(B)$ will be excluded
subsequently, in Lemma~\ref{lem:only-confinement-separation}.

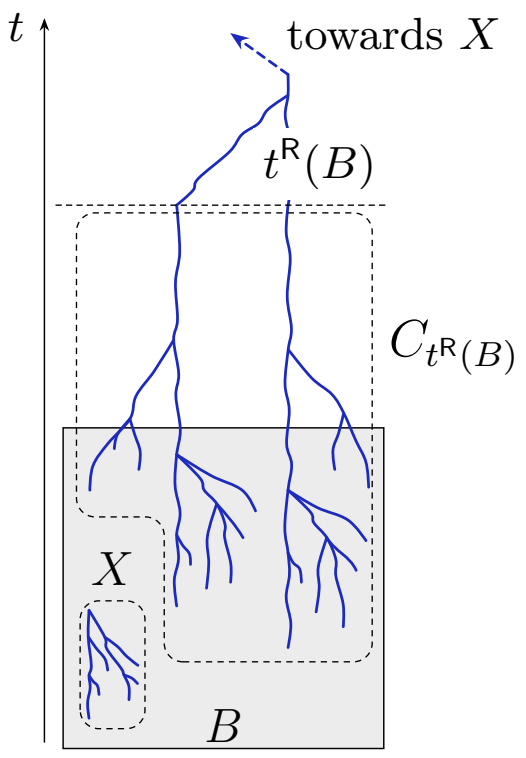
\begin{figure}[!thbp]
\centering
\resizebox{0.42\columnwidth}{!}{\input{figure-stability-syndrome.tex}}
\caption{Syndrome lines from $B$. The lines in $X$ fuse to vacuum early, while other lines survive until the reconnecting time $\RT(B)$ and belong to $C_{\RT(B)}$. The dashed arrow illustrates a possible change of base point towards $X$, which Claim~\ref{claim:only-stable-base} rules out.}
\label{fig:base-point-stability}
\end{figure}

\begin{claim}[Stability of the base point within $\Cong(B)$]
\label{claim:only-stable-base}
Let $B\in\mathcal B_k$, and let $T_{\mathrm{cap}}\ge \RT(B)$. Suppose that, at every decoder time up to $T_{\mathrm{cap}}$, every decoder cluster overlapping the syndrome in $\Cong(B)$ contains only syndrome from $\Cong(B)$. 

If the full time-$\RT(B)$ slice of the syndrome lines from $B$ is nonempty, there is a decoder cluster $C_{\RT(B)}$ with the following property. For $\RT(B)\le t\le T_{\mathrm{cap}}$, let $C_t$ be the decoder cluster containing $C_{\RT(B)}$. Then
\begin{equation} \label{eq:Ct}
\bigl(\Cong(B)\cap S_{=t}\bigr)\setminus\GB_{<k}\subseteq C_t,
\end{equation}
and the decoder assigns a fixed base point $b$ for $C_t$ at times $\RT(B)\le t\le T_{\mathrm{cap}}$ satisfying
\begin{equation} \label{eq:b}
b\in\BN_{31r_k}(\overline B).
\end{equation}
\end{claim}

\begin{proof}
Recall that $\beta(r)=3r$ and $\RT(B)=\mathbb T_+(\overline B)+127r_k$.

\textbf{A surviving syndrome line identifies $C_t$.} Choose a syndrome line whose full slice at $\RT(B)$ is nonempty, and follow it to its first crossing of that time; call such a point $p$. If the crossing lies inside $\GB_{<k}$, take $p$ instead to be the syndrome point immediately before the line enters the guard-box component containing the crossing.  Let $s$ be the starting point of this syndrome line.

By burst separation, $s$ must be outside $\GB_{<k}$, and by Claim~\ref{claim:only-separated-boxes}, we have $\mathbb T(p)\ge\RT(B)-ar_{k-1}-2$. By Lemma~\ref{prop:only-guard-box-property}, $s$ and $p$ belong to one decoder cluster, which we denote $C_{\RT(B)}$. Since $\mathbb T(s)\le\mathbb T_+(\overline B)+1$,
\begin{equation}
2\rad_\Box(C_{\RT(B)})
\ge127r_k-ar_{k-1}-3.
\label{eq:only-stability-main-size}
\end{equation}
Every later cluster $C_t\supseteq C_{\RT(B)}$ inherits this bound.

To prove Eq.~\eqref{eq:Ct}, we follow the same logic as above for  $\RT(B)\le t\le T_{\mathrm{cap}}$. Let $D$ be a decoder cluster at time $t$ containing a time-$t$ syndrome point of $\Cong(B)$ outside $\GB_{<k}$. We will show that it must coincide with $C_t$. 
By Lemma~\ref{prop:only-guard-box-property}, $D$ contains a starting point adjacent to $B$. This starting point has time at most $\mathbb T_+(\overline B)+1$, so $D$ has time span at least $t-\mathbb T_+(\overline B)-1 \geq 127r_k-1$. The starting points in $D$ and $C_t$ are both in $\BN_1(\overline B)$ and thus are separated by at most $2r_k+2$. By Eq.~\ref{eq:only-stability-main-size} and the analogous statement $2\rad_\Box(D) \geq 127 r_k$, 
$D$ and $C_t$ are mutually $\beta$-near, so  $D=C_t$.

\textbf{The base point and any potential replacement lie in $\BN_{31r_k}(\overline B)$.} Set $W=\BN_{31r_k}(\overline B)$ and $b=b(C_{\RT(B)})$. Assume that point $q$ could replace $b$ by time $T_{\mathrm{cap}}$. By the assumption that any cluster containing syndrome from $\Cong(B)$ contains only syndrome from $\Cong(B)$, $q$ must be in $\Cong(B)$, and by definition of the base point
\begin{equation} \label{eq:Tq}
\mathbb T(q) \le \mathbb T(b) \le \mathbb T(s) \le \mathbb T_+(\overline B)+1.
\end{equation}
Follow a syndrome line from $B$ to $q$, starting immediately after its last encounter with $B$. Its starting point is adjacent to $B$, so if $q$ is later in time than the starting point, their time difference is at most $2r_k+2$, but it could also be earlier in time. The bounds in Lemma~\ref{lem:only-motion-clearing} give
\begin{equation} \label{eq:only-early-point-bound}
d_\Box(q,\overline B)
\le1+\frac{2r_k+2}{v_\infty}+E_{k-1}
<31r_k.
\end{equation}
Thus, any possible replacement of $b$ must be in $W$. Additionally, the argument above also applies to $b$, so $b$ must be in $W$ as well.

\textbf{The $\beta$-neighborhoods of clusters containing possible replacements for the base point lie before $\RT(B)$ and outside $C_{\RT(B)}$.}
The base point $b$ is a point of $C_t$ as $C_{\RT(B)} \subseteq C_t$. The base point of $C_t$ can change only if the cluster acquires a syndrome point $q$ preceding $b$ in the decoder's lexicographic order. The assumption that any cluster containing syndrome from $\Cong(B)$ contains only syndrome from $\Cong(B)$ restricts such points to $\Cong(B)$. Every such $q$ must lie outside $C_{\RT(B)}$, and in the previous part we showed that it lies in $W$ and by Eq.~\eqref{eq:Tq}, $\mathbb T(q) \leq \mathbb T_+(\overline B)+1 < \RT(B)$.

To rule out later mergers involving the cluster containing $q$, enlarge $W$ to $\mathscr Z_{k-1}(W)$ from Definition~\ref{def:11}.
We will show that every decoder cluster $A\ne C_{\RT(B)}$ at time $\RT(B)$ containing syndrome from $\Cong(B)$ and intersecting $\mathscr Z_{k-1}(W)$ remains unchanged through $T_{\mathrm{cap}}$. Since $q\in W\subseteq\mathscr Z_{k-1}(W)$, the cluster containing $q$ is among them; proving this will therefore prevent $q$ from entering $C_t$ and replacing $b$. 

Let us first show that if a decoder cluster $A\ne C_{\RT(B)}$ at time $\RT(B)$ contains only syndrome from $\Cong(B)$ and intersects $\mathscr Z_{k-1}(W)$, then its $\beta$-neighborhood must lie entirely in $\mathscr Z_{k-1}(W)$. Since $A$ contains syndrome from $\Cong(B)$, the claim assumes that all its syndrome belongs to $\Cong(B)$. 

First consider the case $A\not\subseteq\GB_{<k}$. Then $A$ contains a point in $\overline B$ or, by Lemma~\ref{prop:only-guard-box-property}, a starting point adjacent to $B$. Choose $s_A\in A\cap\BN_1(\overline B)$. Since $d_\Box(s_A,s)\le2r_k+2$, by Eq.~\eqref{eq:only-stability-main-size}, $s_A$ is in $\BN_\beta(C_{\RT(B)})$. If $2\rad_\Box(A)\ge2r_k+2$, then also $s\in\BN_\beta(A)$, contradicting  $A\ne C_{\RT(B)}$ .
Thus $2\rad_\Box(A)\le2r_k+1$. For every $x\in\BN_\beta(A)$, we have $d_\Box(x,s_A)\le4\rad_\Box(A)-1$, and thus 
\begin{equation}
\BN_\beta(A)\subseteq\BN_{4r_k+2}(\overline B)\subseteq W.
\end{equation}

If instead $A\subseteq\GB_{<k}$, then $\BN_\beta(A)$ must belong to a connected component of $\BN^{\F}_{<k}$ by Claim~\ref{claim:only-fixed-regions}. Since $A$ intersects $\mathscr Z_{k-1}(W)$, this connected component either intersects $W$ or coincides with a connected component included in $\mathscr Z_{k-1}(W)$ because its distance from $W$ is at most one. In both cases,
\begin{equation}
\BN_\beta(A)\subseteq\mathscr Z_{k-1}(W).
\label{eq:only-stability-region-neighborhood}
\end{equation}

We next show that every point in $\BN_\beta(A)$ lies before time $\RT(B)$ and is outside $C_{\RT(B)}$. Claim~\ref{claim:only-F-components} and App.~\ref{app:constants} bound the enlargement of $W$ to $\mathscr Z_{k-1}(W)$ by less than $r_k$; at $k=0$, $\mathscr Z_{k-1}(W) = W$.  Together with Eq.~\eqref{eq:only-stability-main-size} and $s\in\BN_1(\overline B)$, this gives
\begin{align}
\mathscr Z_{k-1}(W)
\subseteq\BN_{32r_k}(\overline B)
\subseteq\BN_\beta(C_{\RT(B)}),
\label{eq:only-stability-region-covered}
\end{align}
and
\begin{equation}
    \max_{x\in\mathscr Z_{k-1}(W)}\mathbb T(x) \le\mathbb T_+(\overline B)+32r_k<\RT(B).
\label{eq:only-stability-region-past}
\end{equation}
Since $\BN_\beta(A) \subseteq \mathscr Z_{k-1}(W)$, it lies in time before $\RT(B)$. Moreover, $A\subseteq\BN_\beta(C_{\RT(B)})$, so separation of the two clusters forces
\begin{equation}
C_{\RT(B)}\cap\BN_\beta(A)=\varnothing.
\label{eq:only-stability-no-contact}
\end{equation}

\textbf{Clusters containing possible base point replacements never merge with $C_t$, so $b$ remains fixed.}
We will now show that every decoder cluster $A\ne C_{\RT(B)}$ at time $\RT(B)$ containing syndrome from $\Cong(B)$ and intersecting $\mathscr Z_{k-1}(W)$ never merges through $T_{\mathrm{cap}}$. Call the collection of all such clusters $\mathcal A$. Since $q\in W\subseteq\mathscr Z_{k-1}(W)$, this includes the cluster containing any potential base point replacement candidate $q$.
So far, we showed that at time $\RT(B)$, $A$ is a separate decoder cluster. A later merge partner $Y$ could contain syndrome already recorded at $\RT(B)$, and possibly include newer syndrome, or could consist entirely of newer syndrome. We will exclude both possibilities.

By independence of the merging order, we may continue merging from the clusters at $\RT(B)$, adding newly recorded syndrome points as single-point clusters. 
Thus an existing cluster grows only through mergers. 
Suppose one of the clusters in $\mathcal A$ merges by $T_{\mathrm{cap}}$, and choose the first such merge event; we suppose this occurs at time $t$. Denote the participating initial cluster by $A \in \mathcal A$, and its partner by $Y_t$. 
The merge event requires a point $x\in Y_t\cap\BN_\beta(A)$. The earlier part showed that
\begin{align}
x\in Y_t\cap\BN_\beta(A)
\subseteq\mathscr Z_{k-1}(W)\setminus C_{\RT(B)}
\end{align}
and $\mathbb T(x)<\RT(B)$.
Thus $Y_t$ must contain a syndrome point before $\RT(B)$, and thus $Y_t$ cannot contain only new syndrome (so this excludes the second possibility). Since the merged cluster $A \cup Y_t$ contains syndrome from $\Cong(B)$ in $A$, the claim's assumption demands that $x\in\Cong(B)$.

Let $Y_{\RT(B)}$ be the ancestor decoder cluster of $Y_t$ containing $x$ at $\RT(B)$. 
Since $x\in\Cong(B)\cap\mathscr Z_{k-1}(W)$ and $x\notin C_{\RT(B)}$, $Y_{\RT(B)}$ contains syndrome from $\Cong(B)$, intersects $\mathscr Z_{k-1}(W)$, but differs from $C_{\RT(B)}$. Thus, $Y_{\RT(B)}  \in \mathcal A$ and, because we assumed that the merge event between $A$ and $Y_t$ is the first merge event involving any cluster in $\mathcal A$, $Y_{\RT(B)} = Y_t$. However, by assumption, $A$ and $Y_{\RT(B)}$ are separate clusters, and they cannot merge. 

Consequently, none of the clusters in $\mathcal A$ can merge through $T_{\mathrm{cap}}$, and thus, no potential replacement $q$ can enter $C_t$. Thus $b(C_t)=b$ at times $\RT(B)\le t\le T_{\mathrm{cap}}$ with $b$ satisfying Eq.~\eqref{eq:b}.
\end{proof}

The preceding claim assumes that the decoder does not cluster syndrome from $\Cong(B)$ with syndrome from outside it. We now prove this assumption for conglomerates of dominant bursts.  Statement (1) of the following lemma establishes the main statement of this subsection, Lemma~\ref{thm:cong_in_syndrome_box}.

\begin{lemma}[Syndrome boxes and decoder clusters]
\label{lem:only-confinement-separation}
For every $k\ge0$ and every dominant burst $B\in\mathcal B_k$, the following statements hold.
\begin{itemize}
\item[(1)] (Containment in the syndrome box). We have 
\begin{equation} \label{eq:lemma51}
    \Cong(B)\subseteq\SB(B)
\end{equation} 
All the anyons originating from $\Cong(B)$ fuse to vacuum, and all reached cluster boxes end before 
\begin{equation} \label{eq:Tcap-lem5}
    T_{\mathrm{cap}} = \mathbb T_+(\overline B)+263r_k
\end{equation}
\item[(2)] (Separation of decoder clusters). At every time step, a cluster containing syndrome of $\Cong(B)$ contains only syndrome from $\Cong(B)$. In particular, it cannot also contain syndrome reached from a burst of higher level than $k$.
\end{itemize}
\end{lemma}
\begin{proof}
Fix a dominant $B \in \mathcal B_k$; we use induction on the level $k$ of $B$. Assume $(1)$ and $(2)$ for all $j<k$. 
As in Claim~\ref{claim:only-stable-base}, set $W=\BN_{31r_k}(\overline B)$.

The proof consists of 7 parts; parts 1-5 use a temporary assumption that no syndrome line from another level-$k$ burst enters $\CB(B)$. Part 6 removes this temporary assumption.

\textbf{1. Syndrome recorded up to $T_{\mathrm{cap}}$ is in $\SB(B)$}. 
Consider any point $p$ on a syndrome line starting from $B$ with $\mathbb T(p)\le T_{\mathrm{cap}}$. Any interrupting cluster box is included in full, even if it extends beyond $T_{\mathrm{cap}}$. If the line encounters further errors in $B$ before reaching $p$, apply Lemma~\ref{lem:only-motion-clearing} to the segment from the syndrome point immediately after the last such encounter to $p$.
Since the syndrome line must start next to an error in $B$, Eqs.~\eqref{eq:only-line-growth} and~\eqref{eq:only-backward-displacement} of Lemma~\ref{lem:only-motion-clearing}, applied at level $k-1$, give
\begin{equation}\label{eq:lemma5-growth}
d_{\Box}(\mathbb X(p),\mathbb X(\overline B))
\le 1+\frac{265r_k+1}{v_\infty}+E_{k-1}
<534r_k.
\end{equation}

The third statement of Lemma~\ref{lem:only-motion-clearing} at level $k-1$ bounds backward time displacement by $E_{k-1}$. Since the starting point has time at least $\mathbb T_-(\overline B)-1$, we have $$\mathbb T(p)\ge\mathbb T_-(\overline B)-1-E_{k-1}.$$ These spatial and temporal bounds apply up to the first encounter with an error in another burst of level at least $k$. Burst separation excludes such an encounter in this region, so the bounds hold up to $T_{\mathrm{cap}}$.

Including cluster boxes adds distance $2\lambda^{\C}r_{k-1}$ in space or time. For $k\ge1$, the inequality $1+E_{k-1}+2\lambda^{\C}r_{k-1}<2r_k$ ensures that every point in an interrupting cluster box still has time greater than $\mathbb T_-(\overline B)-2r_k$.  After including cluster boxes, the earlier bounds are modified to
\begin{align}
d_{\Box}\bigl(\mathbb X(p),\mathbb X(\overline B)\bigr)
&<534r_k+2\lambda^{\mathsf C}r_{k-1},\\
\mathbb T_-(\overline B)-2r_k
&<\mathbb T(p)
\le T_{\mathrm{cap}}+2\lambda^{\mathsf C}r_{k-1}.
\end{align}
Both of them lie within $\SB(B)$; at $k=0$ there are no smaller boxes to include.

\textbf{2. Clusters meeting $\Cong(B)$ contain no outside syndrome up to $T_{\mathrm{cap}}$.} Suppose a decoder cluster $C$ at time $t\le T_{\mathrm{cap}}$ meets $\Cong(B)$ but is not contained in it.
By Proposition~\ref{prop:only-merge-crossing-cut}, the merge history of $C$ contains a merge between clusters $A\subseteq C\cap\Cong(B)$ and $D\subseteq C\setminus\Cong(B)$. The previous part gives $A\subseteq\SB(B)$, so the merge condition yields a point
\begin{equation}
\label{eq:only-first-mixed-merge}
p\in D\cap\BN_\beta(A)\subseteq\BN_\beta(\SB(B))\subseteq\CB(B).
\end{equation}
The point $p$ lies on a syndrome line from a dominant burst $Y\in\mathcal B_j$, $Y\ne B$ (every burst is dominant or feeds a line of a dominant burst). Every case is impossible:
\begin{itemize}
\item $r_j>r_k$ contradicts dominance of $B$.
\item $r_j<r_k$ contradicts statement (2) at level $j$, since $C$ contains syndrome of $\Cong(Y)$ together with syndrome reached from the higher-level burst $B$.
\item $r_j=r_k$ contradicts the extra assumption that no syndrome line from another $k$-burst enters $\CB(B)$.
\end{itemize}
This proves statement (2) up to $T_{\mathrm{cap}}$ under our temporary assumption.

\textbf{3. Base point is in $W$ and syndrome lines remain near $W$}. 
To determine the movement of syndrome from $\Cong(B)$, we first locate the base points toward which the decoder moves it. Assuming that we are running the decoder at $t \leq T_{\mathrm{cap}}$, consider a syndrome point $p\in\Cong(B)\setminus\GB_{<k}$ and let $C$ be its decoder cluster.
By Lemma~\ref{prop:only-guard-box-property}, $C$ contains the starting syndrome point of the syndrome line from $B$ to $p$. This starting point neighbors an error in $B$, so its time is at most $\mathbb T_+(\overline B)+1$, and the base point of $C$ must be at most at this time. By part 2, this base point belongs to $\Cong(B)$, so Eq.~\eqref{eq:only-early-point-bound} in the proof of Claim~\ref{claim:only-stable-base} places it in $W=\BN_{31r_k}(\overline B)$.

Thus all anyon worldline segments outside the smaller guard boxes move towards the spatial square $\mathbb X(W)$. Let $q$ be any syndrome point recorded from $B$ up to $T_{\mathrm{cap}}$, or any point in a cluster box reached by then. We apply Eqs.~\eqref{eq:only-line-contraction} and~\eqref{eq:only-backward-displacement} of Lemma~\ref{lem:only-motion-clearing} at level $k-1$ to the syndrome lines starting from $B$, with target square $H=\mathbb X(W)$:
\begin{equation}
\label{eq:only-slice-near-W}
d_{\Box}\bigl(\mathbb X(q),\mathbb X(W)\bigr)\le E_{k-1}.
\end{equation}

\textbf{4. Shrinking and fusion after reconnecting time.} We show that all surviving anyons are brought to a common point and fused at a time satisfying
\begin{equation}\label{eq:lemma5-fusion-time}
t_{\mathrm{fuse}}
\le \RT(B)+\left\lceil\frac{64r_k+2E_{k-1}}{v_\infty}\right\rceil+\tau_{k-1}
<T_{\mathrm{cap}} - r_k,
\end{equation}
with no cluster-box sections at that time. The last inequality follows from the choice of constants. 

If the full time-$\RT(B)$ slice is nonempty, the property proved in part 2 allows us to apply Claim~\ref{claim:only-stable-base}. This gives a common base point $b\in W$ for syndrome outside $\GB_{<k}$, fixed from $\RT(B)$ through $T_{\mathrm{cap}}$. We can therefore apply the contraction estimate with target $\{\mathbb X(b)\}$. To bound the required contraction time, Eq.~\eqref{eq:only-slice-near-W} places the full time-$\RT(B)$ slice within $E_{k-1}$ of $\mathbb X(W)$. Since $b\in W$ and $\mathbb X(W)$ has diameter at most $64r_k$, every point of this slice is within $64r_k+E_{k-1}$ of $\mathbb X(b)$.

By Eq.~\eqref{eq:only-line-contraction} of Lemma~\ref{lem:only-motion-clearing}, applied at level $k-1$, contraction for $\lceil(64r_k+2E_{k-1})/v_\infty\rceil$ rounds brings the full slice within $E_{k-1}$ of $\mathbb X(b)$. The resulting slice lies in a square centered at $\mathbb X(b)$ of diameter $2E_{k-1}\le20\lambda^{\G}r_{k-1}$. Lemma~\ref{lemma:terminal_time} at level $k-1$ therefore fuses all surviving anyons at $b$ within another $\tau_{k-1}$ rounds. Adding these times gives Eq.~\eqref{eq:lemma5-fusion-time}.

\textbf{5. Fusing $\Cong(B)$ gives vacuum}. To apply Proposition~\ref{prop:only-damage-neutrality}, we first remove the damage of the other lower-level dominant bursts in increasing level order, using statement (1) at each level $j<k$ and the local replacement argument in the proof of Proposition~\ref{prop:only-contractible-no-fault}.

Consider $Y$ a dominant $j$-burst being removed. No line from a higher-level burst can enter $\BN_1(\SB(Y))\subseteq\CB(Y)$, by dominance. Burst separation excludes errors from other bursts of level at least $j$. If a smaller burst $Z$ meets $\BN_1(\SB(Y))$, then
\begin{equation}\label{eq:lemma5-small-box}
\lambda^{\S}r_j+1+(\lambda^{\C}+1)r_{j-1}<\lambda^{\C}r_j
\end{equation}
and therefore, $\CB(Z)\subseteq\CB(Y)$. 
If $Z$ belonged to another remaining conglomerate of level greater than $j$, a line from that conglomerate would enter $\CB(Y)$ by Eq.~\eqref{eq:lemma5-small-box}, contradicting dominance. By statement (1) at level $j$ and burst separation, the damage of other dominant $j$-bursts does not meet $\BN_1(\SB(Y))$. Thus each removal has a surrounding depth-1 layer containing no errors or anyon worldlines, and Proposition~\ref{prop:only-contractible-no-fault} can be applied.

The same argument applies to $B$, using our temporary assumption excluding lines from other level-$k$ bursts. Here the enclosing region is $\SB(B)$ truncated just after the fusion time $t_{\mathrm{fuse}}$ from part 4. After the preceding removals, its depth-1 padding is error-free, and the only possible nontrivial worldline crossing this padding is the outgoing fusion outcome. By Proposition~\ref{prop:only-damage-neutrality}, the fusion gives the vacuum outcome. 

The full time-$t_{\mathrm{fuse}}$ slice is therefore empty, so correction is complete by $t_{\mathrm{fuse}}<T_{\mathrm{cap}}$. Together with part 1, this proves statement (1) at level $k$ under our temporary assumption.

\textbf{6. No syndrome line from another level-$k$ burst enters $\CB(B)$.} Here, we show that the assumption that no syndrome line from another level-$k$ burst enters $\CB(B)$ holds for every dominant burst.

We first show that if a syndrome line starting from a level-$k$ burst $B$ intersects the cluster box of another level-$k$ burst $D$, then $\mathbb T_-(\overline D)>\mathbb T_-(\overline B)$. Suppose, for the sake of contradiction, such a line enters $\CB(D)$ with $\mathbb T_-(\overline D)\le\mathbb T_-(\overline B)$. Let $s$ be the starting syndrome point of the line from $B$ and $p$ its first entry point into $\CB(D)$. Their time difference $\Delta t=\mathbb T(p)-\mathbb T(s)$ satisfies
\begin{equation}\label{eq:lemma5-entry-time}
-E_{k-1}\le\Delta t\le(\lambda^{\C}+3)r_k.
\end{equation}
Take an error $e\in B$ neighboring $s$ and an error $e'\in D$. Their time separation is at most $(\lambda^{\C}+2)r_k+E_{k-1}<\gamma_k$. Since the two bursts are separated by more than $\gamma_k$, their spatial separation must be larger than $\gamma_k$. Therefore
\begin{align}
d_{\Box}(\mathbb X(s),\mathbb X(p))
&>\gamma_k-(\lambda^{\C}+3)r_k\notag\\
&>\frac{(\lambda^{\C}+3)r_k}{v_\infty}+E_{k-1}.
\label{eq:lemma5-entry-distance}
\end{align}
Equation~\eqref{eq:lemma5-entry-distance} contradicts Lemma~\ref{lem:only-motion-clearing} at level $k-1$: for $\Delta t\ge0$, combine Eq.~\eqref{eq:lemma5-entry-time} with Eq.~\eqref{eq:only-line-growth}; for $\Delta t<0$, use Eq.~\eqref{eq:only-backward-displacement}. Thus $\mathbb T_-(\overline D)>\mathbb T_-(\overline B)$ whenever a syndrome line from $B$ enters $\CB(D)$.

Suppose a syndrome line from $Y\in\mathcal B_k$ enters $\CB(D)$ for a dominant $D\in\mathcal B_k$, with $Y\ne D$. Choose $D$ with $\mathbb T_-(\overline D)$ minimal among such candidates. Then, there can be the following cases:
\begin{itemize}
\item $Y$ is nondominant. Choose $Z\in\mathcal B_j$, $j>k$, with $Y\subseteq\Cong(Z)$. We can formally group a line from $Z$ reaching $Y$ with the given line from $Y$ entering $\CB(D)$. Together with $D$, by Definition~\ref{def:only-syndrome-line}, this is a valid syndrome line from $Z$, since $k<j$. Therefore
\begin{equation}
D\subseteq\Cong(Z),
\end{equation}
contradicting the dominance of $D$.

\item $Y$ is dominant. The argument following Eq.~\eqref{eq:lemma5-entry-distance} gives $\mathbb T_-(\overline Y)<\mathbb T_-(\overline D)$. By the minimal choice of $D$, no syndrome line from another level-$k$ burst enters $\CB(Y)$. Parts 1-5 therefore apply to $Y$ and give $\Cong(Y)\subseteq\SB(Y)$, as in Eq.~\eqref{eq:lemma51}. Since $\SB(Y)\cap\CB(D)=\varnothing$ by burst separation, this contradicts the chosen line from $Y$ entering $\CB(D)$.
\end{itemize}
Thus, every dominant level-$k$ burst, including the fixed $B$, satisfies the temporary assumption. Statement (1) therefore holds for $B$ without that assumption.

\textbf{7. Proof of statement (2) at level $k$.} By statement (1), the argument of part 2 of the proof now applies unconditionally, since the same-level case is now excluded by part 6. Thus every decoder cluster meeting $\Cong(B)$ contains only its syndrome.

Finally, a line from a higher-level burst cannot share a syndrome point with $\Cong(B)$: by statement (1) at level $k$, that point would lie in $\SB(B)\subseteq\CB(B)$, contradicting dominance of $B$. This proves both claims of statement (2) at level $k$ and completes the induction.
\end{proof}

\begin{proof}[Proof of Lemma~\ref{thm:cong_in_syndrome_box}]
This is simply statement (1) of Lemma~\ref{lem:only-confinement-separation}.
\end{proof}

Finally, Lemma~\ref{thm:cong_in_syndrome_box} implies Theorem~\ref{thm:only-main-deterministic}.

\begin{proof}[Proof of Theorem~\ref{thm:only-main-deterministic}]
Fix a circuit realization of nonzero probability satisfying the maximal-burst bound.

The conglomerates of dominant bursts cover the damage $D$, and Lemma~\ref{thm:cong_in_syndrome_box} gives
\begin{equation}
D=\bigcup_{B\text{ dominant}}\Cong(B)
\subseteq\bigcup_{B\text{ dominant}}\SB(B).
\end{equation}
We use Definition~\ref{def:only-admissible-scale-schedule}, the constants in Lemma~\ref{lem:only-auxiliary-constraints}, and the assumption of the Theorem in Eq.~\ref{eq:maximal}. For every $B\in\mathcal B_k$, they give
\begin{align}
\lambda^{\S}r_k+1
&<\lambda^{\C}r_k
<\lambda^{\C}Q\,\rad_{\Box}(B)\notag\\
&\le\frac{L-2}{800}.
\end{align}
Thus every dominant conglomerate is contained in a syndrome box that remains proper after depth-1 padding. To apply the local removal argument, we must also establish that the surrounding layer is free of damage.

Parts 5-6 of the proof of Lemma~\ref{lem:only-confinement-separation} show that the syndrome box of a dominant $k$-burst has a damage-free surrounding layer of depth 1 once the damage of lower-level dominant bursts has been removed. Applying the local removal argument in the proof of Proposition~\ref{prop:only-contractible-no-fault} in increasing level order therefore removes all damage without changing the logical action.

It remains to check that the decoder completes the correction within the additional $L$ noise-free recovery rounds. All errors occur by time $T$, so $\mathbb T_+(\overline B)\le T+2r_k$. By statement~(1) of Lemma~\ref{lem:only-confinement-separation}, all anyons of $\Cong(B)$ fuse to vacuum before
\begin{align}
\mathbb T_+(\overline B)+263r_k
&\le T+265r_k\notag\\
&<T+\frac{265(L-2)}{800\lambda^{\C}}
\\
&<T+L.
\end{align}
Thus $L$ noise-free recovery rounds are sufficient, and no logical fault occurs with probability one.
\end{proof}

\subsection{Non-burst-decomposable error configurations are unlikely}
\label{sec:probabilistic}

Thus far, we showed in Theorem~\ref{thm:only-main-deterministic} that if the noise satisfies the burst decomposition assumption in Def.~\ref{def:only-burst-decomposition}, the damage is contractible and thus, no logical error can occur by Proposition~\ref {prop:only-contractible-no-fault}. 
In this section, we prove that the probability for an error configuration to be not burst-decomposable under an $\epsilon$-bounded noise model vanishes quasi-stretched-exponentially with the system size $L$ if $\epsilon$ is below some $O(1)$ threshold value. This statement is the same as G\'acs's sparsity bound, see Ref.~\cite{Gacs1983,gacs,durand-romashchenko-shen}. We provide the statement here for the specific separation function $\gamma(r)$ for the burst decomposition that we introduced in Def.~\ref{def:only-burst-decomposition}.

Let $E\subseteq\intG$ be the random set of error locations. Let $\mathcal{B}$ be its burst decomposition with $\gamma(r)$ for the burst decomposition that we introduced in Def.~\ref{def:only-burst-decomposition}. The $\epsilon$-bounded noise assumption means that, for any fixed set of locations $F\subseteq\intG$, the probability that errors occur at all of them satisfies
\begin{equation}
\Pr[F\subseteq E]\le \epsilon^{|F|}.
\end{equation}

\begin{lemma}[Large-burst probability]
\label{lem:only-large-burst-probability-bound}
Define the large-burst event%
\[
\LB_R=
\{\exists B \in \mathcal{B}: \rad_{\Box}(B)> R
\}.
\]
Then, there exist constants $c > 0$, $R_* > e$, and $\epsilon_* >0$, such that for every \(R\ge R_*\)
\begin{align}
\Pr[\LB_R]\le
\abs{\intG} \left ( \frac{\epsilon}{\epsilon_*} \right)^{\exp \left(c\frac{\log R}{\log\log R}\right)}.
\end{align}
\end{lemma}
The proof is deferred to App.~\ref{app:sparsity-proof}.

\subsection{Full threshold theorem}

Theorem~\ref{thm:only-main-deterministic} shows that a maximal burst size at most $(L-2)/(800\lambda^{\C}Q)$ gives no logical fault. Let $\Fail$ denote the event of an actual logical fault.

\begin{definition}[Large-damage event]
\label{def:only-large-damage-events}
Let $\LD$ be the event that some burst $B\in\mathcal B_k$ satisfies
\[
\lambda^{\C}r_k\ge(L-2)/800.
\]
This is the failure of the bound in Theorem~\ref{thm:only-main-deterministic}.
\end{definition}

If $\LD$ does not occur, the proof of Theorem~\ref{thm:only-main-deterministic} bounds the damage in syndrome boxes and removes it in increasing level order. Thus no logical fault occurs, and $\Fail\subseteq\LD$.

\begin{theorem}[Main threshold theorem]
\label{thm:only-main-threshold}
There exist constants $\epsilon_*>0$, $c>0$, and $R_*>e$ such that, for every $\epsilon<\epsilon_*$ and every $L$ satisfying
\[
R(L)=\frac{L-2}{800\lambda^{\C}Q}\ge R_*,
\]
the logical failure probability satisfies
\begin{align*}
\Pr[\Fail]
\le\Pr[\LD]
\le\abs{\intG}
\left(\frac{\epsilon}{\epsilon_*}\right)^{
\exp\left(c\frac{\log R(L)}{\log\log R(L)}\right)}.
\end{align*}
\end{theorem}
\begin{proof}
If $\LD$ occurs, some $B\in\mathcal B_k$ has
\[
\rad_{\Box}(B)>r_{k-1}=r_k/Q
\ge\frac{L-2}{800\lambda^{\C}Q}=R(L).
\]
Thus $\LD\subseteq\LB_{R(L)}$. Applying Lemma~\ref{lem:only-large-burst-probability-bound} at $R=R(L)$ proves the bound.
\end{proof}

\section{Threshold for logical operations}
\label{sec:logical_operation}
In this section, we outline how our circuit, decoder, and threshold proof can be generalized to include arbitrary topological logical operations, enabling universal quantum computation in addition to memory.

\subsection{Logical gates as spacetime manifolds}

Historically, the idea of ``topological quantum computation'' more commonly referred to a logical operation where the spacetime 3-manifold itself has uninteresting topology but contains a complicated braid of computational-anyon-worldline defects.
However, a logical operation can be much more general, and can involve things like deforming boundaries, braiding twist defects, or interfacing different topological phases along domain walls~\cite{Bombin2010,horsman, Brown2016,Cong2017, huang2025generatinglogicalmagicstates,davydova2025universalfaulttolerantquantum,huang2025generatinglogicalmagicstates,Sajith2025}.
As visual examples,
\begin{equation*}\label{eqn:examples}
\begin{tikzpicture}[scale=0.4]
\coordinate (x) at (3,0);
\coordinate (y) at (2,1);
\coordinate (z) at (0,3.5);
\draw[dashed] (y)--++(x) (y)--++($(z)$) (0,0)--++(y);
\draw[very thick,purple,preaction={draw,white,line width=0.2cm}] ($0.75*(x)+0.5*(y)$)to[out=90,in=-90]($0.75*(x)+0.5*(y)+0.5*(z)$)to[out=90,in=-90]coordinate[pos=0.2](an2)($0.5*(x)+0.5*(y)+1*(z)$);
\draw[very thick,purple,preaction={draw,white,line width=0.2cm}] ($0.25*(x)+0.5*(y)$)to[out=90,in=-90]($0.5*(x)+0.5*(y)+0.5*(z)$)to[out=90,in=-90]($0.75*(x)+0.5*(y)+1*(z)$);
\draw[very thick,purple,preaction={draw,white,line width=0.2cm}] ($0.5*(x)+0.5*(y)$)to[out=90,in=-90]coordinate[pos=0.8](an1)($0.25*(x)+0.5*(y)+0.5*(z)$)to[out=90,in=-90]($0.25*(x)+0.5*(y)+1*(z)$);
\draw (0,0)--++(x)--++(y)--++($(z)$)--++($-1*(y)$)--++($-1*(z)$) (0,0)--++($(z)$)--++(y)--++(x) ($(z)$)--++(x);
\end{tikzpicture}
\;, \quad
\begin{tikzpicture}[scale=0.4]
\coordinate (x) at (3,0);
\coordinate (y) at (2,1);
\coordinate (z) at (0,2.2);
\fill[orange,fill opacity=0.2] ($0.5*(z)$)--++(x)--++(y)--++(z)--++($-1*(x)$)--++($-1*(y)$)--++($-1*(z)$);
\draw (0,0)--++(x)--++(y)--++($2*(z)$)--++($-1*(y)$)--++($-2*(z)$) (0,0)--++($2*(z)$)--++(y)--++(x) ($2*(z)$)--++(x);
\draw[dashed] (y)--++(x) (y)--++($2*(z)$) (0,0)--++(y);
\draw[purple,pattern=north west lines,pattern color=purple] ($0.5*(z)$)--++(x)--++(y)--++($-1*(x)$)--++($-1*(y)$);
\draw[purple,pattern=north west lines,pattern color=purple] ($1.5*(z)$)--++(x)--++(y)--++($-1*(x)$)--++($-1*(y)$);
\end{tikzpicture}
\;, \quad 
\begin{tikzpicture}[scale=0.4]
\coordinate (x) at (2,0);
\coordinate (y) at (0,2);
\coordinate (z) at (1.6,0.8);
\draw[very thick,purple] ($(x)+(z)$)--++($0.5*(y)$)--++($0.4*(x)$)--++($-0.5*(y)$);
\draw[very thick,purple] ($2*(y)+(x)+(z)$)--++($-0.5*(y)$)--++($0.4*(x)$)--++($0.5*(y)$);
\draw[very thick,purple] (z)--++($2*(y)$) ($2.4*(x)+(z)$)--++($2*(y)$);
\draw[very thick,purple] (x)--++($0.5*(y)$)--++($0.4*(x)$)--++($-0.5*(y)$);
\draw[very thick,purple] ($2*(y)+(x)$)--++($-0.5*(y)$)--++($0.4*(x)$)--++($0.5*(y)$);
\draw[very thick,purple] (0,0)--++($2*(y)$) ($2.4*(x)$)--++($2*(y)$);
\draw (0,0)--++(x)--++(z) ($1.4*(x)$)--++(x)--++(z) ($2*(y)$)--++(x)--++(z) ($1.4*(x)+2*(y)$)--++(x)--++(z) ($2*(y)$)--++(z)--++(x) ($1.4*(x)+2*(y)$)--++(z)--++(x);
\draw[dashed] (0,0)--++(z)--++(x) ($1.4*(x)$)--++(z)--++(x) ($1.4*(x)+0.5*(y)$)--++(z) ($1*(x)+0.5*(y)$)--++(z) ($1.4*(x)-0.5*(y)+2*(y)$)--++(z) ($1*(x)-0.5*(y)+2*(y)$)--++(z);
\fill[cyan,opacity=0.4] (0,0)--++(z)--++($2*(y)$)--++($-1*(z)$)--cycle;
\fill[cyan,opacity=0.4] ($2.4*(x)$)--++(z)--++($2*(y)$)--++($-1*(z)$)--cycle;
\fill[cyan,opacity=0.4] (x)--++(z)--++($0.5*(y)$)--++($-1*(z)$)--cycle;
\fill[cyan,opacity=0.4] ($1.4*(x)$)--++(z)--++($0.5*(y)$)--++($-1*(z)$)--cycle;
\fill[cyan,opacity=0.4] ($(x)+0.5*(y)$)--++(z)--++($0.4*(x)$)--++($-1*(z)$)--cycle;
\fill[cyan,opacity=0.4] ($2*(y)+(x)$)--++(z)--++($-0.5*(y)$)--++($-1*(z)$)--cycle;
\fill[cyan,opacity=0.4] ($2*(y)+1.4*(x)$)--++(z)--++($-0.5*(y)$)--++($-1*(z)$)--cycle;
\fill[cyan,opacity=0.4] ($(x)+1.5*(y)$)--++(z)--++($0.4*(x)$)--++($-1*(z)$)--cycle;
\end{tikzpicture}
\;,
\end{equation*}
schematically illustrate spacetime diagrams for braiding computational anyon worldlines, interfacing different bulk phases (orange and transparent) via domain walls, and lattice surgery of patches with two different boundary conditions (blue and transparent) separated by corner/line defects (purple lines).
The logical action of the protocol is fully determined by the defect-decorated 3D spacetime manifold, as well as a topological phase or super-selection sector assigned to each bulk and to each topological defect.

More precisely, a topological logical protocol can be specified by a 3D spacetime manifold whose input and output 2D manifolds are at the bottom and top time boundaries.
In practice, instead of using 3D spacetimes and 2D input/output manifolds with complicated intrinsic topology, we often construct logical protocols by embedding lower-dimensional defects (which are submanifolds) in a 3D spacetime with a simple topology.  Examples of 2-dimensional defects are boundaries and domain walls.  Examples of 1-dimensional defects are ``computational'' (as opposed to noise-caused) anyon worldlines, corner defects separating different boundaries, and twist defects where domain walls terminate.
Examples of 0-dimensional defects are fusion events where different computational anyon worldlines meet.  Due to the addition of lower-dimensional defects, the QEC circuit will necessarily change; a general formalism for constructing such circuits is absent in the literature.  Here, we provide a systematic way of generating them.

The goal of this section is to describe how to implement these kinds of topological protocols fault-tolerantly.
We will focus on topological boundaries as the main example for a topological defect, but the ideas readily generalize.

\subsection{QEC circuit}
\label{sec:defect_circuit}
We start by sketching how to construct a circuit implementing a 3-manifold with topological boundaries. We choose a cellulation of a 3-manifold and all of its lower-dimensional defects. The cellulation of the boundary can be chosen to be simply its two-dimensional triangulation compatible with the bulk cellulation. As in Section~\ref{sec:circuit}, each circuit round consists of fusion-space recovery, syndrome extraction, and correction.

The fault distance $L$ of the logical protocol is the minimum cellular length of a non-contractible path, which can either be a loop or have endpoints on lower-dimensional defects.
For example, the 3D spacetime manifold could be $S_1\times [0,1]\times [0,1]$ where the last $[0,1]$ interval is time, corresponding to a memory block storing logical information in a spatial annulus $S_1\times [0,1]$.
Then a minimal path could either be an $S_1\times \{\frac12\} \times \{\frac12\}$ loop winding around the annulus at a constant time, or a $\{0\}\times [0,1]\times \{\frac12\}$ path with two endpoints connecting the different sides of the annulus at a constant time.
We can decrease the logical error rate arbitrarily by increasing $L$.

\textbf{Boundary tensor.}
There is a canonical path-integral construction for any topological defect which assigns tensors to all the $i$-cells of an $i$-dimensional defect in spacetime.
We assign an $L$-tensor to each triangle of the boundary triangulation. We depict the triangle as thickened into a triangular prism, with thick edges on the boundary face and thin edges on the face to be glued to the bulk:
\begin{equation}
\begin{tikzpicture}
\atoms{cod1vertex}{{0/p={-30:1.8},dlab={d}{-30:0.25}}, {1/p={90:1.8},dlab={e}{90:0.25}}, {2/p={-150:1.8},dlab={f}{-150:0.25}}}
\atoms{vertex}{0x/p={-30:0.9}, 1x/p={90:0.9}, 2x/p={-150:0.9}}
\draw[cod1ind] (0)edge[or](1) (1)edge[or](2) (0)edge[or](2);
\draw (0x)edge[or,sglab=a](1x) (1x)edge[or,sglab=b](2x) (0x)edge[or,sglabr=c](2x);
\draw[extend map] (0)--(0x) (1)--(1x) (2)--(2x);
\node at (0,0){$\alpha$};
\node at (30:0.65){$\beta$};
\node at (150:0.65){$\gamma$};
\node at (-90:0.65){$\delta$};
\end{tikzpicture}
\quad\rightarrow\quad
{L^{abf}_d}_{e\beta\gamma}^{c\alpha\delta}\;.
\end{equation}
The $L$-tensor has 10 indices as shown above.
The indices $a,b,c$ are the string labels on the three edges of the prism's bulk-facing triangular face, and $\alpha$ is the fusion label on that face. When assembling the path integral, $a,b,c$ are identified with the corresponding edge labels of the incident bulk $F$-tensors, and summed over. When the triangular face is glued to a bulk tetrahedron, $\alpha$ is contracted with the corresponding fusion index of its $F$-tensor.
The labels $d,e,f$ are \emph{boundary vertex labels}, which have \emph{quantum dimensions} $d_{d/e/f}$. The labels $\beta,\gamma,\delta$ are \emph{boundary edge labels}, whose allowed ranges depend on the vertex labels at their endpoints and the string label on the corresponding bulk edge. For example, $\delta$ ranges over $0\leq\delta<M_{fcd}$.
Mathematically, if we think of the $F$-tensor as a fusion category, then the $L$-tensor is a \emph{module category} of this fusion category.

\textbf{Hilbert space and fusion-space recovery.}
The vector space at one time step consists of degrees of freedom distributed over the 2D spatial slice at that time.
For a 3-manifold with boundaries, a time slice is a 2D cellulation with boundary, and the degrees of freedom correspond to string labels, fusion labels, boundary edge labels, and boundary vertex labels.
The boundary edge labels correspond to qudits of dimension $\max_{a,b,c} M_{abc}$.
The fusion-space recovery consists of local channels that map the state back to the fusion subspace determined by the constraint $0\leq \delta<M_{fcd}$ for a boundary edge label $\delta$ with adjacent boundary vertex and bulk string labels $f,c,d$.

\textbf{Isometric boundary gluings.}
The syndrome extraction layer corresponds to a thickness-1 layer of the 3D spacetime cellulation separating two time slices.
Each tensor becomes an operator in the circuit.
For example, we can interpret the $L$-tensor as an operator mapping from the bulk triangle and one boundary edge to the other two boundary edges,
\begin{equation}
\begin{gathered}
\mathbf L^{(22)}_{(a,b,f,d,e,\beta,\gamma),(a',b',f',d',c,\delta,\alpha)}\\
= \delta_{aa'}\delta_{bb'}\delta_{ff'}\delta_{dd'}
{L^{abf}_d}_{e\beta\gamma}^{c\alpha\delta} \sqrt{d_c} \sqrt{d_e}
\;.
\end{gathered}
\end{equation}
The deformation of the 2D cellulation due to this gluing is given by
\begin{equation}
\label{eq:2d_boundary_invariance}
\begin{tikzpicture}
\atoms{vertex}{0/, 2/p={1.6,0}, 1/p={0.8,0.7}}
\atoms{cod1vertex}{0x/p={$(0)+(90:-0.3)$}, 2x/p={$(2)+(90:-0.3)$}}
\draw[extend map] (0)--(0x) (2)--(2x);
\draw[cod1ind] (0x)edge[or](2x);
\draw (0)edge[or](2) (0)edge[or](1) (1)edge[or](2);
\end{tikzpicture}
\quad\rightarrow\quad
\begin{tikzpicture}
\atoms{vertex}{0/, 2/p={2,0}, 1/p={1,0}}
\atoms{cod1vertex}{0x/p={$(0)+(-90:0.3)$}, 2x/p={$(2)+(-90:0.3)$}, 1x/p={$(1)+(-90:0.3)$}}
\draw[extend map] (0)--(0x) (2)--(2x) (1)--(1x);
\draw[cod1ind] (0x)edge[or](1x) (1x)edge[or](2x);
\draw (0)edge[or](1) (1)edge[or](2);
\end{tikzpicture}
\;.
\end{equation}
This operator is an isometry restricted to the fusion subspace.
This follows from the topological invariance, irreducibility, and normal form axioms that hold for the boundary, which we do not show here.

\textbf{Boundary anyons and quantum instruments.}
Any implementation of a proper thickness-1 slice of a 3D triangulation as a sequence of gluings of tetrahedra and boundary triangles necessarily involves gluing operations that remove a boundary vertex from the 2D spatial triangulation with boundary.
The corresponding interpretation of the $L$-tensor, as a map from the bulk-facing triangle and two boundary edges to the single remaining boundary edge, is not an isometry on the fusion subspace. We therefore use an alternative gluing involving a \emph{boundary anyon}, which provides a quantum instrument analogous to the bulk construction.
We consider cellulations with a boundary and anyon worldlines inside the 2D boundary.
The anyon worldlines are cellulated, and to each edge of this 1D cellulation we assign the following tensor:
\begin{equation}
\label{eq:boundaryanyon_edge_bdiag}
\begin{tikzpicture}
\atoms{vertex}{{0/p={-0.1,0}}, {1/p={-0.1,1.5}}}
\atoms{cod1vertex}{{0a/p={$(0)+(-0.4,0.5)$},dlab={\alpha}{180:0.25}}, {0b/p={$(0)+(0.8,0.5)$},dlab={\alpha'}{0:0.25}},{1a/p={$(1)+(-0.4,0.5)$},dlab={\beta}{180:0.25}}, {1b/p={$(1)+(0.8,0.5)$},dlab={\beta'}{0:0.25}}}
\atoms{cod2vertex}{{0x/p={$(0)+(0.4,1)$},vback,dlab={\mu}{-80:0.25}}, {1x/p={$(1)+(0.4,1)$},dlab={\nu}{90:0.25}}}
\draw[extend map] (0)--(0a) (0)edge[mark=bar](0b) (0x)edge[backedge](0) (0x)edge[backedge](0a) (0x)edge[backedge](0b) (1)--(1a) (1)edge[mark={bar,p=0.3}](1b) (1x)--(1) (1x)--(1a) (1x)--(1b);
\draw (0)edge[or=0.6,mark={slab=$i$,p=0.6}](1);
\draw[cod1ind] (0a)edge[or](1a) (0b)edge[or](1b);
\draw[cod2ind,backedge] (0x)edge[or=0.6](1x);
\node at (-0.3,1.1) {$\pi$};
\node at (0.5,1.3) {$\pi'$};
\end{tikzpicture}
\quad\rightarrow\quad
{\kappa}_{\alpha\beta\pi\mu}^{i\alpha'\beta'\pi'\nu}
\;.
\end{equation}
The dashed thick vertical edge in the background is the actual boundary-anyon edge.
The two semi-thick vertical edges on the sides are identified with boundary edges of two adjacent boundary triangles.
The two boundary edges are associated with the same bulk edge, which is the thin vertical edge at the front.
Geometrically, we may imagine the above as a cube-like volume that is wedged between two thickened boundary triangles, and attached to the sides of the thickened boundary triangles via the vertical rectangles in front.
The index $i$ of the associated tensor $\kappa$ is a string label shared with the two adjacent $L$-tensors and all adjacent $F$-tensors in the bulk.
$\alpha,\beta,\alpha',\beta'$ are boundary vertex labels shared with the neighboring boundary triangles.
$\pi$ and $\pi'$ are boundary edge labels contracted with the single neighboring boundary triangle on each side.
There are two new types of labels $\mu$ and $\nu$, which are contracted with the same label of the adjacent boundary-anyon edge on each side.
Its dimension depends on the configuration of the adjacent boundary vertex labels, for example, the range of $\mu$ depends on the configuration of $\alpha$ and $\alpha'$.

We interpret this tensor as an operator from the bottom boundary-anyon vertex and two boundary edges at the sides to the top boundary-anyon vertex, parameterized by the boundary-anyon type $\phi$, the boundary vertex labels $\alpha$ and $\alpha'$, and the labels $\mu$ and $\nu$,
\begin{equation}
\begin{aligned}
&\mathbf \kappa^{\phi,\alpha'',\alpha''',\mu,\nu}_{(\beta,\beta'),(\beta'',\beta''',\alpha,\alpha',\pi,\pi')}\\
=& \delta_{\beta\beta''}\delta_{\beta'\beta'''}\delta_{\alpha\alpha''}\delta_{\alpha'\alpha'''} (\kappa^\phi)_{\alpha\beta\pi\mu}^{i\alpha'\beta'\pi'\nu}\sqrt{d_id_\alpha d_{\alpha'}}\sqrt{\frac{d_\phi}{D}}\;.
\end{aligned}
\end{equation}
This set of operators forms a valid set of Kraus operators.
This follows from the topological invariance, irreducibility, and normal form axioms for the boundary-anyon worldline, which we will not discuss explicitly here.

The effect of the corresponding gluing on the 2D spatial cellulation with boundary is to replace two boundary edges with a single boundary vertex,
\begin{equation}
\label{eq:boundary_anyon_move}
\begin{tikzpicture}
\atoms{vertex}{0/, 0x/p={0,1}}
\atoms{cod1vertex}{1/p={0.3,-0.3},2/p={-0.3,-0.3}, 1x/p={0.3,0.7}, 2x/p={-0.3,0.7}}
\atoms{cod2vertex}{3/p={0,-0.6}}
\draw (0)edge[or](0x);
\draw[cod1ind] (1)edge[or](1x) (2)edge[or](2x);
\draw[extend map] (0)--(1) (0)--(2) (0)--(3) (1)--(3) (2)--(3) (0x)--(1x) (0x)--(2x);
\end{tikzpicture}
\quad\rightarrow\quad
\begin{tikzpicture}
\atoms{vertex}{0/}
\atoms{cod1vertex}{1/p={0.3,-0.3},2/p={-0.3,-0.3}}
\atoms{cod2vertex}{3/p={0,-0.6}}
\draw[extend map] (0)--(1) (0)--(2) (0)--(3) (1)--(3) (2)--(3);
\end{tikzpicture}
\;.
\end{equation}

\textbf{Syndrome-extraction layer on a cubic spacetime lattice with spatial boundary.}
As a concrete example, we show how to implement a spacial boundary of a 2D spatial square lattice (located on the right below).
To implement the path integral via isometries and measurements, we need to decorate each bulk vertex with an anyon 1-gon, and each boundary vertex with a boundary anyon location,
\begin{equation}
\label{eq:compactification}
\begin{tikzpicture}
\clip (1.2,1)rectangle++(5,4);
\foreach \x in {0,1,2}{
\foreach \y in {0,1,2,3,4,5}{
\atoms{vertex}{0\x\y/p={1.5*\x,1.5*\y}}
\atoms{cod2vertex}{1\x\y/p={1.5*\x+0.5,1.5*\y}}
\draw (0\x\y)edge[or,bend right=45]++(0:1.5) ($(0\x\y)+(1.5,0)$)edge[or]++(90:1.5);
\draw[or] (0\x\y)to[out=-45,in=-90]++(0:0.8)to[out=90,in=45](0\x\y);
\draw (0\x\y)edge[or,bend left=45]++(0:1.5) (0\x\y)edge[or]++(1.5,1.5);
\draw[extend map] (0\x\y)--(1\x\y);
}}
\foreach \y in {0,1,2,3,4,5}{
\atoms{vertex}{0\y/p={4.5,1.5*\y}}
\atoms{cod2vertex}{2\y/p={5.1,1.5*\y}}
\atoms{cod1vertex}{3\y/p={4.8,1.5*\y-0.3},4\y/p={4.8,1.5*\y+0.3}}
\draw[extend map] (0\y)--(2\y) (0\y)--(3\y) (0\y)--(4\y) (3\y)--(2\y) (4\y)--(2\y);
\draw[cod1ind,or] (4\y)--++(90:0.9);
}
\end{tikzpicture}
\end{equation}
As the corresponding 3D spacetime cellulation we take the product of the 2D cellulation with a 1D cellulation, i.e., a sequence of edges.
As before, we triangulate each resulting triangle prism with three tetrahedra, and additionally, we triangulate each resulting boundary square with two boundary triangles.
Now imagine sequentially gluing the tetrahedra and boundary triangles of the depth-1 layer.
The intermediate time slices are different from the 2D cellulation above:
They contain triangles and boundary edges whose spatial projections are in one-to-one correspondence with the triangles and boundary edges in Eq.~\eqref{eq:compactification}, but the slices additionally contain triangles and boundary edges that are oriented time-parallel in the 3D cellulation.
In each gluing, we glue a tetrahedron or boundary triangle to exactly one of the non-time-parallel edges or boundary triangles, and additionally some number (sometimes zero) of time-parallel triangles or boundary edges.
So we can depict the ordering of gluings by marking each spatial triangle or boundary edge by the time steps in which a tetrahedron or boundary triangle is glued to them.
Inside the bulk, the ordering is the same as in Eq.~\eqref{eq:syndrome_extraction_layer}, and at the boundary we extend it as follows:
\begin{equation}
\label{eq:compactification}
\begin{tikzpicture}
\clip (2.7,1)rectangle++(2.8,2.2);
\foreach \x in {0,1,2}{
\foreach \y in {0,1,2}{
\atoms{vertex}{0\x\y/p={1.5*\x,1.5*\y}}
\atoms{cod2vertex}{1\x\y/p={1.5*\x+0.5,1.5*\y}}
\draw (0\x\y)edge[or,bend right=45]++(0:1.5) ($(0\x\y)+(1.5,0)$)edge[or]++(90:1.5);
\draw[or] (0\x\y)to[out=-45,in=-90]++(0:0.8)to[out=90,in=45](0\x\y);
\draw (0\x\y)edge[or,bend left=45]++(0:1.5) (0\x\y)edge[or]++(1.5,1.5);
\draw[extend map] (0\x\y)--(1\x\y);
}}
\foreach \y in {0,1,2}{
\atoms{vertex}{0\y/p={4.5,1.5*\y}}
\atoms{cod2vertex}{2\y/p={5.1,1.5*\y}}
\atoms{cod1vertex}{3\y/p={4.8,1.5*\y-0.3},4\y/p={4.8,1.5*\y+0.3}}
\draw[extend map] (0\y)--(2\y) (0\y)--(3\y) (0\y)--(4\y) (3\y)--(2\y) (4\y)--(2\y);
\draw[cod1ind,or] (4\y)--++(90:0.9);
}
\node[red,font=\small]at (3.4,2.4){035};
\node[red,font=\small]at (4.1,2.2){146};
\node[red,font=\small]at (4.1,1.5){278};
\node[red,font=\small]at (3.25,1.5){9};
\node[red,font=\small,align=center]at (4.65,2.25){3\\5};
\node[red,font=\small]at (4.8,1.5){9};
\end{tikzpicture}
\end{equation}
The boundary gluings in steps 3 and 5 are of the type in Eq.~\eqref{eq:2d_boundary_invariance} corresponding to the operator $\mathbf L^{(22)}$, and the final gluing in step 9 is of the type in Eq.~\eqref{eq:boundary_anyon_move} corresponding to the operator $\mathbf\kappa$.

\textbf{Corrections.}
Finally, to describe the correction layer, we need additional tensors corresponding to (1) vertices in the boundary where three boundary anyons fuse, and (2) vertices in the boundary where a bulk anyon meets a boundary anyon.
We will not discuss these tensors in detail here, but bulk-to-boundary anyon meeting points are described for group-cohomology models in Ref.~\cite{Magdalena2023}.

\textbf{Other topological defects.}
In general, the 3D bulk or any 2D topological defect needs its own kind of anyons used for syndrome extraction and correction: Ordinary anyons for the bulk, boundary anyons for boundaries, ``domain wall anyons" for domain walls, etc.
1D topological defects such as corners separating different boundary types or computational anyon worldlines can be implemented directly.
In particular, computational anyon worldlines require no change to the syndrome-extraction circuit of Section~\ref{sec:circuit}. Instead, they change which measurement outcomes the decoder regards as nontrivial. Along a prescribed worldline of type $\phi$, the outcome $\phi$ is treated as trivial, while every other type, now including the vacuum, will be treated as a nontrivial syndrome point that enters into the clustering algorithm. 

\myparagraph{Other spacetime orientations}
The boundary circuit considered above was for a spatial boundary.
Fortunately, the same path-integral $L$ tensors can be constructed to define boundary circuits for different spacetime orientations.
Let us briefly discuss a time-perpendicular boundary facing into the past direction.
When put on a spatial torus, for example, such a protocol would fault-tolerantly prepare a specific torus logic state depending on the topological phase of the boundary.
When gluing the boundary together from thickened past-facing boundary triangles, we first spawn disconnected boundary components, then join these up, and finally close any remaining holes in the boundary.
In the first step, we use the $L$-tensors as operators from nothing to all of their indices, or in other words, we just view the $L$-tensor as a state vector and prepare that state.
However, the last step closing holes would require us to apply some $L$-tensor as an operator from all of its three edges to the bulk-facing triangle.
This operator is not an isometry.
To solve this problem, we can use condensation defects, which are 0-dimensional defects where an anyon worldline terminates at a boundary.
Geometrically, such a point is represented by a ``thickened 1-gon'' volume,
\begin{equation}
\label{eq:anyon_condensation_vertex_bdiag}
\begin{tikzpicture}
\atoms{vertex}{0/}
\atoms{cod1vertex}{{0x/p={0,-0.5},dlab={\alpha}{180:0.25}}}
\atoms{cod2vertex}{{0y/p={0.6,0},dlab={\mu}{0:0.25}}}
\atoms{cod3vertex,vback}{0z/p={0.6,-0.5}}
\draw (0)edge[mark={arr,e},out=-45,in=-90,mark={slab=$a$,e,r}]++(0:1.2) (0)edge[out=45,in=90]++(0:1.2);
\draw[cod1ind] (0x)edge[mark={arr,p=0.8},out=-45,in=-90]++(0:1.2) (0x)edge[backedge,out=45,in=90]++(0:1.2);
\draw[extend map] (0)--(0y) (0x)edge[backedge](0z) (0y)edge[midedge](0z) (0)--(0x);
\node at (1,-0.5){$\nu$};
\end{tikzpicture}
\quad\rightarrow\quad
S^{a,\alpha,\mu\nu}
\;,
\end{equation}
with an anyon 1-gon at the top identified with the top or bottom 1-gon of an anyon cylinder, and the side identified with a thickened edge of a thickened boundary triangle ``winding around'' the 1-gon.
As shown, we can associate to this a tensor with 4 indices that are identified with those of the neighboring anyon cylinders and boundary triangles.
Gluing the looping thickened boundary edge does close a hole in the boundary and leaves an anyon 1-gon behind.
When we accordingly view the $S$-tensor as an operator from the thickened edge on the side to the anyon 1-gon at the top, this operator can be extended to a set of Kraus operators parameterized by the anyon type $\phi$ of the adjacent anyon worldline and the multiplicity label $\mu$,
\begin{equation}
\mathbf S^{(\phi\mu)}_{(a),(a'\alpha\nu)}
= \delta_{aa'} S^{a,\alpha,\mu\nu}
\sqrt{d_\alpha}\;.
\end{equation}

Unfortunately, while the circuit above performs the correct state preparation when we post-select every measurement outcome to the trivial one, and has a distance scaling with $L$, it is not decodable.
The reason is that the anyon type measured in the $S$-type measurements is not deterministically trivial even in the absence of noise.
So this constant-depth preparation circuit will produce a random high-density configuration of anyons that break the decoder as they do not obey a sparse cluster decomposition.
So we cannot implement a time-perpendicular past-facing boundary fault tolerantly.
However, we can tilt the boundary in spacetime such that it moves/expands at an $O(1)$ speed.
A spacetime with such a boundary can be glued together without having to fill holes, so without $S$-type measurements, such that in the absence of errors all measurement outcomes are deterministically trivial.
The same principle also rules out time-perpendicular computational-anyon worldline from being implemented fault-tolerantly.
We will discuss this issue in more detail in Section~\ref{sec:causality} after generalizing decoding to logical protocols.

\subsection{Decoder}
We now adapt the decoder of Section~\ref{sec:decoding} by clustering nontrivial measurement outcomes both in the bulk and on topological defects. The defect geometry also changes how the base point of each cluster is chosen.

In order to determine the anyon movements once clusters are identified, we need to define a direction for time, especially since the 3D cellulation may be irregular.
For each vertex in a time-$t$ slice, we choose a neighboring successor vertex in the time-$t+1$ slice, such that a line following a sequence of consecutive successor vertices is regarded as stationary in space.
Then the corrections included in the circuit must be such that they move an anyon with $O(1)$ speed relative to this stationary time direction.
Furthermore, the spatial projection of the base point of a cluster is obtained by following a stationary line to the present time.

There is also a subtlety when it comes to choosing the base point:
For each cluster, we consider the lowest-dimensional topological defect that intersects with its $\beta$-box (the box used for clustering).
If there are multiple intersecting lowest-dimensional defects, we arbitrarily pick one of them.
Then we determine the base point on that lowest-dimensional defect by choosing a deterministic way of ``projecting" the oldest point of the cluster onto that defect.
The procedure is summarized in Algorithm~\ref{alg:defect_decoding}.
To illustrate why we need to project the base point onto the lowest-dimensional nearby defect, we again look at a boundary as an example for a topological defect.
Every boundary type allows a subset of anyon types to ``condense'' there, meaning that anyon worldlines of that type can terminate at the boundary.
When an error burst touches such a boundary, it can create a single one of these condensing anyons.
If the base point was in the bulk, the condensing anyon would just move to that point and remain there forever without being annihilated.
However, if the base point is inside the boundary, the condensing anyon worldline will move to the boundary and terminate (condense) there.

As another example, consider the case where the lowest-dimensional nearby defect is a computational $\tau$ anyon worldline included inside a double-Fibonacci circuit.
An error burst overlapping with the $\tau$ anyon worldline can ``split off" a single $\tau$ anyon from that worldline, due to the $\tau\in\tau\times \tau$ fusion channel.
The only way to remove the extra $\tau$ anyon worldline caused by noise is to fuse it back to the computational $\tau$ line, and choosing the spatial base point on the computational $\tau$ anyon will exactly do that.

\begin{algorithm}[t]
\caption{One step of decoding in the presence of topological defects (informal)}
\label{alg:defect_decoding}
\begin{algorithmic}[1]
\Require History of all syndrome measurement outcomes up to time $t$
\Ensure Movement directions for all anyons at time $t$
\State Compute the cluster decomposition of all non-trivial measurement results, using the separation function $\beta(r)$.
Along a 1D topological defect of anyon type $\phi$, a measurement outcome is non-trivial if it is not equal to $\phi$.
\For{each cluster}
\State Determine a ``preliminary base point'': the oldest point in the cluster with lexicographically smallest coordinate.
\State Determine lowest-dimensional topological defect overlapping with the cluster's $\beta$-box. If there are none, return ``bulk''.
\State Call $I$ the intersection of the cluster's $\beta$-box and the lowest-dimensional topological defect.  Define the base point to be the vertex in $I$ closest to the preliminary base point. 
\EndFor
\State Direct anyons detected in the last round one step towards the spatial projection of the base point of their clusters.
\end{algorithmic}
\end{algorithm}

\subsection{Causality constraints}
\label{sec:causality}
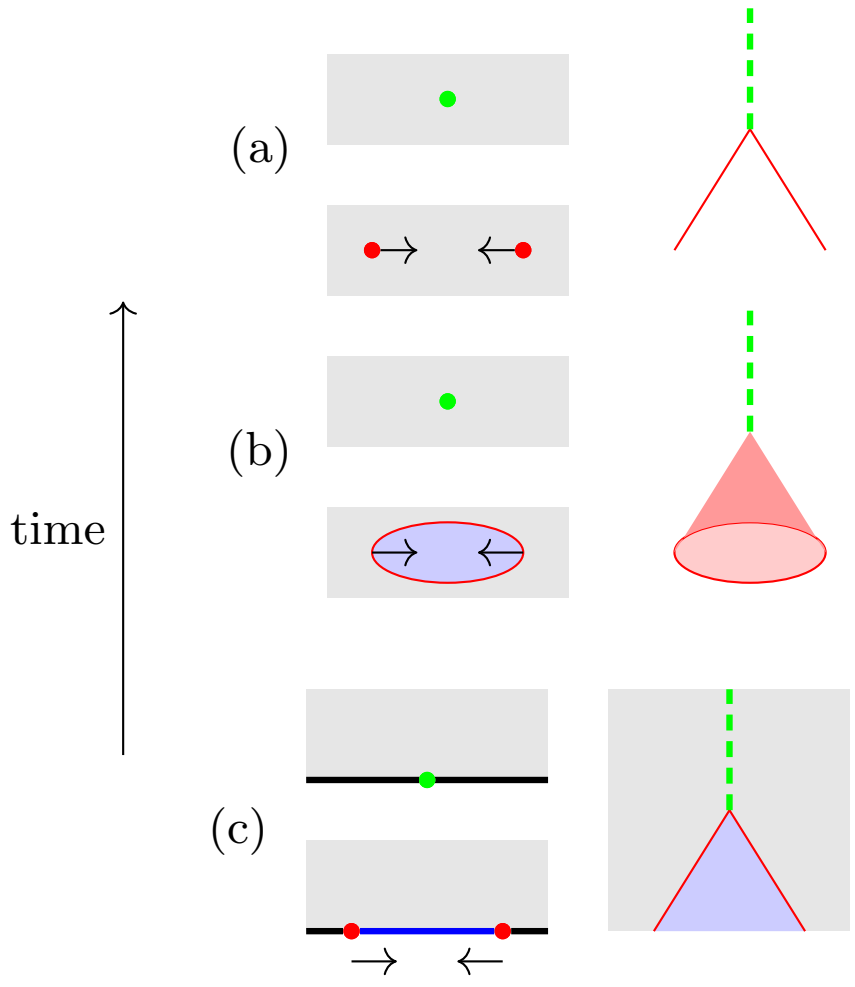
\begin{figure}
    \centering
\begin{tikzpicture}
\draw (-3,-4)edge[->]node[midway,left]{time}++(90:3);
\node (0)at (0,0){
\begin{tikzpicture}
\draw[red] (-0.5,0)--(0,0.8) (0.5,0)--(0,0.8);
\draw[densely dashed,very thick,green] (0,0.8)--++(90:0.8);
\fill[black!10] (-2.8,-0.3)rectangle++(1.6,0.6);
\atoms{vertex,astyle=red}{a/p={-2.5,0}, b/p={-1.5,0}}
\draw (a)edge[->]++(0:0.3) (b)edge[->]++(180:0.3);
\fill[black!10] (-2.8,0.7)rectangle++(1.6,0.6);
\atoms{vertex,astyle=green}{c/p={-2,1}}
\end{tikzpicture}
};
\node (1)at (0,-2){
\begin{tikzpicture}
\draw[red,fill=red!20] (0,0)ellipse(0.5cm and 0.2cm);
\fill[red!40] (-0.5,0)arc(180:0:0.5cm and 0.2cm)--(0,0.8)--cycle;
\draw[densely dashed,very thick,green] (0,0.8)--++(90:0.8);
\fill[black!10] (-2.8,0.7)rectangle++(1.6,0.6);
\fill[black!10] (-2.8,-0.3)rectangle++(1.6,0.6);
\atoms{vertex,astyle=green}{c/p={-2,1}}
\draw[red,fill=blue!20] (-2,0)ellipse(0.5cm and 0.2cm);
\draw (-2.5,0)edge[->]++(0:0.3) (-1.5,0)edge[->]++(180:0.3);
\end{tikzpicture}
};
\node (2)at (0,-4.5){
\begin{tikzpicture}
\fill[black!10] (-0.8,0)rectangle++(1.6,1.6);
\fill[blue!20] (-0.5,0)--(0,0.8)--(0.5,0)--cycle;
\draw[red] (-0.5,0)--(0,0.8) (0.5,0)--(0,0.8);
\draw[densely dashed,very thick,green] (0,0.8)--++(90:0.8);
\fill[black!10] (-2.8,0)rectangle++(1.6,0.6);
\atoms{vertex,astyle=red}{a/p={-2.5,0}, b/p={-1.5,0}}
\draw[very thick,black] (a)--++(180:0.3) (b)--++(0:0.3);
\draw[very thick,blue] (a)--(b);
\draw ($(a)+(-90:0.2)$)edge[->]++(0:0.3) ($(b)+(-90:0.2)$) edge[->]++(180:0.3);
\fill[black!10] (-2.8,1)rectangle++(1.6,0.6);
\draw[very thick,black] (-2.8,1)--++(1.6,0);
\atoms{vertex,astyle=green}{c/p={-2,1}}
\end{tikzpicture}
};
\node[anchor=east] at (0.west){(a)};
\node[anchor=east] at (1.west){(b)};
\node[anchor=east] at (2.west){(c)};
\end{tikzpicture}
\caption{Disallowed constrictions in 3D spacetime diagrams with topological defects with respect to time flowing upwards:
(a) A $\wedge$-like bend in a computational anyon worldline (in red), corresponding to two anyons fusing and disappearing.
The right side shows the anyon-worldline configuration embedded into the 3D spacetime, and the left side shows two spatial slices of the protocol.
This can be fixed by adding a computational anyon worldline (in dashed green) going in the time direction from the constriction point, which is simply saying that in general we cannot deterministically guarantee a trivial fusion outcome when fusing the two anyons.
(b) An inverse-cone-shaped domain wall membrane corresponding to a domain wall loop in space separating two bulk phases (blue and gray), contracting to a point and disappearing.
The contraction of the loop may result in a non-trivial anyon charge, and the protocol can be fixed by adding an explicit computational anyon worldline that captures this charge.
(c) Two corners separating two different types of boundaries (blue and gray) meeting at a point.
Again, this can be fixed by adding a green anyon worldline, which in this case is actually a boundary anyon.
}
\label{fig:constriction_examples}
\end{figure}

To be compatible with the decoding method described above, the 3-manifold diagrams with topological defects must obey the following two constraints with regard to the flow of time:
\begin{enumerate}
\item A topological defect must not move at speed $\geq O(1)$ relative to the previously defined direction of time, and must not be oriented in a time-perpendicular way (which could be regarded as moving at infinite speed).
\item Following the time direction, there cannot be an event where some configuration of defects in the spatial 2D manifold contracts to a point and disappears.  We call this a \emph{constriction event}.
See Figure~\ref{fig:constriction_examples} for examples of this.\footnote{Intuitively, if we smooth out the constricted defect in spacetime, then it is oriented time-perpendicularly at the constriction point. Vice versa, we can locally deform a time-perpendicular defect into a ``squiggly'' line or ``egg-carton'' surface such that it becomes non-time-perpendicular except for constriction points at the maxima.
This shows that two constraints are closely related.}
\end{enumerate}

Let us now explain the first constraint.
We already discussed in Section~\ref{sec:defect_circuit} that a defect that is time-perpendicular or moving at speed $\geq O(1)$ cannot be implemented with isometries and measurements whose outcome is deterministically trivial in the absence of noise.
In addition to that, if a defect moving at speed $\geq 1$ is affected by an error burst that creates an individual condensing anyon, then this anyon will need to be fused back into that defect.
However, since corrections in the circuit move at speed at most $1$, the anyon can never manage to catch up with the defect.
The decoder will correctly put the base point inside the defect, and the anyon will chase after that base point, but never reach it.

Let us now explain the second constraint.
A constriction behaves analogously to ``closing a hole'' when creating a past-facing time-perpendicular boundary as discussed at the end of Section~\ref{sec:defect_circuit}:
In order to implement it as a circuit we need at least one measurement at the constriction whose outcome is not deterministically trivial even in the absence of errors.
So the constriction can randomly create an anyon worldline.
To get rid of that anyon worldline we would have to fuse it into the constricted defect where it was created, which is impossible as this is in the past.
If there is noise, this might create additional anyon worldlines from the constricted defect near but not directly at the constriction.
Also these will be impossible to get rid of if the decoder does not manage to fuse them back into the defect before the defect disappears at the constriction event.

Constriction points can be fixed by simply declaring the non-deterministic measurement at the constriction as the beginning of a computational worldline of indefinite type.
The type of this worldline then corresponds to a logical measurement result in the protocol.
Of course, in order to fault-tolerantly determine the type, we have to keep measuring the indefinite-type worldline for $O(L)$ time.
Noise-caused anyon worldlines emerging from the defect near its constriction can then be removed by fusing them into the indefinite-type worldline.

Finally, we note that the two constraints above do not need to hold for abelian topological protocols, as these can be decoded even if measurement outcomes are random in the absence of errors.
More generally, if only abelian anyons can condense on the future side of a time-perpendicular or $>O(1)$-moving or constricting defect, then the conditions do not need to hold for that defect either.
This is for example the case for the time-perpendicular domain wall used in the D4 protocols in Ref.~\cite{davydova2025universalfaulttolerantquantum}, which can only condense abelian charges on the future sides.
It also generally allows future-facing time-perpendicular boundaries, as the future side is vacuum and has no non-abelian anyons.

\subsection{Fault tolerance}
Given the above considerations, the fault tolerance proof generalizes fairly straightforwardly to general spacetime 3-manifold logical operations.
The burst decomposition of the errors is still defined in the same way, just that the underlying lattice is no longer an $L\times L\times T$ cubic lattice but possibly some less regular spacetime cellulation.
It is still true that bounded damage means no logical fault, since the irreducibility axiom of the path integral is not only enforced for the bulk but also for the topological defects.
The probability bound for a large-burst event still works in the same way, as long as the cellulation has a bounded geometry property.
The main proof idea that anyon worldlines move towards their cluster base point at speed 1 if they are uninterrupted by error bursts, and at a reduced but bounded speed $v_k$ if they are interrupted by bursts up to level $k$, also generalizes straightforwardly.

\section{Conclusion}
\myparagraph{Summary}
We have given a concrete syndrome extraction circuit and a
classical decoder for fault tolerantly maintaining quantum memory in an arbitrary 2D non-chiral intrinsic
topological phase.  We proved the following threshold theorem: for any
\(\epsilon\)-bounded (local stochastic) circuit-level noise model there is a threshold
value \(\epsilon_*\) such that for \(\epsilon<\epsilon_*\) the
logical error rate decays quasi-stretched-exponentially with the
system size \(L\).  The circuit is based on a spacetime Turaev--Viro path
integral corresponding to the topological phase, for which the spatial version would be 
the Levin--Wen string-net model. At each step, the circuit measures the syndrome, runs a global classical decoding algorithm, and applies a correction. The decoder run at each timestep uses a clustering
algorithm on the recorded spacetime history of the noise-caused anyon worldlines up to this timestep, and determines the correction whose goal is to bring the syndrome points belonging to the same cluster to a common base point. The correction moves anyons by at most a lattice step towards the associated base point each time. 

Beyond memory, we have outlined how the path-integral construction and decoding scheme extend to logical operations involving computational anyons and other topological defects. We described the corresponding circuit and decoder modifications, identified causality constraints on the defect evolution, and sketched how the threshold argument can be adapted.  The topological path integral formalism nicely captures these generalizations.

\myparagraph{Future directions}
A natural question for further investigation would be to find new universal sets of logical protocols implemented by spacetime 3-manifolds with defects.
Section~\ref{sec:logical_operation} provides a general formalism for how to do this, and there are far richer classes of protocols than those studied in the literature.
For example, it would be interesting to see what gates we get from interfacing different non-solvable topological orders along domain walls, analogous to what was studied in 
Ref.~\cite{davydova2025universalfaulttolerantquantum} for the solvable $D_4$ phase.

A further direction is to construct shallow and hardware-optimized circuits for specific non-abelian anyon models such as double-Fibonacci.
We have already seen that the spacetime path-integral perspective gives us more freedom in designing such circuits.
An interesting question is whether one can use other types of microscopic defects in place of noise-caused anyon worldlines.
From the path-integral perspective, the only reason why noise-caused anyons show up in the circuit is due to turning a non-unitary path-integral operator into a set of Kraus operators corresponding to a measurement instrument.
Using anyon tensors corresponding to a tube-algebra representation allows us to show directly from the path-integral axioms that the set of Kraus operators forms a trace-preserving channel.
However, it is in general not difficult to extend a single non-unitary operator to a set of Kraus operators, and there are many possibilities to do so.
Non-trivial measurement outcomes would then no longer correspond to anyon 1-gons with a nice TQFT description, but be some different kind of microscopic ``defect'' (to be distinguished from a computational/topological defect).
All we need to be able to do is to clean/move these defects at an $O(1)$ speed.
Our decoding algorithm and threshold proof work independently of whether the noise-caused microscopic defects are anyon worldlines or any other kind of alteration to the post-selected circuit:
All of these are considered ``damage'' that can be repaired by padding it with a layer of damage-free circuit/path integral.
Using different microscopic defects has potential to simplify the microscopic circuit, but this would likely have to be a model-specific construction rather than one that works for all Turaev-Viro path integrals.

In fact, Ref.~\cite{davydova2025universalfaulttolerantquantum} gives circuits for the type-III twisted Dijkgraaf-Witten model which is an example of a Turaev-Viro path integral, but not all non-trivial measurement outcomes correspond to anyons as in our present construction.
Instead, the microscopic defects corresponding to non-trivial $Z$-type stabilizer measurement outcomes are ``bare fluxes'' in the model, which can be moved by bare Pauli-$X$ operators instead of more complicated tube-algebra-based string operators.
This makes the circuit much simpler than in our general construction, and it also allows for Floquet or measurement-based protocols where syndrome-measurement outcomes are non-deterministic even in the absence of noise.
In our general method, it is crucial that there are no non-trivial syndrome measurements in the absence of noise, and it would be interesting to see for which classes of models there are fault-tolerant implementations that relax this requirement.

Another natural direction would be to search for model-specific optimizations of the decoder that lead to a larger threshold or lower logical error rate.  For instance, it may be the case that the decoder in Ref.~\cite{Dauphinais_2017} could be directly combined with our circuit and path-integral formulation to provide an alternative circuit-level threshold proof for non-cyclic anyon models.

Finally, it would be interesting to see how our methods generalize to other types of topological phases, such as 3D topological phases, intrinsically fermionic phases, and fracton phases.
For 3+1D topological phases, there is a similar path integral based on fusion 2-categories.
In this case, it appears that the syndrome will have to consist both of point-like 0+1D anyon worldlines and string-like 1+1D excitations, in analogy with the 3+1D toric code.
In these cases, we believe the decoder and proof will be largely unchanged, as they are based on clustering and removing any kind of ``damage'' in the path integral, whether it is point-like, string-like, or defined on fermionic degrees of freedom.

\section*{Acknowledgements}
All the proof ideas and proof sketches were developed by the authors prior to July 2024.  The authors decided to write the paper in 2026, and ChatGPT 6 was used, but only for checking and further simplifying the arguments, and for generating some figures. We thank Tyler Ellison, Julio Magdalena de la Fuente, Anasuya Lyons, and Dominic Williamson for discussions. Y.R. worked on this project during his PhD studies at MIT and was supported by the U.S. Department of Energy, Office of Science, National Quantum Information Science Research Centers, Quantum Systems Accelerator, under Grant number DOE DE-SC0012704. S.B. and M.D. were supported by the Walter Burke Institute for Theoretical Physics at Caltech and by the Institute for Quantum Information and Matter, an NSF Physics Frontiers Center (NSF Grant PHY-2317110).
A.B. was supported by the U.S. Army Research Laboratory and the U.S. Army Research Office under contract/grant number W911NF2310255, and by the U.S. Department of Energy, Office of Science, National Quantum Information Science Research Centers, and the Co-design Center for Quantum Advantage (C2QA) under contract number DE-SC0012704.
\bibliography{main}%

\appendix

\section{Properties of cluster decomposition}
\label{app:only-component-closure-facts}

In this appendix, we prove a simple properties of the clustering algorithm, namely that it provides a unique cluster decomposition.

\begin{lemma}[Monotonicity of nearness]
\label{lem:only-nearness-monotone}
If \(A\sim_g B\), \(A\subseteq A'\), and \(B\subseteq B'\), then
\(A'\sim_g B'\).
\end{lemma}
\begin{proof}
First, we observe that \(H(B)\subseteq H(B')\).  Since \(g(r)-r\) is an increasing function of $r$,
\[
\BN_g(B)
\subseteq\BN_g(B').
\]
Then, any point in $A \cap \mathscr{B}_g(B)$ is also in $A \cap \mathscr{B}_g(B') \subseteq A' \cap \mathscr{B}_g(B') $ and any point in $B \cap \mathscr{B}_g(A)$ is also in $B \cap \mathscr{B}_g(A') \subseteq B' \cap \mathscr{B}_g(A')$.  This proves the lemma.
\end{proof}

\begin{lemma}
\label{prop:only-merge-crossing-cut}
Let \(B\) be an intermediate cluster identified during a run of
Algorithm~\ref{alg:clustering}, and let
\(\varnothing\ne A\subsetneq B\).  In the merge history of \(B\),
there exist nonempty clusters \(P\subseteq A\) and
\(Q\subseteq B\setminus A\) that were merged with \(P\sim_g Q\).
\end{lemma}
\begin{proof}
Represent the merge history of the cluster as a tree whose nodes are clusters and edges denote merging events.  The leaves of the merge history are single points, and can be partitioned into leaves in $A$ and in $B \setminus A$.  Move up the tree and identify the first merge point whose two children clusters are in $A$ and in $B \setminus A$ and whose merged cluster has support in both $A$ and $B \setminus A$.  By definition of the merge procedure, these two sets, which are $P$ and $Q$ in the lemma statement, must be mutually \(g\)-near.
\end{proof}

\begin{proposition}[Termination of the merge process]
\label{prop:only-clustering-termination}
Running Algorithm~\ref{alg:clustering} on input $S$
terminates after at most \(\max\{\abs S-1,0\}\) merges, the final cluster decomposition is independent of the merge order, and the final cluster decompositio obeys conditions (1) and (2) of Def.~\ref{def:only-g-cluster-closure}.
\end{proposition}
\begin{proof}
During each merge, the total number of clusters decreases by 1.  Since each point starts out as its own cluster (so there are a total of $|S|$ clusters) and there is at least one cluster in the end, there must be $\leq |S|-1$ merges.

Let $\mathcal Q$ be any cluster decomposition of $S$ whose clusters are not mutually $g$-near. We claim that every intermediate cluster during a merge process is contained in one of the clusters of $\mathcal Q$. This holds initially since all clusters are single points. Suppose it holds before merging $C_0\sim_g C_1$, and choose $Q_0,Q_1\in\mathcal Q$ with $C_i\subseteq Q_i$. If $Q_0\ne Q_1$, Lemma~\ref{lem:only-nearness-monotone} gives $Q_0\sim_g Q_1$, contradicting the assumption on $\mathcal Q$. Thus $Q_0=Q_1$, and the merged cluster continues to be contained in a single cluster in $\mathcal Q$.

Suppose the algorithm produces two different cluster decompositions $\mathcal Q$ and $\mathcal Q'$ under different merge orders, both of which satisfy condition (1) of Def.~\ref{def:only-g-cluster-closure}.  Since $\mathcal Q$ refines cluster decompositions at all steps regardless of the merge order, it must refine $\mathcal Q'$.  Similarly, $\mathcal Q'$ must refine $\mathcal Q'$.  This implies $\mathcal Q' = \mathcal Q$, a contradiction, proving uniqueness of the final decomposition.

Finally, let $C$ be a cluster in $\mathcal Q$ and $\varnothing\ne D\subsetneq C$. By Lemma~\ref{prop:only-merge-crossing-cut}, there exist clusters $U\subseteq D$ and $V\subseteq C\setminus D$ in the merge history of $C$ with $U\sim_g V$. Lemma~\ref{lem:only-nearness-monotone} then gives $D\sim_g C\setminus D$, proving condition (2) of Def.~\ref{def:only-g-cluster-closure}. Hence the cluster decomposition satisfies all properties of Def.~\ref{def:only-g-cluster-closure}.
\end{proof}

\section{Proof of the sparsity bound}\label{app:sparsity-proof}
\begin{proof}[Proof of Lemma~\ref{lem:only-large-burst-probability-bound}]
We will group the original bursts in $\mathcal B$ into an auxiliary hierarchy with size scales $a_n$ determined by the separation function $\gamma$, and introduce an auxiliary procedure that removes isolated bursts of errors at increasing radii $a_n$. We will show that any error remaining after $n$ levels of such removal will be bounded in probability as a function of $n$. Stopping this procedure at the largest $n$ such that $a_n \leq R$ will then show the Lemma statement. 

\textbf{Removing isolated bursts.}
Define a new sequence of burst radii:
\begin{equation}
\label{eq:ivd-auxiliary-scales}
a_0 = 0, \quad a_1=1,
\quad a_{n+1}=16 \gamma(a_n),
\end{equation}
which we use to define the auxiliary hierarchy for $n \geq 1$:
\begin{equation}
\label{eq:ivd-auxiliary-burst-levels}
\widetilde{\mathcal B}_n
=\{B\in\mathcal B:a_{n-1}<\rad_{\Box}(B)\le a_n\}.
\end{equation}
Until the end of this proof, a burst's level refers to this auxiliary hierarchy. The bursts within this hierarchy are isolated in the sense that no two distinct bursts are mutually $\gamma$-near. Both unions
\begin{equation}
\label{eq:ivd-same-bursts}
\mathcal B=\bigcup_{n\ge1}\widetilde{\mathcal B}_n
=\bigcup_{k\ge0}\mathcal B_k.
\end{equation}
contain the same collection of bursts; the only thing that differs is the assignment to the level. Therefore, if the auxiliary hierarchy contains no burst of radius greater than $R$, then neither does the hierarchy $\bigcup_{k\ge0}\mathcal B_k$ with scales $r_k=Q^k$.

The rest of the proof will bound the probability that $\bigcup_{j>n}\widetilde{\mathcal B}_j$ is nonempty. When $n$ is the largest integer such that $a_n \leq R$, we have $\LB_R \subseteq \LB_{a_n}$. 

\textbf{Minimal error size required for a large burst.}
Consider a burst $B$ in $\widetilde{\mathcal B}_i$ for $i > n$, so that $\rad_{\Box}(B)>a_n$. 
Consider any point $x \in B$; we will show that there exists $y\in B$ such that
\begin{equation}
\label{eq:ivd-partner-annulus}
\frac{a_n}{2} < d_{\Box}(x,y) \leq \gamma(a_n).
\end{equation}
To show this, consider $C=B\cap\BN_{a_n/2}(x)$; clearly, we have $\rad_{\Box}(C)\le a_n$. Since $x\in C$ and $\rad_{\Box}(B)>a_n$, $C$ must be nonempty and $B \setminus C$ must be nonempty as well.  
Suppose that no $y\in B$ satisfies \eqref{eq:ivd-partner-annulus}. Then the distance between $x$ and $B\setminus C$ is larger than $\gamma(a_n)$. On the other hand, for every $z\in\BN_\gamma(C)$, simple geometry gives
\begin{equation}
\begin{aligned}
d_{\Box}(x,z)
\le\gamma\bigl(\rad_{\Box}(C)\bigr)+\rad_{\Box}(C)-1,
\end{aligned}
\end{equation}
and applying monotonicity of $\gamma(r) - r$ gives 
\begin{align*}
    \gamma(\rad_{\Box}(C)) + \rad_{\Box}(C) 
    \leq \gamma(a_n) - a_n + 2 \rad_{\Box}(C),
\end{align*}
so
\begin{align}
d_{\Box}(x,z)
&\le\gamma(a_n)-a_n+2\rad_{\Box}(C)-1 \nonumber
\\
&\le\gamma(a_n).
\end{align}
Thus $(B\setminus C)\cap\BN_\gamma(C)=\varnothing$, so $C$ and $B\setminus C$ are not mutually $\gamma$-near. This contradicts the definition of a burst: every nonempty proper subset of a burst must be mutually $\gamma$-near its complement.

For $j\le n$, $\rad_{\Box}(B)>a_n\ge a_j$, and the argument above applies to every point of $B$ at each of these scales.
We will apply this argument in decreasing order for $j=n,n-1,\ldots,1$. $x$ will be the root of the tree; applying the argument at scale $n$ gives two points, $x$ and $y$, satisfying eq.~\eqref{eq:ivd-partner-annulus}; applying it again at scale $n-1$ will give $(x,x')$ and $(y,y')$ where each pair satisfies an analogous condition at one scale below, and so on. This way, we construct a binary tree with $2^n$ leaves at level 0. All the points corresponding to the leaves of the tree are in $B$.

Let us show that the leaf points are all distinct. A leaf point at level 0 originating from a node at level $j-1$ can be at most distance
\begin{equation}
\label{eq:ivd-scale-separation}
\sum_{i=1}^{j-1}\gamma(a_i)
=\frac{1}{16}\sum_{i=1}^{j-1}a_{i+1}
\le\frac{1}{15}a_j
\end{equation}
from its parent. Now suppose that two distinct leaf nodes have the same coordinate. Call them $v$ and $v'$, and assume their paths first separate at a node labeled $z$ at some level $j \geq 1$.
Splitting this node gives two children at level $j-1$, labeled $z$ and $z'$. By eq.~\eqref{eq:ivd-partner-annulus}, $d_{\Box}(z,z') > a_j/2$. From this, applying eq.~\eqref{eq:ivd-scale-separation} and the triangle inequality, $v$ and $v'$ must satisfy $d_{\Box}(v,v') > a_j/2- 2/15 \, a_j > a_j/4$, which contradicts the assumption that they are the same point.

Thus, for $\widetilde{\mathcal B}_{i}$ with $i > n$ to be nonempty, the error $E$ must contain at least $2^n$ points. We will call this subset of points the level-$n$ witness of the error $E$. 

\textbf{Counting the possible choices.}
The event $\LB_{a_n}$ must contain a level-$n$ witness. To bound the probability of the event, note that under an $\epsilon$-bounded error model, a fixed witness has probability at most $\epsilon^{2^n}$. The probability of $\LB_{a_n}$ can be loosely upper-bounded by the total number of possible witnesses of size $2^n$ times $\epsilon^{2^n}$.

We will count the number of possible witnesses by counting the number of possible associated binary trees with allowed separations between coordinates of their nodes. Fix the root location $x$; there is a total of $|\intG|$ choices for this. For a node at scale $j$ of the binary tree, the first child repeats its parent's location, and the partner is within distance $\gamma(a_j)$. The spacetime volume from which one can choose the second point is at most 
\begin{equation}
\label{eq:ivd-branch-count}
M_j=8(\gamma(a_j)+1)^3 = 8\left (\frac{1}{16} a_{j+1} +1 \right )^3,
\end{equation}
where we used the definition of $a_{j+1}$.

There are $2^{n-j}$ nodes at stage $j$, so the total of all possible trees is at most 
\begin{equation}
    \prod_{j=1}^n M_j^{2^{n-j}}
\end{equation}
A union bound over all trees and all root locations therefore gives
\begin{equation}
\label{eq:ivd-tree-count}
\begin{aligned}
\Pr[\LB_{a_n}]
&\le |\intG| \,\epsilon^{2^n}\prod_{j=1}^n M_j^{2^{n-j}}\\
&= |\intG|  \epsilon^{2^n} \left (\exp\left (\sum_{j=1}^n \frac{\log M_j}{2^j} \right) \right )^{2^n}.
\end{aligned}
\end{equation}

\textbf{Final bound.}
Recall that
\begin{equation}
    \gamma(r) = C_\gamma r \log^\zeta(e+r),
\end{equation}
where $\zeta = 3/2$. 

Since $a_j\to\infty$, we have for $j\ge2$,
\begin{align}
\frac{\log a_{j+1}}{\log a_j}
&= \frac{\log (16 \gamma(a_j))}{\log a_j}
\\
&=1+\frac{\log(16C_\gamma)+\zeta\log\log(e+a_j)}{\log a_j}
\rightarrow 1. \nonumber
\end{align}
 Therefore, using the definition of $M_j$:
\begin{align}
\frac{(\log M_{j})/2^{j}}{(\log M_{j-1})/2^{j-1}} = \frac{1}{2}\frac{ \log a_{j+1} + O(1)}{ \log a_{j} + O(1)} 
\rightarrow \frac 1 2.
\end{align}
Therefore, by the ratio test, the series $\sum_{j\ge1}(\log M_j)/2^{j}$ converges. Its sum depends only on $\gamma$, so we may define an $O(1)$ constant
\begin{equation}
\label{eq:ivd-threshold}
\epsilon_*:= \exp\left(-\sum_{j=1}^\infty \frac{\log M_j}{2^j}\right)>0.
\end{equation}
Since every partial sum is bounded by the full sum, plugging this into \eqref{eq:ivd-tree-count} gives
\begin{equation}
\label{eq:ivd-burst-depth-bound}
\Pr[\LB_{a_n}]
\le|\intG|\left(\frac{\epsilon}{\epsilon_*}\right)^{2^n}.
\end{equation}

\textbf{Going from $n$ to $R$.} For sufficiently large $R$, choose the largest integer $n\ge1$ such that $a_n\le R$, so that
\begin{equation}
\label{eq:ivd-level-at-radius}
a_n\le R<a_{n+1}.
\end{equation}
Since $a_1=1$ and $a_j\le R$ for every $1\le j\le n$, multiplying the scale ratios gives
\begin{equation}
\begin{aligned}
R<a_{n+1}
&=\prod_{j=1}^n\frac{a_{j+1}}{a_j}
=\prod_{j=1}^n16C_\gamma[\log(e+a_j)]^\zeta\\
&\le\left(16C_\gamma[\log(e+R)]^\zeta\right)^n.
\end{aligned}
\end{equation}
Taking logarithms gives
\begin{equation}
\log R<n\left(\log(16C_\gamma)+\zeta\log\log(e+R)\right).
\end{equation}
Then,
\begin{equation}
\frac{\log R}{\log \log R}<n \frac{\left(\log(16C_\gamma)+\zeta\log\log(e+R)\right)}{\log \log R}.
\end{equation}
When $R > R_* = \exp(32 C_\gamma)$, using that $\log(16 C_\gamma) > \max(\zeta, e)$, we have
\begin{equation}
\frac{\left(\log(16C_\gamma)+\zeta\log\log(e+R)\right)}{\log \log R} \leq 2 \zeta,
\end{equation}
and thus
\begin{equation}
n\ge\frac{1}{2\zeta}\frac{\log R}{\log\log R}.
\end{equation}
Therefore,
\begin{equation}
\label{eq:fin}
2^n\ge\exp\!\left(c\frac{\log R}{\log\log R}\right),
\qquad c=\frac{\log2}{2\zeta}>0.
\end{equation}
Combining \eqref{eq:fin} with \eqref{eq:ivd-burst-depth-bound} and $\LB_R\subseteq\LB_{a_n}$ proves the claimed bound for $\epsilon/\epsilon_*<1$.
\end{proof}

\section{Choice of constants}\label{app:constants}
We now choose the constants used in the proof and check the required inequalities.
\begin{lemma}[Choice of constants]
\label{lem:only-auxiliary-constraints}
Set
\[
\begin{aligned}
C_\beta&=3, & \zeta&=\tfrac32, & v_\infty&=\tfrac12,\\
Q&=10^{6}, & C_\gamma&=10^{19},\\
\lambda^{\R}&=128, & \lambda^{\S}&=1024,\\
\lambda^{\C}&=7173, & \lambda^{\G}&=28692,\\
\lambda^{\D}&=143460, & \lambda^{\T}&=5738400.
\end{aligned}
\]
These constants satisfy the following constraints:
\begin{itemize}
\item[(a)] \emph{Lower bound on $C_\gamma$:}
\(C_\gamma\ge12\lambda^{\D}Q^2/v_\infty+512\lambda^{\D}Q+2\).
\item[(b)] \emph{Separation of boxes at the same level:}
\((\lambda^{\C}+\lambda^{\S}+2)Qr<\gamma(r)-r\) for every \(r\ge\tfrac12\).
\item[(c)] \emph{Comparison of consecutive levels:}
\(64\lambda^{\D}r_{k-1}+3\le\lambda^{\D}r_k\) for \(k\ge1\).
\item[(d)] \emph{Bounds for guard boxes and fusion times:}
\(Q\ge65\), \(Q>960(1+2C_\beta)\),
\(\lambda^{\G}(1-8/(Q-1))\ge3\lambda^{\C}+2\),
\(\lambda^{\T}+20\lambda^{\G}/v_\infty\le240\lambda^{\G}\), and
\(20\lambda^{\G}/v_\infty\) is an integer.
\item[(e)] \emph{Positive velocities:}
\(\sum_{j\ge0}\epsilon_j\le1-v_\infty\), where
\(\epsilon_j=2\lambda^{\D}r_j/\gamma_j=10\lambda^{\G}r_j/\gamma_j\).
Thus the velocities in Eq.~\eqref{eq:only-motion-constants} satisfy \(v_k\ge v_\infty>0\).
\item[(f)] \emph{Separation between bursts:}
\[
256\lambda^{\D}r_j+6\lambda^{\D}r_{j+1}/v_\infty+1\le\gamma_j/2
\quad(j\ge0).
\]
\item[(g)] \emph{Bounds on $\lambda^{\D}$:}
\((1+1/v_\infty)(\lambda^{\C}+3)\le\lambda^{\D}\) and \(\lambda^{\D}\ge4\).
\item[(h)] \emph{Syndrome box:}
\(\lambda^{\S}\ge\lambda^{\R}+2+4/v_\infty+240\lambda^{\G}/Q\).
\item[(i)] \emph{Ratio between levels:} \(Q\ge C_\beta+1\).
\item[(j)] \emph{Reconnecting time:} \(\lambda^{\R}\ge C_\beta+4\).
\item[(k)] \emph{Cluster box:}
\(\lambda^{\C}=(1+2C_\beta)\lambda^{\S}+C_\beta+2\).
\item[(l)] \emph{Guard boxes and fusion times:}
\(\lambda^{\G}=4\lambda^{\C}\), \(\lambda^{\D}=5\lambda^{\G}\), and
\(\lambda^{\T}=100\lambda^{\G}/v_\infty\).
\end{itemize}
The bounds on connected components, syndrome lines, and fusion times in
Claims~\ref{claim:only-separated-boxes}--\ref{claim:only-stable-base} and
Lemmas~\ref{lem:only-motion-clearing}--\ref{lem:only-confinement-separation}
also hold with these constants. In particular, for \(k\ge1\),
\begin{equation}
\label{eq:only-small-scale-margins}
E_{k-1}<r_k,\qquad
24\lambda^{\G}r_{k-1}+2<r_k.
\end{equation}
\end{lemma}
\begin{proof}
Rows (b)--(d) and (g)--(l) follow by substitution. The right-hand sides in (a) and (h) are respectively less than $4\times10^{18}$ and $145$. To check the inequalities that must hold at every level, we use
\[
\gamma_0\ge(C_\gamma-1)/2,\qquad
\gamma_j\ge(C_\gamma-1)r_{j-1}\quad(j\ge1).
\]
These verify row (f), treating $j=0$ separately. More precisely,
\[
\gamma(r)-r\ge(C_\gamma-1)r[\log(e+r)]^{3/2}.
\]
Using $\log Q\ge1$ and $\sum_{n\ge1}n^{-3/2}\le3$, we obtain
\[
\sum_{j\ge0}\epsilon_j
\le\frac{4\lambda^{\D}+8\lambda^{\D}Q}{C_\gamma-1}
<10^{-6}<1-v_\infty.
\]
This proves row (e).

For the component estimates, put $a=2\lambda^{\G}+8$ and $\lambda^{\F}=8\lambda^{\G}$. The same separation bounds give
\begin{align*}
2(\lambda^{\G}+2)r_j+ar_{j-1}+2&<\gamma_j,\\
2\lambda^{\G}r_j+2ar_{j-1}+4&\le ar_j,\\
2(\lambda^{\F}+1)r_j+3\lambda^{\F}r_{j-1}+2&<\gamma_j,\\
2\lambda^{\F}r_j+6\lambda^{\F}r_{j-1}+4&\le3\lambda^{\F}r_j.
\end{align*}
The lower-level terms can be omitted at $j=0$. Also,
\[
3(\lambda^{\G}r_j+3\lambda^{\F}r_{j-1}+2)+1<\lambda^{\F}r_j,
\]
which verifies the neighborhood bound in Claim~\ref{claim:only-fixed-regions}.

For the shortcuts, $\delta_j=\gamma_j-2ar_j$ satisfies
\[
\frac{ar_j}{\delta_j}\le\frac{5\lambda^{\G}r_j}{2\gamma_j}.
\]
Hence $\theta_k\le(1-v_k)/4<1/8$. With $X_0=\sum_{j=0}^k ar_j$,
\[
\frac{2X_0}{1-2\theta_k}
\le\frac{8a}{3(1-Q^{-1})}r_k
<10\lambda^{\G}r_k=E_k.
\]
For the time needed to fuse the anyons in $\S^4_k$, the separation bounds give
\[
\max\{62\lambda^{\G}+4,\lambda^{\T}+2\lambda^{\G}+4\}r_k<\gamma_k.
\]
The time estimate in $\S^4_k$ follows from
\[
\frac{50\lambda^{\G}}{v_\infty}+2\lambda^{\G}
+\frac{2\lambda^{\T}}{Q}+\frac{4}{r_k}<\lambda^{\T}
\quad(k\ge1),
\]
and is immediate at $k=0$ with $\tau_{-1}=1$.

Finally, $E_{k-1}/r_k<3/10$ for $k\ge1$, and the second ratio in Eq.~\eqref{eq:only-small-scale-margins} is less than $7/10$. These margins give the radii $33r_k$ and $102r_k$ in the base-point argument. They also give
\[
127r_k-ar_{k-1}-3\ge103r_k+1,
\]
and
\[
\lambda^{\R}r_k+
\left\lceil\frac{64r_k+2E_{k-1}}{v_\infty}\right\rceil
+\tau_{k-1}<263r_k.
\]
For $k=0$, the latter left-hand side is $257$, since $E_{-1}=0$ and $\tau_{-1}=1$. The separation bounds also imply
\[
\gamma_k>(1+1/v_\infty)(\lambda^{\C}+3)r_k+E_{k-1},
\]
as used to show that a line from a level-$k$ burst can only enter another level-$k$ cluster box if that box starts later. This verifies the remaining inequalities. The functions $\beta(r)=3r$ and $\gamma(r)$ satisfy Definition~\ref{def:only-g-nearness}, since their excess over $r$ is nonnegative and nondecreasing.
\end{proof}

\end{document}

%% file: andi_thesis_preamble.tex
\usepackage{tikz}
\usetikzlibrary{decorations,decorations.pathmorphing,3d} 

\usetikzlibrary{spath3}
\ExplSyntaxOn 
\tikzset{
  only~ if~ drawable/.code={
    \tl_set:Nn \l__tikzspath_tikzpath_finish_tl
    {
      \spath_get_current_path:N \l__tikzspath_tmpa_tl
      \spath_bake_round:NV \l__tikzspath_tmpa_tl \l__tikzspath_tmpa_tl
      \spath_bake_shorten:NV \l__tikzspath_tmpa_tl \l__tikzspath_tmpa_tl
      \spath_reallength:NV \l__tikzspath_tmpa_int \l__tikzspath_tmpa_tl
      \int_compare:nNnT {\l__tikzspath_tmpa_int} > {0}
      {
        \tikzset{#1}
      }
    }
  }
}
\ExplSyntaxOff

\tikzset{
cod1ind/.style={line width=1.5}, 
cod2ind/.style={line width=3}, 
preticked/.style={draw,postaction={sharp corners,draw,semithick,only if drawable = {decorate},decoration={ticks,segment length=0.08cm,amplitude={0.04cm},pre=curveto,pre length=0.07cm,post=curveto,post length=0.03cm}}},
ticked/.style={preticked, every edge/.style=preticked},
sparsedashed/.style={dashed, dash pattern = on 3pt off 1pt},
premyzigzag/.style={postaction={draw,only if drawable = {decorate},decoration={snake,amplitude=1pt,segment length=3pt}}},
myzigzag/.style={premyzigzag, every edge/.style=premyzigzag},
}

\input{network_drawing.tex}

\usetikzlibrary{patterns.meta}

\def\gluecol{red}

\definecolor{col2}{rgb}{0,0.6,1}
\definecolor{col3}{rgb}{1,0,0}
\setmarks{
ar/.style={arr,asty=fill},
arbar/.style={arrbar,asty=fill},
vor/.style={hbar,s,poff=0.07}, 
}
\tikzset{
ind/.style={mark={lab=$#1$,a}}, 
startind/.style={mark={lab=$#1$,b}}, 
sind/.style={mark={lab=${\scriptstyle #1}$,a}}, 
srind/.style={mark={lab=$\textcolor{red}{{\scriptstyle #1}}$,a}}, 
enddots/.style={mark={three dots, a}},
redlab/.style={mark={slab=$\textcolor{red}{\scriptscriptstyle #1}$}}, 
redlabr/.style={mark={slab=$\textcolor{red}{\scriptscriptstyle #1}$,r}},
sglab/.style={mark={slab=$\scriptstyle #1$}}, 
sglab/.default=0.5,
sglabr/.style={mark={slab=$\scriptstyle #1$,r}},
sglabr/.default=0.5,
glab/.style={mark={slab=$#1$}}, 
glab/.default=0.5,
glabr/.style={mark={slab=$#1$,r}},
glabr/.default=0.5,
glue/.style={\gluecol, densely dotted, line width=0.15cm, opacity=0.3}, 
fermionpar/.style={mark={diamond,f,p=#1}},
fermionpar/.default=0.5,
}

\setelements{
vertex/.style={scale=0.3,circ,all},
cod1vertex/.style={scale=0.5,circ,all},
cod2vertex/.style={circ,scale=0.7,all},
cod3vertex/.style={circ,all},
drlab/.style 2 args={lab={t=$\textcolor{red}{#1}$,p=#2:0.25}},
pdrlab/.style 2 args={lab={t=$\textcolor{red}{#1}$,p=#2}},
dlab/.style 2 args={lab={t={$#1$},p={#2}}},
ldlab/.style 2 args={lab={t={$#1$},p={#2}},postdecstyle=lightback},
vback/.style={decastyle={opacity=0.6}}, 
vmid/.style={decastyle={opacity=0.4}}, 
corner/.style={postdecstyle={line width=1.2},arc=#1:0.2},
omega/.style={postdecstyle=red,postdec={scale=0.5}{shadecirc}},
dorientin/.style={small,circ,dcross}, 
dorientout/.style={small,circ,dot},
regionvertex/.style={circ,tiny}, 
glueedge/.style={void,postdecstyle=\gluecol,postdec={scale=0.5}{shadecirc}}, 
cyclevertex/.style={postdecstyle=#1,postdec={scale=0.5}{shadecirc}},
cyclevertex/.default=cyclecol,
vca/.style={astyle=red}, 
vcb/.style={style=darkgreen},
vcc/.style={style=blue},
vcd/.style={style=darkyellow},
}

\tikzset{
backedge/.style={dashed},
midedge/.style={dotted},
back/.style={opacity=0.4}, 
mid/.style={opacity=0.2}, 
extend map/.style={gray},
or/.style={mark={p=#1,arr}}, 
or/.default=0.5,
ior/.style={mark={p=#1,arr,-}}, 
ior/.default=0.5,
weight/.style={mark={p=#1,bar}}, 
weight/.default=0.5,
hweight/.style={mark={p=#1,bar,asty=gray}}, 
hweight/.default=0.5,
fav/.style={mark={p=#1,hcirc}},
fav/.default=0.5,
rfav/.style={mark={p=#1,hcirc,r}},
rfav/.default=0.5,
cycleedge/.style={postaction={draw,#1,dash pattern=on 1pt off 2pt,line width=3}},
cycleedge/.default=cyclecol,
cycleface/.style={postaction={pattern=north west lines, pattern color=#1}},
cycleface/.default=cyclecol,
actualedge/.style={line width=2}, 
edge behind/.style={orange}, 
worldline/.style={red,line width=5,opacity=0.3,line join=round},
enddots/.style={mark={three dots,a}},
anyonfill/.style={pattern={Dots[radius=0.025cm]}},
lightback/.style = {}, 
front/.style={preaction={draw,white,line width=3}}, 
anyon1gon/.style={fill=red,fill opacity=0.2,text opacity=1}, 
}

\tikzset{
mke0/.style args={#1}{},
mke0/.default = {},
mke1/.style={mark={dot,asty={fill=white},#1}},
mke1/.default = {},
mke2/.style={mark={dot,f}},
}
\newcommand{\markedtetrahedron}[6]{
\atoms{void}{0/p={0,0}, 1/p={-150:0.6}, 2/p={-30:0.6}, 3/p={90:0.6}};
\draw (0) edge[#1] (1) (0) edge[#2] (2) (0) edge[#3] (3) (1) edge[#4] (2) (2) edge[#5] (3) (1) edge[#6] (3);
}

\newcommand{\markedcylinder}[4]{
\begin{tikzpicture}
\atoms{void}{0/, {1/p={0,0.8}}}
\draw (0)edge[#1={p=1},out=-45,in=-90]++(0:0.6) (0)edge[backedge,out=45,in=90]++(0:0.6);
\draw (1)edge[#2={p=1},out=-45,in=-90]++(0:0.6) (1)edge[out=45,in=90]++(0:0.6);
\draw (0)edge[#3](1) (0)edge[out=20,in=-90,#4={p=1}](0.6,0.4) (0.6,0.4)edge[backedge,out=90,in=-20](1);
\end{tikzpicture}
}

%% file: network_drawing.tex
\usepackage{tikz}
\usetikzlibrary{decorations,decorations.pathmorphing,decorations.markings,calc,snakes,positioning,fixedpointarithmetic,shapes,shapes.geometric,patterns}

\usepackage{etoolbox}
\AtBeginEnvironment{tikzpicture}{\catcode`$=3 } 

\tikzset{alignmid/.style={baseline={([yshift=-.5ex]current bounding box.center)}}} 
\tikzset{every picture/.append style=alignmid}

\tikzset{
bottomzigzag/.style={postaction={draw,decorate, decoration={zigzag,amplitude=1pt,segment length=3pt,raise=1pt}}},
zigzag/.style={draw,decorate, decoration={zigzag,amplitude=1pt,segment length=3pt}},
rc/.style=rounded corners,
}

\tikzset{
    -|/.style={to path={-| (\tikztotarget)}},
    |-/.style={to path={|- (\tikztotarget)}},
}

\usepackage{ifthen}
\usepackage{xparse,pgffor}

\tikzset{
mark/.code={
\tikzset{postaction={/network/mark/.cd,#1,/tikz/.cd,decorate,decoration={name=markings,mark=at position \netmarkpos with{
\begin{scope}[netmarktrafo]
\netmarkcode
\end{scope}
}}}}
\def\netmarkpos{0.5}
},
}

\def\netmarkpos{0.5}
\def\netmarkposoff{0cm}
\def\netmarkcode{}

\tikzset{
netmarktrafo/.style={},
netmarkstyle/.style={solid,semithick,sharp corners},
}

\pgfkeys{
/network/mark/.cd,
sty/.code={
\tikzset{netmarkstyle/.style={#1}}
},
asty/.code={
\tikzset{netmarkstyle/.append style={#1}}
},
f/.style={asty=fill},
p/.code={ 
\def\netmarkpos{#1}
},
poff/.code={ 
\def\netmarkposoff{#1cm}
},
e/.code={ 
\def\netmarkpos{\pgfdecoratedpathlength-0.005cm-\netmarkposoff}

\tikzset{netmarktrafo/.append style={shift={(-\netmarkwidth,0)}}}
},
s/.code={ 
\def\netmarkpos{0.005cm+\netmarkposoff}

\tikzset{netmarktrafo/.append style={shift={(\netmarkwidth,0)},xscale=-1,yscale=-1}}
},
a/.code={ 
\def\netmarkpos{\pgfdecoratedpathlength-0.005cm}

\tikzset{netmarktrafo/.append style={xscale=-1,shift={(-\netmarkwidth,0)}}}
},
b/.code={ 
\def\netmarkpos{0.005cm}

\tikzset{netmarktrafo/.append style={xscale=-1,shift={(\netmarkwidth,0),yscale=-1}}}
},
-/.code={ 
\tikzset{netmarktrafo/.append style={xscale=-1}}
},
r/.code={ 
\tikzset{netmarktrafo/.append style={yscale=-1}}
},
sideoff/.code={ 
\tikzset{netmarktrafo/.append style={shift={(0,#1)}}}
},
slab/.code={ 
\def\netmarkwidth{0}
\def\netmarkcode{
\node[inner sep=0.04cm,netmarkstyle,draw=none] (mylabelwidthtest) at (0,0){\phantom{#1}};
\path let \p1=(mylabelwidthtest.north east), \p2=(mylabelwidthtest.south east), \n1 = {max(abs(\y1),abs(\y2))} in node[inner sep=0.04cm,netmarkstyle] at (0,\n1) {#1};
}
},
lab/.code={ 
\def\netmarkwidth{0}
\def\netmarkcode{
\node[inner sep=0.04cm,anchor=\netmarkanchor] (mylabelwidthtest) at (0,0) {\phantom{#1}};
\draw[white] (mylabelwidthtest.\pgfdecoratedangle)--(mylabelwidthtest.\pgfdecoratedangle+180);
\node[inner sep=0.04cm,anchor=\netmarkanchor,netmarkstyle] at (0,0) {#1};
}
},
rotlab/.code={ 
\def\netmarkwidth{0}
\def\netmarkcode{
\node[inner sep=0.04cm,fill=white,transform shape,rotate=90,anchor=\netmarkrotanchor,netmarkstyle] (mydecorationnodename) at (0,0) {#1};
}
},
mrotlab/.style={ 
rotlab={$\scriptstyle{#1}$}
},
arr/.code={ 
\def\netmarkwidth{0.04}
\def\netmarkcode{\draw[netmarkstyle] (-0.04,0.08)--(0.04,0)--(-0.04,-0.08);}
},
arrbar/.code={ 
\def\netmarkwidth{0.08}
\def\netmarkcode{\draw[netmarkstyle] (-0.08,0.08)--(0,0)--(-0.08,-0.08) (0.04,0.08)--(0.04,-0.08);}
},
rarr/.code={ 
\def\netmarkwidth{0.04}
\def\netmarkcode{\draw[netmarkstyle] (-0.04,-0.08)arc(90-180:90:0.08);}
},
dot/.code={ 
\def\netmarkwidth{0.08}
\def\netmarkcode{\draw[netmarkstyle] (0,0)circle(0.08);}
},
flag/.code={ 
\def\netmarkwidth{0.06}
\def\netmarkcode{\draw[netmarkstyle] (-0.06,0)--(0,0.09)--(0.06,0)--cycle;}
},
darr/.code={ 
\def\netmarkwidth{0.08}
\def\netmarkcode{\draw[netmarkstyle] (-0.04,0)--(0.04,0)--(-0.04,0.08)--cycle;}
},
circ/.code={
\def\netmarkwidth{0.1}
\def\netmarkcode{\draw[netmarkstyle] (0,0) circle (0.1);}
},
hcirc/.code={ 
\def\netmarkwidth{0.1}
\def\netmarkcode{\draw[netmarkstyle] (-0.1,0) arc (180:0:0.1);}
},
diamond/.code={
\def\netmarkwidth{0.1}
\def\netmarkcode{\draw[netmarkstyle] (-0.1,0)--(0,-0.1)--(0.1,0)--(0,0.1)--cycle;}
},
bar/.code={ 
\def\netmarkwidth{0.05}
\def\netmarkcode{
\draw[netmarkstyle] (0,-0.08cm-0.5*\pgflinewidth)--(0,0.08cm+0.5*\pgflinewidth);
}
},
dbar/.code={ 
\def\netmarkwidth{0.13}
\def\netmarkcode{
\draw[netmarkstyle] (-0.04cm,-0.08cm-0.5*\pgflinewidth)--(-0.04cm,0.08cm+0.5*\pgflinewidth) (0.04cm,-0.08cm-0.5*\pgflinewidth)--(0.04cm,0.08cm+0.5*\pgflinewidth);
}
},
hbar/.code={ 
\def\netmarkwidth{0.05}
\def\netmarkcode{
\draw[netmarkstyle] (0, 0.5*\pgflinewidth)--++(0,0.12);
}
},
mill/.code={
\def\netmarkwidth{0.16}
\def\netmarkcode{
\draw[netmarkstyle] (0,-0.5*\pgflinewidth)--++(-0.08,-0.08)--++(0,0.08);
\draw[netmarkstyle] (0,0.5*\pgflinewidth)--++(0.08,0.08)--++(0,-0.08);
}
},
three dots/.code={
\def\netmarkwidth{0.2}
\def\netmarkcode{
\fill (-0.12,0) circle (0.5*0.05) (0,0) circle (0.5*0.05) (0.12,0) circle (0.5*0.05);
}
},
}

\tikzset{wid/.style={minimum width=#1cm}}
\tikzset{hei/.style={minimum height=#1cm}}

\tikzset{sx/.style={xshift=#1cm}}
\tikzset{sy/.style={yshift=#1cm}}

\tikzset{box/.style={draw,rectangle}}
\tikzset{fbox/.style={draw,rectangle, line width=1.1}}
\tikzset{roundbox/.style={draw,rectangle,rounded corners}}
\tikzset{froundbox/.style={draw,rectangle, rounded corners, line width=1.1}}
\tikzset{rounddiamond/.style={draw,diamond,rounded corners}}
\tikzset{dot/.style={draw, shape=circle, fill=black, scale=0.5}}

\tikzset{
netbox/.code={
\node[draw,netbdstyle] (\atomname) at (0,0) {#1};
\coordinate (\atomname-r) at (\atomname.east);
\coordinate (\atomname-l) at (\atomname.west);
\coordinate (\atomname-t) at (\atomname.north);
\coordinate (\atomname-b) at (\atomname.south);
\coordinate (\atomname-tr) at (\atomname.north east);
\coordinate (\atomname-br) at (\atomname.south east);
\coordinate (\atomname-tl) at (\atomname.north west);
\coordinate (\atomname-bl) at (\atomname.south west);
},
}

\tikzset{bdlw/.code={\tikzset{mybdstyle/.style={draw, line width=#1}}}}
\tikzset{bdcol/.code={\tikzset{mybdstyle/.append style={#1}}}}

\newcommand\setelements[1]{
\pgfkeys{/network/atom/.cd,#1}
}

\newcommand\setmarks[1]{
\pgfkeys{/network/mark/.cd,#1}
}

\newcommand\atoms[2]{
\foreach \name/\keys in {#2}{
\expandafter\atom\expandafter{\keys,#1}{\name}
}
}

\newcommand\atom[2]{
\def\atomname{#2}
\tikzset{
nettrafo/.style={},
netatompos/.style={},
netdeco/.style={},
netpostdeco/.style={},
}

\pgfkeys{/network/atom/.cd,#1}

\begin{scope}[netatompos] 
\begin{scope}[nettrafo] 
\netshapecoords 
\fill[netbackstyle] \netshapepath;
\clip \netshapepath;
\tikzset{netdeco}
\draw[netbdstyle] \netshapepath;
\end{scope}
\tikzset{netpostdeco} 
\end{scope}

}

\pgfkeys{
/network/atom/.cd,
square/.code={

\def\netshapepath{(-\tempsize,-\tempsize)rectangle (\tempsize,\tempsize)}
\def\netshapecoords{
\node[rectangle,wid=2*\tempsize,hei=2*\tempsize,inner sep=0,transform shape](\atomname)at(0,0){};
\coordinate(\atomname-c) at (0,0);
\coordinate(\atomname-r) at (\tempsize,0);
\coordinate(\atomname-l) at (-\tempsize,0);
\coordinate(\atomname-t) at (0,\tempsize);
\coordinate(\atomname-b) at (0,-\tempsize);
\coordinate(\atomname-br) at (\tempsize,-\tempsize);
\coordinate(\atomname-tr) at (\tempsize,\tempsize);
\coordinate(\atomname-bl) at (-\tempsize,-\tempsize);
\coordinate(\atomname-tl) at (-\tempsize,\tempsize);
}},
circ/.code={

\def\netshapepath{(0,0)circle(\tempsize)}
\def\netshapecoords{
\node[circle,wid=2*\tempsize,hei=2*\tempsize,inner sep=0,transform shape](\atomname)at(0,0){};
\coordinate(\atomname-c) at (0,0);
\coordinate(\atomname-r) at (\tempsize,0);
\coordinate(\atomname-l) at (-\tempsize,0);
\coordinate(\atomname-t) at (0,\tempsize);
\coordinate(\atomname-b) at (0,-\tempsize);
}},
triang/.code={

\def\netshapepath{(-30:\tempsize)--(90:\tempsize)--(-150:\tempsize)--cycle}
\def\netshapecoords{
\node[regular polygon,regular polygon sides=3,wid=2*\tempsize,inner sep=0,transform shape](\atomname)at(0,0){};
\coordinate(\atomname-c) at (0,0);
\coordinate(\atomname-cr) at (-30:\tempsize);
\coordinate(\atomname-cl) at (-150:\tempsize);
\coordinate(\atomname-ct) at (90:\tempsize);
\coordinate(\atomname-mb) at (-90:0.5*\tempsize);
\coordinate(\atomname-mr) at (30:0.5*\tempsize);
\coordinate(\atomname-ml) at (150:0.5*\tempsize);
}},
diamond/.code={

\def\netshapepath{(0,-\tempsize)--(\tempsize,0)--(0,\tempsize)--(-\tempsize,0)--cycle}
\def\netshapecoords{
\node[rotate=45,rectangle,wid=sqrt(2)*\tempsize,hei=sqrt(2)*\tempsize,inner sep=0,transform shape](\atomname)at(0,0){};
\coordinate(\atomname-c) at (0,0);
\coordinate(\atomname-r) at (\tempsize,0);
\coordinate(\atomname-l) at (-\tempsize,0);
\coordinate(\atomname-t) at (0,\tempsize);
\coordinate(\atomname-b) at (0,-\tempsize);
}},
pentagon/.code={

\def\netshapepath{(-126:\tempsize)--(-54:\tempsize)--(18:\tempsize)--(90:\tempsize)--(162:\tempsize)--cycle}
\def\netshapecoords{
\node[regular polygon,regular polygon sides=5,wid=2*\tempsize,inner sep=0,transform shape](\atomname)at(0,0){};
\coordinate(\atomname-c) at (0,0);
\coordinate (\atomname-mb)at(-90:{\tempsize*cos(36)});
\coordinate (\atomname-mbr)at(-18:{\tempsize*cos(36)});
\coordinate (\atomname-mtr)at(54:{\tempsize*cos(36)});
\coordinate (\atomname-mtl)at(126:{\tempsize*cos(36)});
\coordinate (\atomname-mbl)at(-162:{\tempsize*cos(36)});
\coordinate (\atomname-cbr)at(-54:\tempsize);
\coordinate (\atomname-cr)at(18:\tempsize);
\coordinate (\atomname-ct)at(90:\tempsize);
\coordinate (\atomname-cl)at(162:\tempsize);
\coordinate (\atomname-cbl)at(-126:\tempsize);
}},
halfcirc/.code={

\def\netshapepath{(\tempsize,0)arc(0:180:\tempsize)--++(0,-0.04)-|cycle}
\def\netshapecoords{
\node[circle,wid=2*\tempsize,hei=2*\tempsize,inner sep=0,transform shape](\atomname)at(0,0){};
\coordinate(\atomname-c) at (0,0);
\coordinate(\atomname-r) at (\tempsize,0);
\coordinate(\atomname-l) at (-\tempsize,0);
\coordinate(\atomname-t) at (0,\tempsize);
\coordinate(\atomname-b) at (0,0);
}},
void/.code={
\def\netshapepath{}
\def\netshapecoords{
\coordinate(\atomname) at (0,0);
\coordinate(\atomname-c) at (0,0);
}},
labbox/.code={
\def\netshapepath{(0,0)}
\def\netshapecoords{}
\tikzset{netpostdeco/.append style={netbox=#1}}
},
}

\pgfkeys{
/network/atom/.cd,
p/.code={\tikzset{netatompos/.append style={shift={(#1)}}}},
xscale/.code={\tikzset{nettrafo/.append style={xscale=#1}}},
yscale/.code={\tikzset{nettrafo/.append style={yscale=#1}}},
scale/.code={\tikzset{nettrafo/.append style={scale=#1}}},
rot/.code={\tikzset{nettrafo/.append style={rotate=#1}}},
hflip/.style={xscale=-1},
vflip/.style={yscale=-1},
big/.style={scale=3/2},
small/.style={scale=2/3},
huge/.style={scale=9/4},
tiny/.style={scale=4/9},
flat/.style={yscale=2/3},
fflat/.style={yscale=4/9},
}

\tikzset{
netbdstyle/.style={line width=0.15em}, 
netdecstyle/.style={},
netpostdecstyle/.style={},
netbackstyle/.style={white},
}
\pgfkeys{
/network/atom/.cd,
bdstyle/.code={\tikzset{netbdstyle/.style={#1}}},
bdastyle/.code={\tikzset{netbdstyle/.append style={#1}}},
decstyle/.code={\tikzset{netdecstyle/.style={#1}}},
decastyle/.code={\tikzset{netdecstyle/.append style={#1}}},
postdecstyle/.code={\tikzset{netpostdecstyle/.style={#1}}},
postdecastyle/.code={\tikzset{netpostdecstyle/.append style={#1}}},
backstyle/.code={\tikzset{netbackstyle/.style={#1}}},
backastyle/.code={\tikzset{netbackstyle/.append style={#1}}},
style/.style={bdstyle=#1, decstyle=#1, postdecstyle=#1},
astyle/.style={bdastyle=#1, decastyle=#1, postdecastyle=#1},
whiteblack/.style={decstyle=white,backstyle=black},
nobd/.style={bdstyle={line width=0}},
}

\tikzset{
netbscope/.code={\begin{scope}[#1]},
netescope/.code={\end{scope}},
}

\pgfkeys{
/network/atom/.cd,
bscope/.code={\tikzset{netdeco/.append style={netbscope={#1}}}},
escope/.code={\tikzset{netdeco/.append style={netescope}}},
dec/.style 2 args={bscope={#1},#2,escope}, 
vbar/.style={dec={rotate=90}{hbar}},
dcross/.style={dec={rotate=45}{hbar},dec={rotate=-45}{hbar}},
cross/.style={hbar,vbar},
sect6/.style={sect=60},
sect8/.style={sect=45},
sect10/.style={sect=36},
}

\def\regdec#1{\pgfkeys{/network/atom/.cd,#1/.code={\tikzset{netdeco/.append style={net#1}}}}}
\regdec{all}\regdec{rhalf}\regdec{rquart}\regdec{brquart}\regdec{dot}\regdec{spiral}\regdec{swirl}\regdec{hstripe}\regdec{hbar}\regdec{rrey}
\pgfkeys{/network/atom/.cd,sect/.code={\tikzset{netdeco/.append style={netsect=#1}}}}

\tikzset{
netall/.code={\fill[netdecstyle] (-0.3,-0.3)rectangle (0.3,0.3);}, 
netrhalf/.code={\fill[netdecstyle] (0,-0.3)rectangle (0.3,0.3);}, 
netrquart/.code={\fill[netdecstyle] (0.075,-0.3)rectangle (0.3,0.3);}, 
netbrquart/.code={\fill[netdecstyle] (0,0)rectangle (0.3,-0.3);}, 
netsect/.code={\fill[netdecstyle] (0,0)--(0,-0.3)arc(-90:-90+#1:0.3)--cycle;}, 
netdot/.code={\fill[netdecstyle] (0,0)circle(0.07);}, 
netspiral/.code={\draw[netdecstyle] plot [variable=\t,domain=0:4] ({0.075*\t*cos(pi*(\t-0.5) r)},{0.075*\t*sin(pi*(\t-0.5) r)});}, 
netswirl/.code={\fill[netdecstyle] plot [variable=\t,domain=0:2] ({0.15*\t*cos(pi*(\t-0.5) r)},{0.15*\t*sin(pi*(\t-0.5) r)}) arc(-90:-450:0.3)--cycle;}, 
nethstripe/.code={\fill[netdecstyle] (-0.3,-0.05)rectangle(0.3,0.05);}, 
nethbar/.code={\draw[netdecstyle] (-0.3,0)--(0.3,0);}, 
netrrey/.code={\draw[netdecstyle] (0,0)--(0.3,0);} 
}

\pgfkeys{
/network/atom/.cd,
postbscope/.code={\tikzset{netpostdeco/.append style={netbscope={#1}}}},
postescope/.code={\tikzset{netpostdeco/.append style={netescope}}},
postdec/.style 2 args={postbscope={#1},#2,postescope}, 
arc/.code={\tikzset{netpostdeco/.append style={netarc=#1}}},
lab/.code={\tikzset{netpostdeco/.append style={netlab={#1}}}},
shadecirc/.code={\tikzset{netpostdeco/.append style=netshadecirc}},
shaderect/.code={\tikzset{netpostdeco/.append style={netshaderect={#1}}}},
debug/.code={\tikzset{netpostdeco/.append style=netdebug}},
markline/.code 2 args={\tikzset{netpostdeco/.append style={netmarkline={#1}{#2}}}},
postcirc/.code={\tikzset{netpostdeco/.append style=netpostcirc}},
}

\tikzset{
netlab/.code={
\pgfkeys{/network/atom/lab/.cd,#1}
\node[netpostdecstyle] at (\ifdefined\netlabpos\netlabpos\else\netlabang:\netlabdist\fi) {\netlabwrap{\netlabtext}};
},
netarc/.code args={#1:#2:#3}{
\draw[netpostdecstyle] (#1:#3) arc (#1:#2:#3);
},
netshadecirc/.code= {
\fill[opacity=0.4,netpostdecstyle] (0,0)circle(0.4);
},
netpostcirc/.code= {
\draw[netpostdecstyle] (0,0)circle(0.15);
},
netshaderect/.code= {
\fill[rc,opacity=0.4,netpostdecstyle] ($-1*(#1)$) rectangle (#1);
},
netdebug/.code= {
\node[red] at (0,0){\atomname};
},
netmarkline/.code 2 args= {
\draw (\atomname)edge[mark={#2}]++(#1);
},
}

\def\netlabwrap#1{#1}
\pgfkeys{
/network/atom/lab/.cd,
t/.code={\def\netlabtext{#1}}, 
p/.code={\def\netlabpos{#1}}, 
ang/.code={\def\netlabang{#1}},
dist/.code={\def\netlabdist{#1}},
wrap/.code={\def\netlabwrap##1{#1}}, 
}

%% file: figure-stability-syndrome.tex
\begingroup
\definecolor{syndromeblue}{RGB}{28,43,190}
\begin{tikzpicture}[
  x=.75cm,y=1cm,
  >=Stealth,
  syndrome/.style={draw=syndromeblue,line width=1.326pt,line cap=round},
  cluster/.style={draw=black,dash pattern=on 3pt off 2.5pt,
    line width=.65pt,rounded corners=9pt},
  every node/.style={font=\small,scale=2.4}
]
  \filldraw[fill=gray!15,draw=black,line width=.7pt]
    (0,0) rectangle (7.2,5.4);
  \node[font=\large,scale=.8] at (3.6,.38) {$B$};
  \draw[->,line width=.7pt] (-.42,.1)--(-.42,12.30);
  \node[anchor=east] at (-.50,12.15) {$t$};

  \draw[syndrome] plot[smooth,tension=.65] coordinates {(0.582,0.51) (0.5496,0.766) (0.6072,1.022) (0.582,1.2292)
    (0.6324,1.3996) (0.564,1.88366) (0.582,2.3268)};
  \draw[syndrome] plot[smooth,tension=.65] coordinates {(0.806,0.91) (0.772,1.054) (0.66912,1.134) (0.582,1.2292)};
  \draw[syndrome] plot[smooth,tension=.65] coordinates {(1.478,0.822) (1.47728,1.046) (1.366,1.2388) (1.30344,1.3492)
    (0.98136,1.67847) (0.93704,1.87331) (0.582,2.3268)};
  \draw[syndrome] plot[smooth,tension=.65] coordinates {(1.67,1.094) (1.544,1.1844) (1.366,1.2388)};
  \draw[syndrome] plot[smooth,tension=.65] coordinates
    {(1.68,1.4) (1.43,1.55) (1.21,1.725) (.93704,1.87331)};
  \draw[syndrome] plot[smooth,tension=.65] coordinates
    {(1.00,1.32) (.94,1.50777) (.73,1.70175) (.564,1.88366)};
  \draw[cluster] (0.382,0.334) rectangle (1.838,2.4868);
  \node[anchor=south,font=\large,scale=.8] at (1.11,2.5368) {$X$};

  \draw[syndrome] plot[smooth,tension=.65] coordinates {(2.55,2.4) (2.496,2.83) (2.5905,3.19) (2.55,3.6)
    (2.6265,3.97) (2.514,4.43) (2.55,4.95) (2.631,5.37)
    (2.532,5.94) (2.595,6.31) (2.4825,6.88) (2.568,7.22)
    (2.523,7.85) (2.6175,8.37) (2.55,9.15)};
  \draw[syndrome] plot[smooth,tension=.65] coordinates {(2.86,3.1) (2.8346,3.28) (2.6755,3.43) (2.55,3.6)};
  \draw[syndrome] plot[smooth,tension=.65] coordinates {(3.76,2.8) (3.765,3.11) (3.6271,3.55) (3.6334,3.78)
    (3.45,4.12) (3.1688,4.35) (3.0766,4.55) (2.55,4.95)};
  \draw[syndrome] (4.04,3.55) .. controls (3.87,3.77) and (3.64,3.96) .. (3.45,4.12);

  \draw[syndrome] plot[smooth,tension=.65] coordinates {(5.05,1.7) (5.113,2.1) (5.0095,2.54) (5.0815,2.93)
    (5.05,3.3) (4.978,3.72) (5.05,4.35) (5.131,4.78)
    (5.0185,5.13) (5.077,5.7) (4.9825,6.14) (5.0545,6.72)
    (5.1265,7.07) (5.032,7.61) (5.0995,8.25) (5.0095,8.69)
    (5.05,9.15)};
  \draw[syndrome] plot[smooth,tension=.65] coordinates {(5.35,2.8) (5.3134,3.02) (5.1646,3.13) (5.05,3.3)};
  \draw[syndrome] plot[smooth,tension=.65] coordinates {(6.29,2.05) (6.2879,2.43) (6.1252,2.9) (6.1218,3.19)
    (5.92,3.55) (5.631,3.8) (5.5548,3.98) (5.05,4.35)};
  \draw[syndrome] (6.58,3) .. controls (6.43,3.25) and (6.13,3.30) .. (5.92,3.55);

  \draw[syndrome] plot[smooth,tension=.65] coordinates
    {(3.22,2.70) (3.27,3.12) (3.24,3.57) (3.45,4.12)};
  \draw[syndrome] plot[smooth,tension=.65] coordinates
    {(4.32,4.00) (4.02,4.35) (3.31,4.60) (2.55,4.95)};
  \draw[syndrome] plot[smooth,tension=.65] coordinates
    {(5.68,2.30) (5.75,2.70) (5.73,3.06) (5.92,3.55)};
  \draw[syndrome] plot[smooth,tension=.65] coordinates
    {(6.80,3.50) (6.39,3.85) (5.70,4.06) (5.05,4.35)};

  \draw[syndrome] plot[smooth,tension=.65] coordinates
    {(.60,4.35) (.67,4.76) (1.00,5.13) (1.50,5.55)
     (1.64,5.94) (2.05,6.36) (2.4825,6.88)};
  \draw[syndrome] plot[smooth,tension=.65] coordinates
    {(1.15,5.05) (1.22,5.29) (1.50,5.55)};
  \draw[syndrome] plot[smooth,tension=.65] coordinates
    {(1.70,4.70) (1.73,5.00) (1.61,5.28) (1.50,5.55)};

  \draw[syndrome] plot[smooth,tension=.65] coordinates
    {(6.86,4.40) (6.80,4.82) (6.57,5.19) (6.30,5.65)
     (6.04,5.96) (5.58,6.24) (5.0545,6.72)};
  \draw[syndrome] plot[smooth,tension=.65] coordinates
    {(6.10,4.70) (6.06,5.03) (6.19,5.32) (6.30,5.65)};

  \draw[syndrome] plot[smooth,tension=.65] coordinates {(2.55,9.15) (2.9158,9.4) (3.0375,9.6) (3.7364,10.05)
    (4.015,10.34) (4.3902,10.53) (4.5971,10.76) (5.05,11)};
  \draw[syndrome] plot[smooth,tension=.65] coordinates {(5.05,9.15) (5.113,9.52) (5.014,9.91) (5.0815,10.26)
    (4.987,10.66) (5.05,11) (5.05,11.35)};

  \draw[syndrome,dash pattern=on 4pt off 3pt,->]
    (5.05,11.35)--(3.75,12.05);
  \node[anchor=west] at (4.65,12.05) {towards $X$};

  \draw[cluster] (2.27,1.46)--(6.94,1.46)--(6.94,9.02)
    --(.30,9.02)--(.30,3.90)--(2.27,3.90)--cycle;
  \node[anchor=west,inner sep=2pt,font=\large,scale={1/1.15}] at (7.10,6.75) {$C_{t^{\mathsf R}(B)}$};
  \draw[densely dashed,line width=.65pt] (-.17,9.15)--(7.30,9.15);
  \node[anchor=south east,fill=white,inner sep=2pt]
    at (7.28,9.23) {$t^{\mathsf R}(B)$};
\end{tikzpicture}
\endgroup

%% file: main.bib
@misc{Kitaev_2003,
      title={Fault-tolerant quantum computation by anyons},
      author={A. Yu. Kitaev},
      year={1997},
      eprint={quant-ph/9707021},
      archivePrefix={arXiv},
      primaryClass={quant-ph},
      doi={10.1016/S0003-4916(02)00018-0},
      url={https://arxiv.org/abs/quant-ph/9707021},
}

@misc{Levin_2005,
      title={String-net condensation: A physical mechanism for topological phases},
      author={Michael A. Levin and Xiao-Gang Wen},
      year={2004},
      eprint={cond-mat/0404617},
      archivePrefix={arXiv},
      primaryClass={cond-mat.str-el},
      doi={10.1103/PhysRevB.71.045110},
      url={https://arxiv.org/abs/cond-mat/0404617},
}

@article{TURAEV1992865,
title = {State sum invariants of 3-manifolds and quantum 6j-symbols},
journal = {Topology},
volume = {31},
number = {4},
pages = {865-902},
year = {1992},
issn = {0040-9383},
doi = {10.1016/0040-9383(92)90015-A},
url = {https://www.sciencedirect.com/science/article/pii/004093839290015A},
author = {V.G. Turaev and O.Y. Viro}
}

@misc{Koenig_2010,
      title={Quantum computation with {Turaev-Viro} codes},
      author={Robert Koenig and Greg Kuperberg and Ben W. Reichardt},
      year={2010},
      eprint={1002.2816},
      archivePrefix={arXiv},
      primaryClass={quant-ph},
      doi={10.1016/j.aop.2010.08.001},
      url={https://arxiv.org/abs/1002.2816},
}

@misc{kirillov2011stringnetmodelturaevviroinvariants,
      title={String-net model of {Turaev-Viro} invariants},
      author={Alexander {Kirillov Jr}},
      year={2011},
      eprint={1106.6033},
      archivePrefix={arXiv},
      primaryClass={math.AT},
      url={https://arxiv.org/abs/1106.6033},
}

@misc{Bonesteel_2012,
      title={Quantum Circuits for Measuring {Levin-Wen} Operators},
      author={N. E. Bonesteel and D. P. DiVincenzo},
      year={2012},
      eprint={1206.6048},
      archivePrefix={arXiv},
      primaryClass={quant-ph},
      doi={10.1103/PhysRevB.86.165113},
      url={https://arxiv.org/abs/1206.6048},
}

@misc{Hu_2018,
      title={Full Dyon Excitation Spectrum in Generalized {Levin-Wen} Models},
      author={Yuting Hu and Nathan Geer and Yong-Shi Wu},
      year={2018},
      eprint={1502.03433},
      archivePrefix={arXiv},
      primaryClass={cond-mat.str-el},
      doi={10.1103/PhysRevB.97.195154},
      url={https://arxiv.org/abs/1502.03433},
}

@misc{schotte2021quantumerrorcorrectionthresholds,
      title={Quantum error correction thresholds for the universal {Fibonacci Turaev-Viro} code},
      author={Alexis Schotte and Guanyu Zhu and Lander Burgelman and Frank Verstraete},
      year={2021},
      eprint={2012.04610},
      archivePrefix={arXiv},
      primaryClass={quant-ph},
      url={https://arxiv.org/abs/2012.04610},
}

@misc{Duclos_Cianci_2010,
      title={Fast Decoders for Topological Quantum Codes},
      author={Guillaume Duclos-Cianci and David Poulin},
      year={2010},
      eprint={0911.0581},
      archivePrefix={arXiv},
      primaryClass={quant-ph},
      doi={10.1103/PhysRevLett.104.050504},
      url={https://arxiv.org/abs/0911.0581},
}

@misc{ducloscianci2013faulttolerantrenormalizationgroupdecoder,
      title={Fault-Tolerant Renormalization Group Decoder for {Abelian} Topological Codes},
      author={Guillaume Duclos-Cianci and David Poulin},
      year={2013},
      eprint={1304.6100},
      archivePrefix={arXiv},
      primaryClass={quant-ph},
      url={https://arxiv.org/abs/1304.6100},
}

@misc{Bravyi_2013,
      title={Analytic and numerical demonstration of quantum self-correction in the {3D Cubic Code}},
      author={Sergey Bravyi and Jeongwan Haah},
      year={2011},
      eprint={1112.3252},
      archivePrefix={arXiv},
      primaryClass={quant-ph},
      doi={10.1103/PhysRevLett.111.200501},
      url={https://arxiv.org/abs/1112.3252},
}

@misc{Wootton_2016,
      title={Active error correction for {Abelian} and non-{Abelian} anyons},
      author={James R. Wootton and Adrian Hutter},
      year={2015},
      eprint={1506.00524},
      archivePrefix={arXiv},
      primaryClass={quant-ph},
      doi={10.1103/PhysRevA.93.022318},
      url={https://arxiv.org/abs/1506.00524},
}

@misc{Dauphinais_2017,
      title={Fault-Tolerant Quantum Error Correction for non-{Abelian} Anyons},
      author={Guillaume Dauphinais and David Poulin},
      year={2017},
      eprint={1607.02159},
      archivePrefix={arXiv},
      primaryClass={quant-ph},
      doi={10.1007/s00220-017-2923-9},
      url={https://arxiv.org/abs/1607.02159},
}

@misc{Brown_2020,
      title={A fault-tolerant non-{Clifford} gate for the surface code in two dimensions},
      author={Benjamin J. Brown},
      year={2021},
      eprint={1903.11634},
      archivePrefix={arXiv},
      primaryClass={quant-ph},
      doi={10.1126/sciadv.aay4929},
      url={https://arxiv.org/abs/1903.11634},
}

@misc{Bauer_2024,
      title={Topological error correcting processes from fixed-point path integrals},
      author={Andreas Bauer},
      year={2024},
      eprint={2303.16405},
      archivePrefix={arXiv},
      primaryClass={quant-ph},
      doi={10.22331/q-2024-03-20-1288},
      url={https://arxiv.org/abs/2303.16405},
}

@misc{Bauer_2025,
      title={Low-overhead non-{Clifford} fault-tolerant circuits for all non-chiral abelian topological phases},
      author={Andreas Bauer},
      year={2025},
      eprint={2403.12119},
      archivePrefix={arXiv},
      primaryClass={quant-ph},
      doi={10.22331/q-2025-03-25-1673},
      url={https://arxiv.org/abs/2403.12119},
}

@misc{bauer2025xyfloquetcodesimple,
      title={The x+y {Floquet} code: A simple example for topological quantum computation in the path integral approach},
      author={Andreas Bauer},
      year={2025},
      eprint={2408.07265},
      archivePrefix={arXiv},
      primaryClass={quant-ph},
      url={https://arxiv.org/abs/2408.07265},
}

@misc{davydova2025universalfaulttolerantquantum,
      title={Universal fault tolerant quantum computation in {2D} without getting tied in knots},
      author={Margarita Davydova and Andreas Bauer and Julio C. Magdalena de la Fuente and Mark Webster and Dominic J. Williamson and Benjamin J. Brown},
      year={2025},
      eprint={2503.15751},
      archivePrefix={arXiv},
      primaryClass={quant-ph},
      doi={10.1103/vrty-qs5h},
      url={https://arxiv.org/abs/2503.15751},
}

@misc{delafuente2026highthresholddecodingnonpaulicodes,
      title={High-threshold decoding of non-{Pauli} codes for {2D} universality},
      author={Julio C. Magdalena de la Fuente and Noa Feldman and Jens Eisert and Andreas Bauer},
      year={2026},
      eprint={2604.02033},
      archivePrefix={arXiv},
      primaryClass={quant-ph},
      url={https://arxiv.org/abs/2604.02033},
}

@misc{Barrett1993,
      title={Invariants of Piecewise-Linear 3-Manifolds},
      author={John W. Barrett and Bruce W. Westbury},
      year={1995},
      eprint={hep-th/9311155},
      archivePrefix={arXiv},
      primaryClass={hep-th},
      url={https://arxiv.org/abs/hep-th/9311155},
}

@phdthesis{thesis,
  author = {Bauer, Andreas},
  title  = {Topological Phases, Fixed-Point Models, Extended {TQFT}, and Fault-Tolerant Quantum Computation -- Tensors in Spacetime},
  school = {Freie Universit\"at Berlin},
  year   = {2023},
  doi    = {10.17169/refubium-45650},
  url    = {https://refubium.fu-berlin.de/handle/fub188/45937},
  note   = {{PDF}: \url{https://refubium.fu-berlin.de/bitstream/handle/fub188/45937/Dissertation_Andreas_Bauer.pdf?sequence=4}},
}

@misc{dennis2001,
      title={Topological quantum memory}, 
      author={Eric Dennis and Alexei Kitaev and Andrew Landahl and John Preskill},
      year={2001},
      eprint={quant-ph/0110143},
      archivePrefix={arXiv},
      primaryClass={quant-ph},
      doi={10.1063/1.1499754},
      url={https://arxiv.org/abs/quant-ph/0110143}, 
}

@misc{horsman,
      title={Surface code quantum computing by lattice surgery}, 
      author={Dominic Horsman and Austin G. Fowler and Simon Devitt and Rodney Van Meter},
      year={2013},
      eprint={1111.4022},
      archivePrefix={arXiv},
      primaryClass={quant-ph},
      doi={10.1088/1367-2630/14/12/123011},
      url={https://arxiv.org/abs/1111.4022}, 
}

@misc{bombin2009quantummeasurementsgatescode,
      title={Quantum Measurements and Gates by Code Deformation}, 
      author={H. Bombin and M. A. Martin-Delgado},
      year={2009},
      eprint={0704.2540},
      archivePrefix={arXiv},
      primaryClass={quant-ph},
      doi={10.1088/1751-8113/42/9/095302},
      url={https://arxiv.org/abs/0704.2540}, 
}

@misc{fowler2012surfacecodespracticallargescale,
      title={Surface codes: Towards practical large-scale quantum computation}, 
      author={Austin G. Fowler and Matteo Mariantoni and John M. Martinis and Andrew N. Cleland},
      year={2012},
      eprint={1208.0928},
      archivePrefix={arXiv},
      primaryClass={quant-ph},
      doi={10.1103/PhysRevA.86.032324},
      url={https://arxiv.org/abs/1208.0928}, 
}

@misc{bravyi2004universalquantumcomputationideal,
      title={Universal Quantum Computation with ideal {Clifford} gates and noisy ancillas}, 
      author={Sergei Bravyi and Alexei Kitaev},
      year={2004},
      eprint={quant-ph/0403025},
      archivePrefix={arXiv},
      primaryClass={quant-ph},
      doi={10.1103/PhysRevA.71.022316},
      url={https://arxiv.org/abs/quant-ph/0403025}, 
}

@misc{freedman2000modularfunctoruniversalquantum,
      title={A modular functor which is universal for quantum computation}, 
      author={Michael Freedman and Michael Larsen and Zhenghan Wang},
      year={2000},
      eprint={quant-ph/0001108},
      archivePrefix={arXiv},
      primaryClass={quant-ph},
      url={https://arxiv.org/abs/quant-ph/0001108}, 
}

@misc{nayak2008nonabeliananyonstopologicalquantum,
      title={Non-{Abelian} Anyons and Topological Quantum Computation}, 
      author={Chetan Nayak and Steven H. Simon and Ady Stern and Michael Freedman and Sankar Das Sarma},
      year={2008},
      eprint={0707.1889},
      archivePrefix={arXiv},
      primaryClass={cond-mat.str-el},
      doi={10.1103/RevModPhys.80.1083},
      url={https://arxiv.org/abs/0707.1889}, 
}

@misc{lyons2026quantumcomputinganyonsfault,
      title={Quantum computing with anyons is fault tolerant}, 
      author={Anasuya Lyons and Benjamin J. Brown},
      year={2026},
      eprint={2602.11258},
      archivePrefix={arXiv},
      primaryClass={quant-ph},
      url={https://arxiv.org/abs/2602.11258}, 
}

@misc{bombin2019,
      title={{2D} quantum computation with {3D} topological codes}, 
      author={Hector Bombin},
      year={2018},
      eprint={1810.09571},
      archivePrefix={arXiv},
      primaryClass={quant-ph},
      url={https://arxiv.org/abs/1810.09571}, 
}

@misc{twisted_color_circuits,
      title={Planar fault-tolerant circuits for non-{Clifford} gates on the {2D} color code}, 
      author={Andreas Bauer and Julio C. Magdalena de la Fuente},
      year={2026},
      eprint={2505.05175},
      archivePrefix={arXiv},
      primaryClass={quant-ph},
      doi={10.1103/p7c9-x1m9},
      url={https://arxiv.org/abs/2505.05175}, 
}

@misc{schotte2022faulttoleranterrorcorrectionuniversal,
      title={Fault-tolerant error correction for a universal non-{Abelian} topological quantum computer at finite temperature}, 
      author={Alexis Schotte and Lander Burgelman and Guanyu Zhu},
      year={2022},
      eprint={2301.00054},
      archivePrefix={arXiv},
      primaryClass={quant-ph},
      url={https://arxiv.org/abs/2301.00054}, 
}

@misc{Wootton2013,
      title={Decoding non-{Abelian} topological quantum memories}, 
      author={James R. Wootton and Jan Burri and Sofyan Iblisdir and Daniel Loss},
      year={2014},
      eprint={1310.3846},
      archivePrefix={arXiv},
      primaryClass={quant-ph},
      doi={10.1103/PhysRevX.4.011051},
      url={https://arxiv.org/abs/1310.3846}, 
}

@misc{Hutter2015,
      title={Continuous error correction for {Ising} anyons}, 
      author={Adrian Hutter and James R. Wootton},
      year={2016},
      eprint={1508.04033},
      archivePrefix={arXiv},
      primaryClass={quant-ph},
      doi={10.1103/PhysRevA.93.042327},
      url={https://arxiv.org/abs/1508.04033}, 
}

@misc{gacs,
      title={Reliable Cellular Automata with Self-Organization}, 
      author={Peter Gacs},
      year={2024},
      eprint={math/0003117},
      archivePrefix={arXiv},
      primaryClass={math.PR},
      doi={10.1023/A:1004823720305},
      url={https://arxiv.org/abs/math/0003117}, 
}

@book{harrington2004analysis,
  title={Analysis of quantum error-correcting codes: symplectic lattice codes and toric codes},
  author={Harrington, James William},
  year={2004},
  publisher={California Institute of Technology}
}

@article{balasubramanian2026local,
  title={A local automaton for the 2D toric code},
  author={Balasubramanian, Shankar and Davydova, Margarita and Lake, Ethan},
  journal={Quantum},
  volume={10},
  pages={2125},
  year={2026},
  publisher={Verein zur F{\"o}rderung des Open Access Publizierens in den Quantenwissenschaften},
  eprint={2412.19803},
  archivePrefix={arXiv},
  primaryClass={quant-ph},
  doi={10.22331/q-2026-06-02-2125},
  url={https://arxiv.org/abs/2412.19803},
}

@inproceedings{Gacs1983,
  author    = {G{\'a}cs, Peter},
  title     = {Reliable Computation with Cellular Automata},
  booktitle = {Proceedings of the Fifteenth Annual {ACM} Symposium on Theory of Computing},
  series    = {STOC '83},
  year      = {1983},
  pages     = {32--41},
  publisher = {Association for Computing Machinery},
  doi       = {10.1145/800061.808730}
}

@article{durand-romashchenko-shen,
  author        = {Durand, Bruno and Romashchenko, Andrei and Shen, Alexander},
  title         = {Fixed-point tile sets and their applications},
  journal       = {Journal of Computer and System Sciences},
  volume        = {78},
  number        = {3},
  pages         = {731--764},
  year          = {2012},
  doi           = {10.1016/j.jcss.2011.11.001},
  eprint        = {0910.2415},
  archivePrefix = {arXiv},
  primaryClass  = {cs.CC}
}

@article{Cong2017,
  title = {Universal Quantum Computation with Gapped Boundaries},
  author = {Cong, Iris and Cheng, Meng and Wang, Zhenghan},
  journal = {Phys. Rev. Lett.},
  volume = {119},
  issue = {17},
  pages = {170504},
  numpages = {5},
  year = {2017},
  month = {Oct},
  publisher = {American Physical Society},
  doi = {10.1103/PhysRevLett.119.170504},
  url = {https://link.aps.org/doi/10.1103/PhysRevLett.119.170504}
}

@misc{huang2025generatinglogicalmagicstates,
      title={Generating logical magic states with the aid of non-Abelian topological order}, 
      author={Sheng-Jie Huang and Yanzhu Chen},
      year={2025},
      eprint={2502.00998},
      archivePrefix={arXiv},
      primaryClass={quant-ph},
      url={https://arxiv.org/abs/2502.00998}, 
}

@article{Bombin2010,
Author = {H. Bombin},
Title = {Topological Order with a Twist: Ising Anyons from an Abelian Model},
Year = {2010},
Eprint = {arXiv:1004.1838},
journal = {Phys.Rev.Lett.},
volume={105},
pages={030403},
Doi = {10.1103/PhysRevLett.105.030403},
}

@article{Brown2016,
Author = {Benjamin J. Brown and Katharina Laubscher and Markus S. Kesselring and James R. Wootton},
Title = {Poking holes and cutting corners to achieve Clifford gates with the surface code},
Year = {2017},
Eprint = {arXiv:1609.04673},
journal = {Phys. Rev. X},
volume= {7},
pages={021029},
Doi = {10.1103/PhysRevX.7.021029},
}

@article{Magdalena2023,
Author = {Julio C. Magdalena de la Fuente and Jens Eisert and Andreas Bauer},
Title = {Bulk-to-boundary anyon fusion from microscopic models},
Year = {2023},
Eprint = {arXiv:2302.01835},
journal = {J. Math. Phys.},
volume={64},
pages={111904},
Doi = {10.1063/5.0147335},
}

@misc{Sajith2025,
Author = {Rohith Sajith and Zijian Song and Brenden Roberts and Varun Menon and Yabo Li},
Title = {Non-Clifford gates between stabilizer codes via non-Abelian topological order},
Year = {2025},
Eprint = {arXiv:2505.18265},
}

@article{gacs1988simple,
  title={A simple three-dimensional real-time reliable cellular array},
  author={G{\'a}cs, Peter and Reif, John},
  journal={Journal of Computer and System Sciences},
  volume={36},
  number={2},
  pages={125--147},
  year={1988},
  publisher={Elsevier}
}

@online{gacsrecife,
  title={Reliably computing cellular automata},
  author={G{\'a}cs, Peter},
  url={https://cs-web.bu.edu/faculty/gacs/papers/Recife_2012_lect.pdf},
  year={2012}
}
